\documentclass[useAMS,usenatbib]{mnras}

\usepackage{graphicx}
\usepackage{color}
\usepackage{amsmath}
\usepackage{amssymb}
\usepackage{caption}

\title[SCUBA-2 imaging of colour-selected \textit{Herschel} sources]{Red, redder, reddest: SCUBA-2 imaging of colour-selected \textit{Herschel} sources}
\author[S. Duivenvoorden et al.]{S. Duivenvoorden$^{1}$\thanks{E-mail: \href{mailto:S.Duivenvoorden@Sussex.ac.uk}{S.Duivenvoorden@Sussex.ac.uk}},  
S. Oliver$^{1}$, 
J. M. Scudder$^{1}$,
J. Greenslade$^{2}$, 
D. A. Riechers$^{3}$, \newauthor  
S. M. Wilkins$^{1}$, 
V. Buat$^{4}$, 
S. C. Chapman$^{5}$,
D. L. Clements$^{2}$,  
A. Cooray$^6$,  \newauthor 
K. E. K. Coppin$^{7}$,  
H. Dannerbauer$^{8,9}$, 
G. De Zotti$^{10}$,  
J. S.  Dunlop$^{11}$,   
S. A. Eales$^{12}$,\newauthor  
A. Efstathiou$^{13}$,  
D. Farrah$^{14}$,   
J. E. Geach$^{7}$, 
W. S. Holland$^{15,11}$,
P. D. Hurley$^{1}$, \newauthor 
R. J. Ivison$^{16,11}$,  
L. Marchetti$^{17,18,19}$,  
G. Petitpas$^{20}$,  
M. T. Sargent$^1$,  
D. Scott$^{21}$, \newauthor  
M. Symeonidis$^{22}$, 
M. Vaccari$^{18,19}$,
J. D. Vieira$^{23}$,  
L. Wang$^{24,25}$,
J. Wardlow$^{26}$  \newauthor  and  
M. Zemcov$^{27}$   
\\ 
Affiliations are listed at the end of the paper
}

\date{Accepted xxx. Received xxx; in original form xxx}
\pubyear{2018}
\begin{document}
\label{firstpage}
\pagerange{\pageref{firstpage}--\pageref{lastpage}} 
\maketitle

\begin{abstract}
High-redshift, luminous, dusty star forming galaxies (DSFGs) constrain the extremity of galaxy formation theories.  The most extreme are discovered through follow-up on candidates in large area surveys. Here we present 850 $\mu$m SCUBA-2 follow-up observations of 188 red DSFG candidates from the \textit{Herschel} Multi-tiered Extragalactic Survey (HerMES) Large Mode Survey, covering 274 deg$^2$. We detected 87 per cent  with a signal-to-noise ratio $>$ 3 at 850~$\mu$m. We introduce a new method for incorporating the confusion noise in our spectral energy distribution fitting by sampling correlated flux density fluctuations from a confusion limited map.  The new 850~$\mu$m data provide a better constraint on the  photometric redshifts of the candidates, with photometric redshift errors decreasing from $\sigma_z/(1+z)\approx0.21$ to $0.15$.  Comparison spectroscopic redshifts also found little bias ($\langle  (z-z_{\rm spec})/(1+z_{\rm spec})\rangle = 0.08 $). The mean photometric redshift is found to be 3.6 with a dispersion of $0.4$ and we identify 21 DSFGs with a high probability of lying at $z > 4$. After simulating our selection effects we find number counts are consistent with phenomenological galaxy evolution models. 
There is a statistically significant excess of WISE-1 and SDSS sources near our red galaxies, giving a strong indication that lensing may explain some of the apparently extreme objects.  Nevertheless, our sample should include examples of galaxies with the highest star formation rates in the Universe ($\gg10^3$ M$_\odot$yr$^{-1}$). 
\end{abstract}

\begin{keywords}
Infrared: galaxies -- galaxies: high-redshift -- galaxies: starburst.
\end{keywords}


\section{Introduction}

Over the last few decades, great progress has been made in understanding the star formation history of the Universe (see e.g. review by \citealt{2014ARA&A..52..415M}).  It has become apparent that observing at UV and optical wavelengths is insufficient as a large fraction of the star formation is obscured, resulting in  dusty star forming galaxies (DSFGs see e.g reviews by \protect\citealt{1984ApJ...287...95L, 1996A&A...315L..32C,1997ApJ...490L...5S,2013A&A...554A..70B} and \citealt{2014PhR...541...45C}). The most extreme forms of obscured star formation at high-redshift still pose serious challenges to galaxy evolution models \citep[e.g.][]{2005MNRAS.356.1191B,2010MNRAS.405....2L,2010MNRAS.401.1613N,2013MNRAS.434.2572H, beth}. The discovery and characterisation of the rarest and most extreme galaxies (star formation rates, SFR, $\gg10^3$ M$_\odot$yr$^{-1}$, number densities $\ll 10^{-4} \,{\rm Mpc}^{-3}$, \citealt{2013MNRAS.432...23G}) is thus an important goal, but requires large volume surveys at long wavelengths.

This is now possible with deep large-area surveys ($\gg10$ deg$^2$) at far infrared (FIR) and sub-mm wavelengths with e.g. the South Pole Telescope \citep[SPT,][]{2011PASP..123..568C} and   the \textit{Herschel Space Observatory} \citep[][]{2010A&A...518L...1P}.   

Follow-up of SPT sources has been very successful in finding high redshift DSFGs \citep{2013Natur.495..344V,2013ApJ...767...88W,2016ApJ...822...80S,2017ApJ...842L..15S}. The SPT source selection at a wavelength of 1.4 mm has however a broader redshift distribution than {\it Herschel}  detected sources \citep{2012ApJ...756..101G} 

{\it Herschel} surveys cover a huge area $\sim1300\,{\rm deg}^2$ (the largest being HerMES \citealt{2012MNRAS.424.1614O} and \textit{H}-ATLAS \citealt{2010PASP..122..499E}) and while  most detections are associated with $z \sim 1-2$ starburst galaxies (e.g, \citealt{2012ApJ...761..139C}; \citealt{2012ApJ...761..140C}) it has  been clearly demonstrated that selecting those with red colours is extremely efficient for identifying a tail extending towards higher redshift ($z > 4$)  \citep{2011ApJ...740...63C,2013Natur.496..329R,2014ApJ...780...75D,2016MNRAS.462.1989A,2016ApJ...832...78I,2017ApJ...850....1R}. The challenge now is using these very large {\em Herschel} surveys to find and systematically study, large, homogeneous samples of rare, extremely luminous, $z > 4$ sources.

\cite{2016MNRAS.462.1989A} probed this high-redshift population in the largest HerMES field, the HerMES Large Mode Survey (HeLMS, covering approximately 300 deg$^2$) by selecting all bright ``500 $\mu$m riser" ($S_{500} > S_{350} > S_{250}$) DSFGs candidates.  This sample was selected over an area a factor of 13 times larger than previous 500 $\mu$m riser HerMES surveys \citep{2014ApJ...780...75D}. 
The number of sources that fulfilled these criteria (477) is an order of magnitude higher than predicted by galaxy evolution models \citep{2011A&A...529A...4B,2012ApJ...757L..23B,2014ApJ...780...75D}

Another large 600 deg$^2$ red DSFGs search in the \textit{H}-ATLAS survey \citep{2016ApJ...832...78I} used a $3.5\sigma$ (30 mJy) detection threshold at $S_{500}$ in combination with $S_{500}/S_{250} \ge 1.5$ and $S_{500}/S_{350} \ge 0.85$ colour selection criteria to obtain a sample of 7961 candidate high-redshift DSFGs. After a visual inspection \citep{2016ApJ...832...78I} a sub-sample of 109 DSFGs, candidates were selected for follow up at longer wavelengths with SCUBA-2 or LABOCA. 

All these red sources are candidates for  high-luminosity sources. Some, particularly those with a flux density at $S_{500} >$ 100 mJy, are likely to be strongly gravitationally lensed \citep{2010Sci...330..800N,2011ApJ...732L..35C,2016ApJ...823...17N,2017MNRAS.465.3558N} others may be blends \citep[e.g.][]{2016MNRAS.460.1119S}. Nevertheless, they are extremely interesting because, those that are not lensed, blended or otherwise boosted  may represent the most active galaxies in cosmic history.

In this work, we present a follow-up study of 188 of  the brightest 200 ($S_{500} > 63$ mJy), of the 477 \cite{2016MNRAS.462.1989A} objects using SCUBA-2 \citep{2013MNRAS.430.2513H} on the James Clerk Maxwell Telescope (JCMT). With the addition of the $S_{850}$ data provided by SCUBA-2 we have a better constraint on both the FIR luminosities and the redshifts of these DSFGs and prepare the way for high resolution follow-up. 

With our sample of 188 galaxies observed by SCUBA-2 we roughly double the number of 500 $\mu$m riser galaxies possessing longer submm wavelength data. 

The format of this paper is as follows. We describe the data in Section \ref{sec:data}. We describe our methods for determining the photometric redshifts, FIR luminosities and star formation rates (SFRs) in Section \ref{sec:param}. The results are described in Section \ref{sec:results}, and the discussion and conclusions in Sections \ref{discussion} and \ref{sec:conclusion}, respectively. We use a standard flat cosmology with $\Omega_{\rm M} = 0.3$ and $H_0 = 70\ \text{km s}^{-1} \text{Mpc}^{-1}$.

\section{Data} \label{sec:data}

\subsection{Selecting high-redshift dusty galaxies in HeLMS}

We use the red HeLMS sample identified in \cite{2016MNRAS.462.1989A} and below follows a short summary of their selection. The area mapped by HeLMS is a 300 deg$^2$ equatorial field which is part of the HerMES project. The observations were performed using the SPIRE instrument \citep{2010A&A...518L...3G} on board the \textit{Herschel Space Observatory}. Some parts of the HeLMS field were masked. Edge effects, along with a ``seagull-shaped" region of strong Galactic cirrus were removed, leaving a useful area of 274 deg$^2$. 

Sources were detected using a map-based search method described in \cite{2016MNRAS.462.1989A}, similar to what was used in \cite{2014ApJ...780...75D}, instead of sources from the HerMES catalogue derived directly from the 250 $\mu$m map (Clarke et al. in prep.). For a description of how the sources were selected and the exact spatial filters adopted we refer the reader to \cite{2016MNRAS.462.1989A}, but we give a brief description here for completeness. 

The SPIRE 250, 350, 500 $\mu$m maps are created with the same pixel size (6 arcsec) and (for source detection only) smoothed to the same resolution using an optimal filter for easy comparison between wavebands \citep{2011MNRAS.411..505C}. The local background is removed by smoothing the maps with a 2D median boxcar filter on 3 arcmin scales to remove any cirrus contamination. The filters are also applied to the error map to find the typical instrumental noise in the smoothed map. The $1\sigma$ instrumental noise values are 7.56, 6.33 and 7.77 mJy, in the 250, 350, and 500 $\mu$m SPIRE bands.

The confusion noise ($\sigma_{\rm {conf}}$) in the SPIRE map is caused by sources which emit at all three SPIRE wavelengths. This causes the confusion noise to be correlated between wavelengths. This information is used to construct a difference map (D) from the SPIRE 500 $\mu$m ($M_{500}$) and SPIRE 250 $\mu$m ($M_{250}$) maps with a reduced confusion limit \citep{2014ApJ...780...75D};  
\begin{equation}
  D = \sqrt{1-k^2}M_{500} -kM_{250}
\end{equation}  
with a \textit{k} value of 0.392 to maximize the \textit{D}/$\sigma_{\rm {conf}}$. This \textit{D}-map has a confusion noise of 3.50 mJy, which is much lower than in the three smoothed SPIRE bands (13.66, 11.21, 6.98  mJy at 250, 350 and 500 $\mu$m, respectively).

The bright peaks in the \textit{D}-map are selected with a 4$\sigma$ cut-off at 34 mJy. At these positions the SPIRE flux densities are determined from the (higher resolution) nominal resolution map while taking into account the positional uncertainty of 6 arcsec (as measured with simulations in \citealt{2016MNRAS.462.1989A}). From these flux densities a catalogue of $S_{500} > S_{350} > S_{250}$ sources is created. There is no requirement for a detection in both 250 and 350 $\mu$m,  in order to avoid biassing the selection against the reddest objects.

The smoothed and raw images are compared with each other within a 30 $\times$ 30 arcsec region around each source to find cosmic rays. All candidate sources with $S_{\rm {raw}}- S_{ \rm {smooth}} > 5 \sigma_{\rm {raw}}$ are removed. The final catalogue is selected to have $S_{500} > 52$ mJy in order to minimize the effect of faint cosmic rays which are not found by the described technique.  All 17 sources with radio fluxes in excess of 1 mJy are removed using the the 21-cm radio catalogues from the NRAO VLA Sky Survey \citep[NVSS,][]{1995ApJ...450..559B} and the Radio Sky at Twenty-cm (FIRST) survey \citep{1998AJ....115.1693C} to avoid contamination by flat spectrum quasars at z $<$ 1. The rejection of NVSS/FIRST sources means that we potentially miss some genuine red sources that are lensed by radio-loud galaxies \citep{2014ApJ...790...46H,2016ApJ...818..196L}. The final \cite{2016MNRAS.462.1989A}  catalogue contains a total of 477 sources.

\subsection{SCUBA-2}

We selected the 200 brightest galaxies i.e. $S_{500}  >$ 63 mJy, of the 477 \cite{2016MNRAS.462.1989A} sources, and we observed a random sub-set of 188 of them for 15 minutes each using the \texttt{DAISY} pattern with the SCUBA-2 camera at the JCMT \citep{2013MNRAS.430.2513H}. The observations were taken in semester 15B between 31-7-2015 and 15-11-2015 with an opacity at 225GHz between 0.05 and 0.12. 

Our integration times were based on the previous observations of 28 red objects from \cite{2014ApJ...780...75D} with almost identical selection criteria as our sample. Those observations were 12.5 minute \texttt{DAISY} observations and 27 out of the 28 where detected. Using the $S_{850}/S_{500}$ colour distribution from these data to simulate the 850 $\mu$m fluxes of the HeLMS sample we estimated that a 1$\sigma$ RMS$_{850}$ = 4.5mJy would detect 70 per cent of our targets at $>$3$\sigma$.

We explored several data reduction methods including the data reduction used for the  SCUBA-2 Cosmology Legacy Survey \citep[][S2CLS]{2017MNRAS.465.1789G}, and the quick pipeline reduction using \texttt{REDUCE SCAN FAINT POINT SOURCES}. We found that the ``zero-mask"  \citep{2017MNRAS.470.3606H} data reduction used in \cite{2016ApJ...832...78I} provided us with the highest signal-to-noise values and a RMS$_{850}$ ranging between 3.2 mJy and 6.4 mJy with a mean of 4.3 mJy where the S2CLS method reaches an average RMS$_{850}$ of 4.9 mJy. The flux densities obtained with the zero-mask method are on average 2.6$\pm$4.0 mJy higher than the S2CLS method. We decided to use the zero-mask data reduction technique for all our observations because of its effectiveness in suppressing large scale noise \citep{2016ApJ...832...78I,2017MNRAS.470.3606H}. 

The zero-mask data reduction uses the Dynamic Iterative Map Maker within the \texttt{SMURF} package \citep{2013MNRAS.430.2545C}. This algorithm assumes that the image is free of significant emission except for a 60 arcsec diameter region centred on our target. Since the positions of our targets are in the centres of our \texttt{DAISY} observations this algorithm is very effective in suppressing large scale noise. This has an advantage over the S2CLS pipeline \citep{2017MNRAS.465.1789G}, which can make no prior assumptions about the positions of the targets. The maps are generated with 1 arcsec $\times$ 1 arcsec pixels. 

We use the same data reduction technique for the SCUBA-2 flux calibrators to get accurate flux conversion factors (FCF). These FCFs, ranging between 658 and 777 Jy pW$^{-1}$ beam$^{-1}$, are used to convert our reduced image to units of Jy/beam. The FCFs are expected to be accurate to within 5 per cent \citep{2013MNRAS.430.2534D}.  

Our prior positions are derived from the \textit{Herschel} data and have a typical positional uncertainty ($\sigma_H$) of 6 arcsec \citep{2016MNRAS.462.1989A}. Another positional uncertainty arises from the JCMT 2-3 arcsec rms pointing accuracy ($\sigma_J$). We combine both uncertainties to obtain the final positional uncertainty ($\sigma_p$):
\begin{equation}
  \sigma_{p} = \sqrt{\sigma_H^2 + \sigma_J^2}.
\end{equation}   
We apply our source extraction by taking the flux density of the brightest pixel within a 20 arcsec radius of our prior position in the beam convolved image.  This 20 arcsec radius corresponds roughly to the 3$\sigma_{p}$ positional uncertainty of our prior source in the SCUBA-2 map. We obtained an average noise level of $4.3$ mJy for our point source extraction.

For the purpose of analysis we divide our sample into three sub-groups with fairly arbitrary signal-to-noise ratio boundaries. Group 1 contains objects that have a clear detection, $S_{850}\ge 5\sigma$. Group 2 consists of detections between $3\sigma\le S_{850}< 5\sigma$. Finally, Group 3 are galaxies for which we do not have a clear detection, $S_{850}< 3\sigma$. (Due to the large uncertainty in position we are unable to obtain a significant detection in the stacked signal for the Group 3 galaxies.) The three groups contain 64, 99 and 25 objects respectively.

\begin{figure} 
  \centering  
  \includegraphics[width = 1.0 \columnwidth]{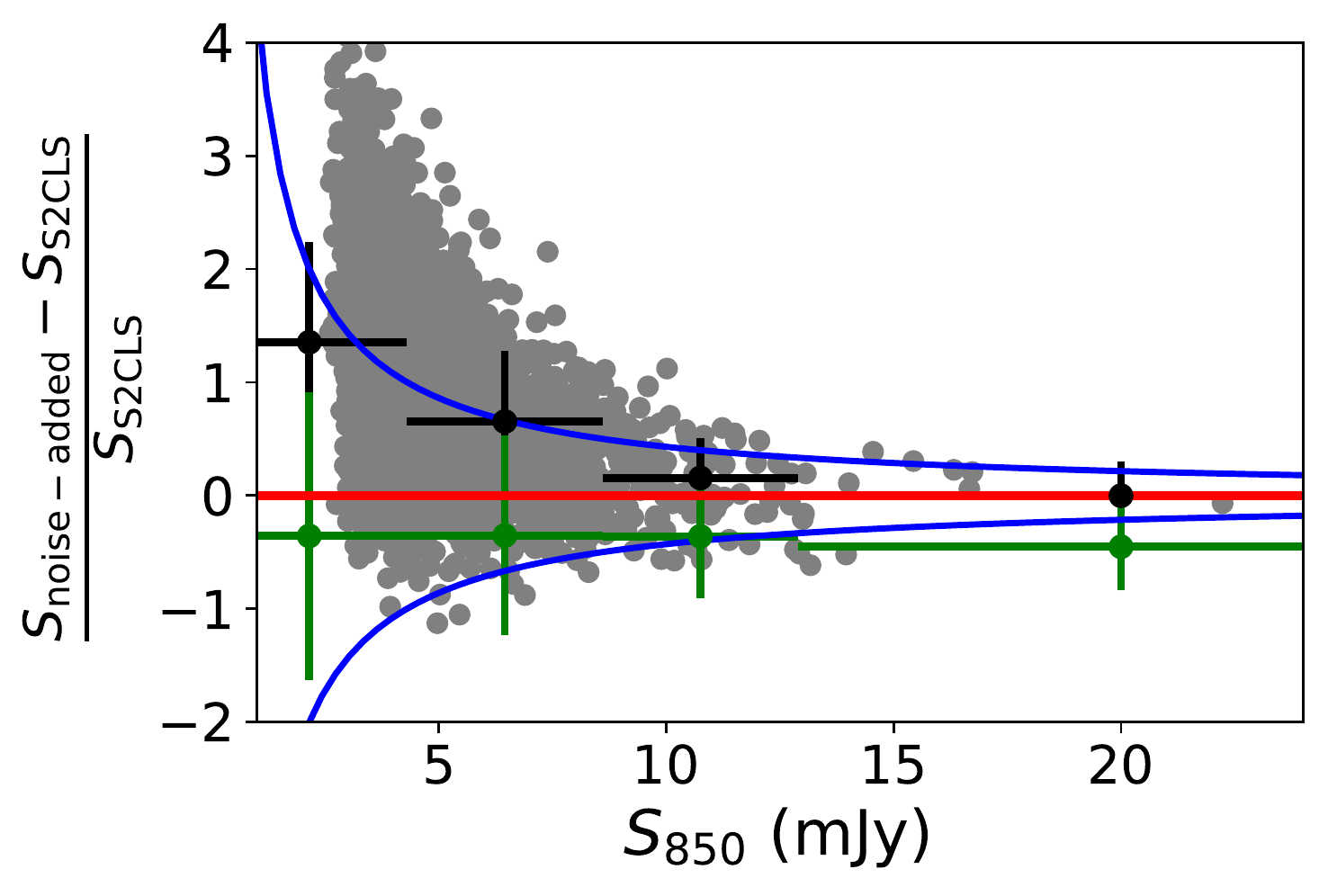}
  \caption{Simulation of our photometric errors and biases. S2CLS maps and catalogues are taken to be the truth and the noise-added fluxes are generated by adding Gaussian noise to mimic our observations ($\sigma=4.3\,{\rm mJy}$).  Flux densities are measured by taking the highest flux density within a 20 arcsec radius from the new S2CLS source position. The new S2CLS positions are generated by adding a random positional error of  $\sigma$ = 7 arcsec to it, which is comparable with the positional error of our data. The fractional difference between the (S2CLS) 850 $\mu$m flux density ``truth" and the measured 850 $\mu$m flux density  are plotted as function of the S2CLS flux density for all sources in grey, the black points show the mean of this measured fraction and the green points show the mean for a nearest pixel source extraction. The red line indicates zero offset and the blue lines indicate 1$\sigma$ (4.3 mJy) bounds.}
  \label{fig:s2cls}
\end{figure}

 As we are considering SCUBA-2 measurements  of {\em Herschel} detected galaxies we are concerned about the accuracy of the flux measurement, rather than the reality of a catalogued source (as we would be with a blank field survey). Nevertheless we would expect random noise fluctuations and confusion noise from galaxies not associated with our original target. 
 Furthermore we are using the brightest pixel, so our flux measurements are biased high \citep{2008MNRAS.384.1597C}.
We quantify this bias using the simulation shown in Figure \ref{fig:s2cls}. This  simulation takes all deep S2CLS fields as the ``truth". 

We add noise to the S2CLS maps by adding extra Gaussian noise to reach a total noise of $\sigma=4.3$~mJy, similar to those of our observations, we call this new maps the noise-added map. We then add positional errors to the S2CLS catalogue with a mean of zero and a standard deviation of 7 arcsec to the S2CLS positions to simulate the positional uncertainty of our DSFGs.  We then apply our photometric measurement at the original S2CLS position and compare with the original S2CLS flux. We repeat this process 5 times to get the results from different random noise simulations.

The comparison shows that for sources with $S_{850}$  below 13 mJy (3$\sigma$) we are (on average) overestimating the flux density, but this overestimation is on average lower than 4.3 mJy (1$\sigma$). We also tested the sources extraction method of picking the nearest pixel to our prior positions and find that this method underestimates the flux density significantly for sources with $S_{850} >$ 13 mJy. We decided to use our brightest pixel sources extraction because we expect that a significant percentage of our sources will lie above $S_{850} >$ 13 mJy given that $S_{500} >$ 63 mJy.


\subsection{Ancillary data} \label{se:ACC}

It is unlikely that our high-redshift galaxy sample will be directly detected in any shallow large-field surveys at optical/NIR wavelengths which are not likely to contain $z > 1$ galaxies without an AGN (Section \ref{sec:quas}). However, low-redshift galaxies can significantly magnify a higher redshift source behind them via gravitational lensing. 

Therefore it is possible to identify a lens using the available low-redshift galaxies from the Wide-field Infrared Survey Explorer \citep[][WISE]{2010AJ....140.1868W} and the Sloan Digital Sky Survey \citep[][SDSS]{2000AJ....120.1579Y}. We examined the SDSS images for possible contamination from large extended nearby galaxies and we found none. However, we do find several SDSS galaxies nearby and within the FWHM area of the SPIRE beam. Due to the large SPIRE/SCUBA-2 beam it will not be possible to unambiguously identify which of the several galaxies within the beam is potentially lensing the DSFG or is the optical/NIR counterpart of the DSFG.

For all our sources (excluding HELMS\_RED\_80 and HELMS\_RED\_421, see AGN Section \ref{sec:quas}) we find a total of 400 WISE detected sources  \citep{2013yCat.2328....0C} within a 20 arcsec radius. Of those sources only one is detected ($> 5\sigma$) in WISE-4 and this source is located 19.8 arcsec away from the SPIRE detection, additionally we find four WISE-3 detections ($> 5\sigma$) near other sources  which are all located $>$11.2 arcsec away from the SPIRE detection. For the numerous detections in the WISE-1 band it is not clear if the WISE source is a random aligned nearby galaxy, associated with our source, is an AGN or is lensing the background DSFG. We therefore did not use WISE data in our SED fit. We can, however, study the statistical excess of galaxies nearby to our sources \citep{2011MNRAS.414..596W}, where we only use WISE-1 sources as all but two WISE-2 galaxies are detected in WISE-1. We use SDSS DR9 \citep{2012ApJS..203...21A} and the \cite{2013yCat.2328....0C} WISE catalogue to select all detected galaxies near the line-of-sight of our targets (see Section \ref{sec:blend}).

Strong gravitational lensing, with a lensing magnification factor ($\mu$) larger than 2, could provide an explanation for our high flux densities. Wide field \textit{Herschel} surveys show that galaxies with a flux density at $S_{500} >$ 100 mJy are likely to be strongly gravitationally lensed \citep{2010Sci...330..800N,2011ApJ...732L..35C,2016ApJ...823...17N,2017MNRAS.465.3558N}. This $S_{500} >$ 100 mJy limit comes from the steep slope in the FIR luminosity function, which causes intrinsically luminous ($S_{500} >$ 100 mJy) sources to be extremely rare. Our sample of 500 $\mu$m riser galaxies contains 9 galaxies with $S_{500} >$ 100 mJy, of which we expect $\ge$ 80 per cent to be strongly lensed \citep{2010Sci...330..800N,2013ApJ...762...59W}. The probability that a DSFG is strongly lensed  declines for $S_{500} <$ 100 mJy, but for galaxies around 70 mJy at $S_{500}$ there is still a significant ($\sim$20 per cent) chance that they are lensed \citep{2015ApJ...812...43B,2016ApJ...823...17N}.

Other follow-up programs have observed part of our sample:
 \begin{itemize}
   \item  Four of the sources  (HELMS\_RED\_3, HELMS\_RED\_4, HELMS\_RED\_6 and HELMS\_RED\_7) were observed at the CSO using MUSIC \citep{2014SPIE.9153E..04S} at four wavelengths, 2.09, 1.4, 1.1 and 0.92 mm. The resulting flux densities can be found in section 6.2 and Table 4 of \cite{2016MNRAS.462.1989A}.
   \item  Two sources (HELMS\_RED\_4, HELMS\_RED\_31) have spectroscopic follow up with the Atacama Large Millimeter Array (ALMA). The resulting spectra can be found in \cite{2016MNRAS.462.1989A}. The redshift of  HELMS\_RED\_4 is 5.162 and the redshift of HELMS\_RED\_31 is 3.798 or 4.997 depending on the line detection being the CO(5-4) or the CO(4-3) line. 
   \item  Two sources (HELMS\_RED\_1, HELMS\_RED\_2) have spectroscopic follow up by the Combined Array for Research in Millimeter-wave Astronomy (CARMA). The detected redshifts are 4.163 and 4.373, respectively (Riechers et al. in prep., Leung et al. in prep).
  \item Five sources (HELMS\_RED\_1, 2, 4, 10, 13) have been observed with the Submillimeter Array (SMA), and will be discussed in detail in  Greenslade et al., in prep. 
    \item Two sources (HELMS\_RED\_1, 3) are detected in the Atacama Cosmology Telescope (ACT) equatorial survey \citep{2017MNRAS.464..968S}. The measured flux densities at 148, 218 and 278 GHz are 12.49$\pm$1.74, 35.11$\pm$2.62 and  72.32$\pm$6.26 mJy for HELMS\_RED\_1 and  6.14$\pm$1.76, 19.50$\pm$2.56 and  35.32$\pm$6.24 mJy for HELMS\_RED\_3.

 \end{itemize}

MUSIC and ACT provide even more data points in the Rayleigh$\text{-}$Jeans part of the spectrum. These additional long wavelength data will improve our SED-fitting process. The spectroscopic redshifts from CARMA and ALMA will be used to help validate our SED-fitting process and to confirm that our selection process does indeed pre-select high-redshift galaxies. We use the preliminary SMA results to get accurate information about the source positions and to determine if any sources are blended.

\section{Modeling the DSFGs} \label{sec:param}

\subsection{SED fitting for photometric redshifts}

\begin{figure}
  \centering  
  \includegraphics[trim = 0mm 0mm 0mm 0mm,width = 1.0\columnwidth]{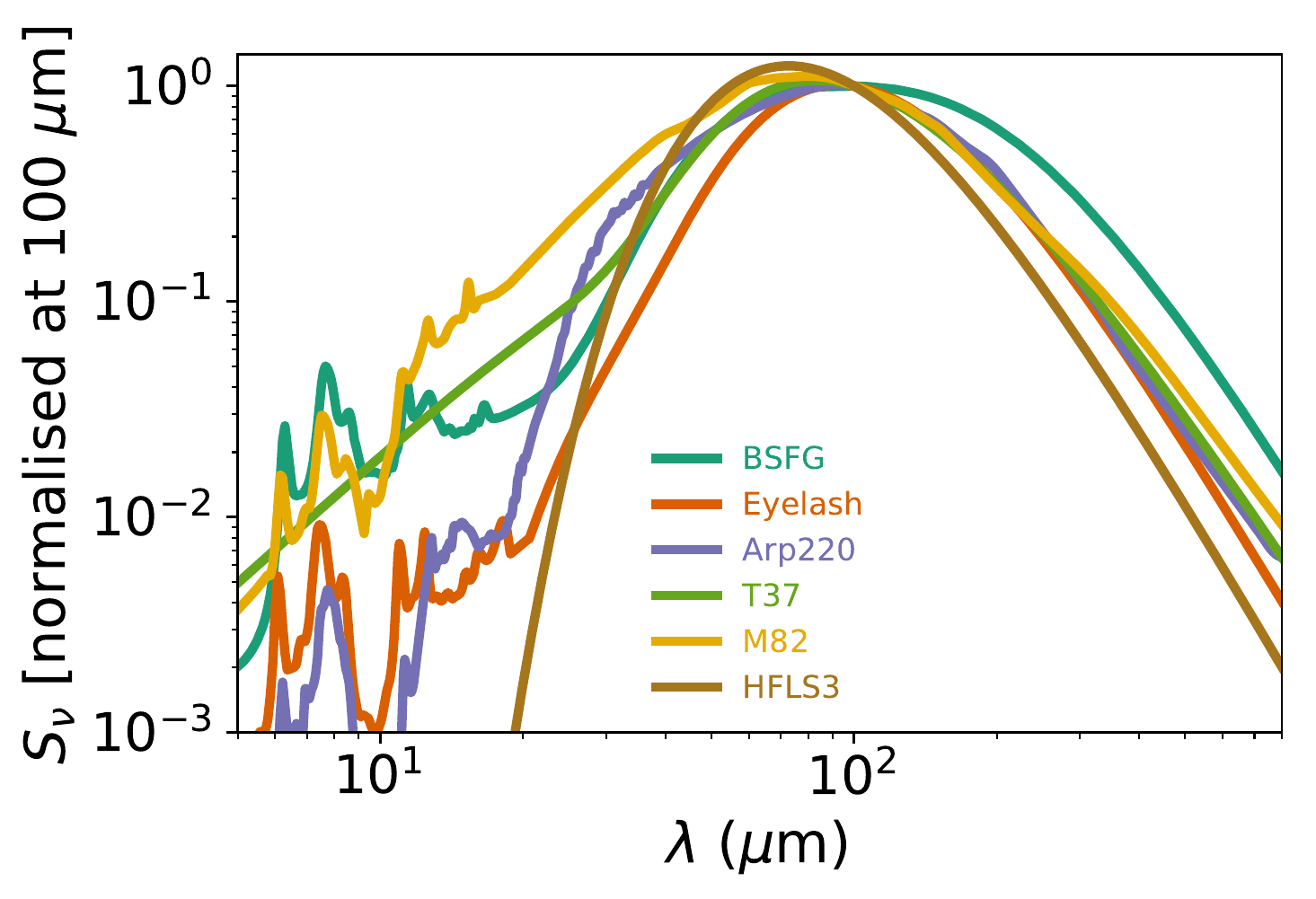}
  \caption{The six spectral energy distribution templates SEDs that we use in our photometric redshift fitting process. These are: broad star-forming -- BSFG  derived by \citet{2013A&A...551A.100B}, cosmic Eyelash and three warm starburst galaxies M82, Arp220 \citep{2007ApJ...663...81P} and HFLS3 \citep{2013Natur.496..329R}.} 
  \label{fig:seds}
\end{figure}

Fits to the FIR/submm spectral energy distributions (SED) to obtain photometric redshifts and integrated properties are performed using the \texttt{EAZY} code \citep{2008ApJ...686.1503B} using a sample of representative FIR/submm templates  \citep[e.g.][]{2003MNRAS.342..759A}. 

The FIR peak of luminous infrared galaxies ($L_{\rm IR} > 10^{10}  {\rm L_\odot}$)  can be, crudely,  characterized by cool dust with average temperatures in the 25-45 K range \citep[e.g.][]{1984ApJ...283L...1S,1997A&A...325L..21K}. The lack of strong features means it is difficult to distinguish between either very cold dust or high-redshift galaxies using only SPIRE photometry. The addition of the $S_{850}$ data enables us to estimate the peak of the FIR emission, and therefore able to place far tighter constraints on the redshift (Section \ref{sec:phys}).  However, since temperature and redshift are degenerate the choice of templates is a critical factor in photometric redshift estimation and so our templates have been carefully chosen to cover a broad range of temperatures. 

Our six templates consist of the broad star-forming galaxy (BSFG) derived by \cite{2013A&A...551A.100B}, cosmic Eyelash and three warm starburst galaxies M82, Arp220 \citep{2007ApJ...663...81P} and HFLS3 \citep{2013Natur.496..329R}. However, these templates have a gap at an effective temperature 37 K so we create an extra SED template from a modified black body (MBB) with a temperature of 37 K, a dust emissivity index ($\beta$) of 1.5 and a MIR power law component ($\alpha$) of 2.0 \citep{2012MNRAS.425.3094C}. These templates are illustrated in Figure~\ref{fig:seds}. With {\sc EAZY} we fit all possible linear combinations of our templates set.

\begin{figure}
  \centering  
  \includegraphics[trim = 0mm 0mm 0mm 0mm,width = 1.0 \columnwidth]{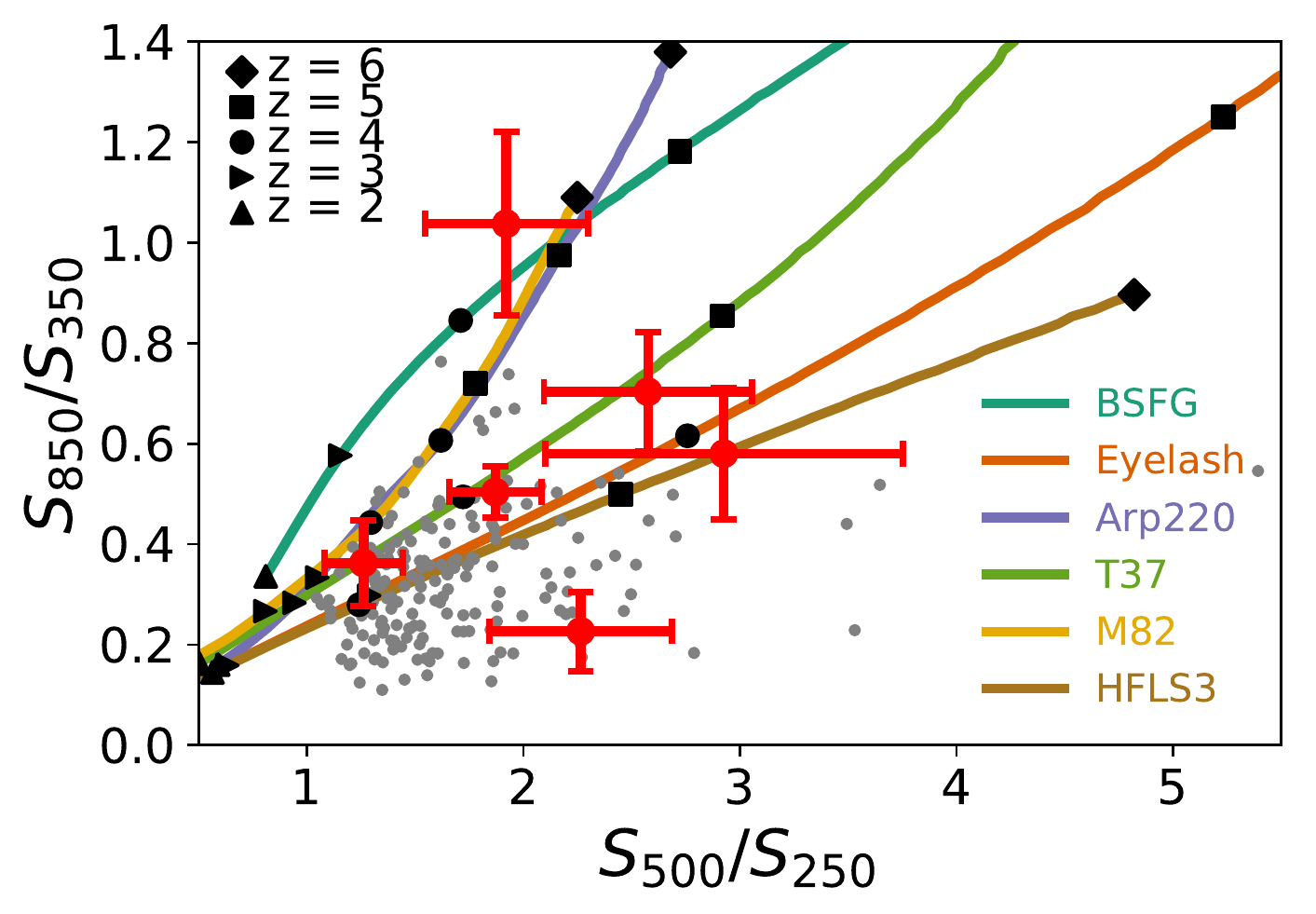}
  \caption{Colour-colour plot of our sample of  DSFGs in grey, with a sub-set of points with representative error bars in red. The coloured lines show the redshift tracks of our SED templates. The crossing of such a line indicates that for a certain colour there are degenerate solutions for the photo-z estimates. The black shapes indicate the colour of a SED template at the indicated redshift. The data points significantly below the HFLS3  line could only be sampled by a non-physical template narrower than a black-body. The presence of the DSFGs in this part of the diagram indicates flux boosting in either $S_{350}$ or $S_{500}$ (see Section \ref{discussion}).} 
  \label{fig:colour}
\end{figure}

In Figure \ref{fig:colour} we show the colour-colour plot of our observations. We overlay the redshift tracks from our sample of SED-templates. Our template set thus contains a wide range of representative DSFGs over a large redshift range. We can exclude very cold ($T \sim 20$ K) galaxies at $z \lesssim 1.7$ as they would not be a 500 $\mu$m riser. Such galaxies at higher redshift could potentially contaminate our sample.  But this type of galaxies are very rare between $0.1<z<2.0$ \citep{2013MNRAS.431.2317S}. Such a cold galaxy would furthermore have a higher $S_{850}/S_{500} $ colour than any of our measured $S_{850}/S_{500} $ colours at $z > 2.5$.

We only use broad-band FIR data, and we neglect the contribution of emission lines. At redshifts of $z\sim4$  FIR lines have a $\sim 6$ per cent effect at 250 $\mu{\rm m}$, however, they have a negligible effect at 350, and 500 $\mu{\rm m}$; at  850 
$\mu{\rm m}$ they have a $\sim 1$ per cent contribution at $z\sim4$ though this rises to $\sim 8$ per cent at $z\sim5$ \citep{2011MNRAS.414L..95S}.

We adjust  {\sc EAZY} to allow for 10 per cent systematic error for the data. This 10 per cent incorporates both the 5 per cent error in the FCF for SCUBA-2 and our use of a different algorithm to reduce the data for SCUBA-2 and SPIRE. The advantage of using this extra 10 per cent systematic error is that it dominates unrealistically small statistical errors for very bright ($>10 \sigma$) sources.

In Section~\ref{sec:testz} we directly compare our method with other methods, other template choices and with spectroscopic redshifts.

\subsection{Noise estimates} \label{sec:noise}

The SPIRE and SCUBA-2 maps contain both confusion and instrumental noise.  Both have to be included in the SED fitting to ensure that the errors on fitted parameters, e.g. photometric redshift  are assigned the appropriate errors. The confused background in the SPIRE band is caused by coincident sources; these contribute in all three wavelength bands. The  instrumental noise can be assumed to be  uncorrelated and included straightforwardly in the $\chi^2$ calculations within \texttt{EAZY}.  However, to incorporate the confusion noise we need to consider that this is correlated noise. 

The confusion noise at $S_{850}$ from SCUBA-2 is significantly lower than the confusion in the SPIRE bands \citep[1 mJy vs. $\sim$ 6-7 mJy;][]{2017MNRAS.465.1789G,2010A&A...518L...5N} due to the smaller beams size of SCUBA-2 and lower number counts. The SCUBA-2 confusion noise is subdominant to the instrumental noise we obtained in the images. We can therefore safely neglect the effects of confusion in our SCUBA-2 flux density estimates.  We can simulate possible values for the SPIRE contribution in the following way.

In a confusion limited map, where the instrumental noise is negligible compared with the confusion noise, the fluctuations in that map can be considered to be caused by confusion noise alone. We can randomly sample such a confusion limited map at the same position in all three bands drawing a 3-tuple of flux density values that represent the confusion noise. These samples automatically include the correlation between the bands\footnote{An alternative, would be to estimate the covariance matrix between the maps, and synthesise correlated flux density values from this assuming Gaussian fluctuations. However, by sampling directly from the map we skip this step and get a more direct model of the correlated confusion noise}.

The HELMS field is not confusion limited so we sample the confusion limited COSMOS \citep{2007ApJS..172....1S} field. COSMOS has a 1$\sigma$ instrumental noise $<$ 2.5 mJy, though small, this residual instrumental noise means we will slightly overestimate the confusion noise values.  We perturb the 3-tuple flux  of each object in our catalogue by one of the sample 3-tuples drawn from  COSMOS.  We then run \texttt{EAZY} on the perturbed catalogue.  We do this simulation exercise 1000 times (however, due to the finite size of the field these are not independent).

\begin{figure}
  \centering  
  \includegraphics[trim = 0mm 0mm 0mm 0mm,width = 1.0\columnwidth]{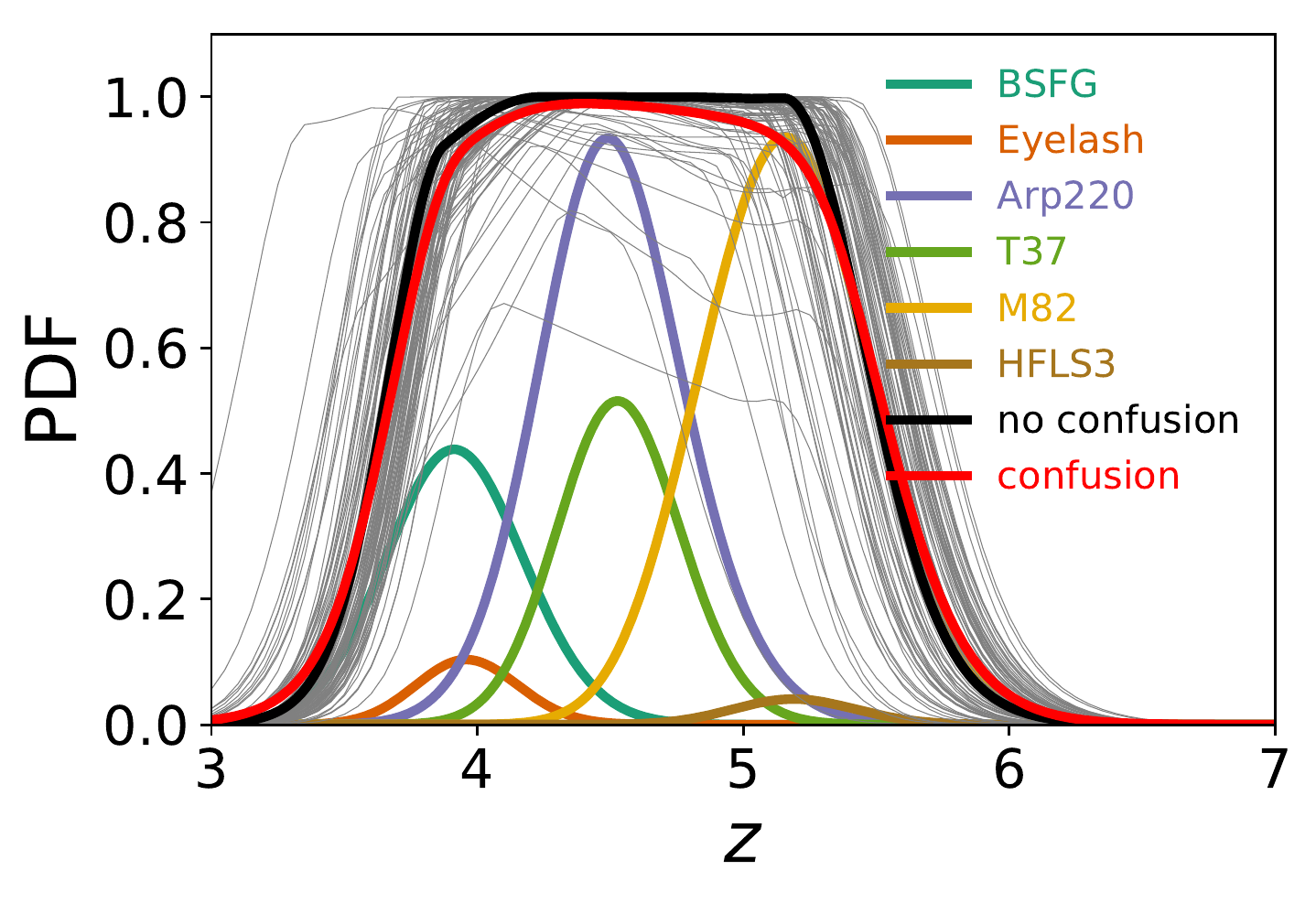}
  \caption{ Redshift Probability Distribution Function (PDF) for a 
  single galaxy, illustrating the contribution from different galaxy templates. 
 Each grey line represent the PDF from a single run with \texttt{EAZY}, perturbed by one particular sample of the confusion noise. The red line represents the average of the 1000 EAZY runs and the black line is the result from the traditional method without confusion noise. The coloured lines shows the contribution to the PDF from each galaxy template used. 
  } 
  \label{fig:eazy}
\end{figure}

We average the redshift probability distribution function (PDF) over all simulation runs to obtain the final PDF for each galaxy. The results of 1000 runs for a single representative galaxy are shown in Figure \ref{fig:eazy}. The resulting PDF is slightly broader than the PDF from the traditional method of not using the confusion noise. This effect would be larger if  the noise in HeLMS had been dominated by  confusion noise.

\begin{figure}
  \centering  
  \includegraphics[trim = 0mm 0mm 0mm 0mm,width = 1.\columnwidth]{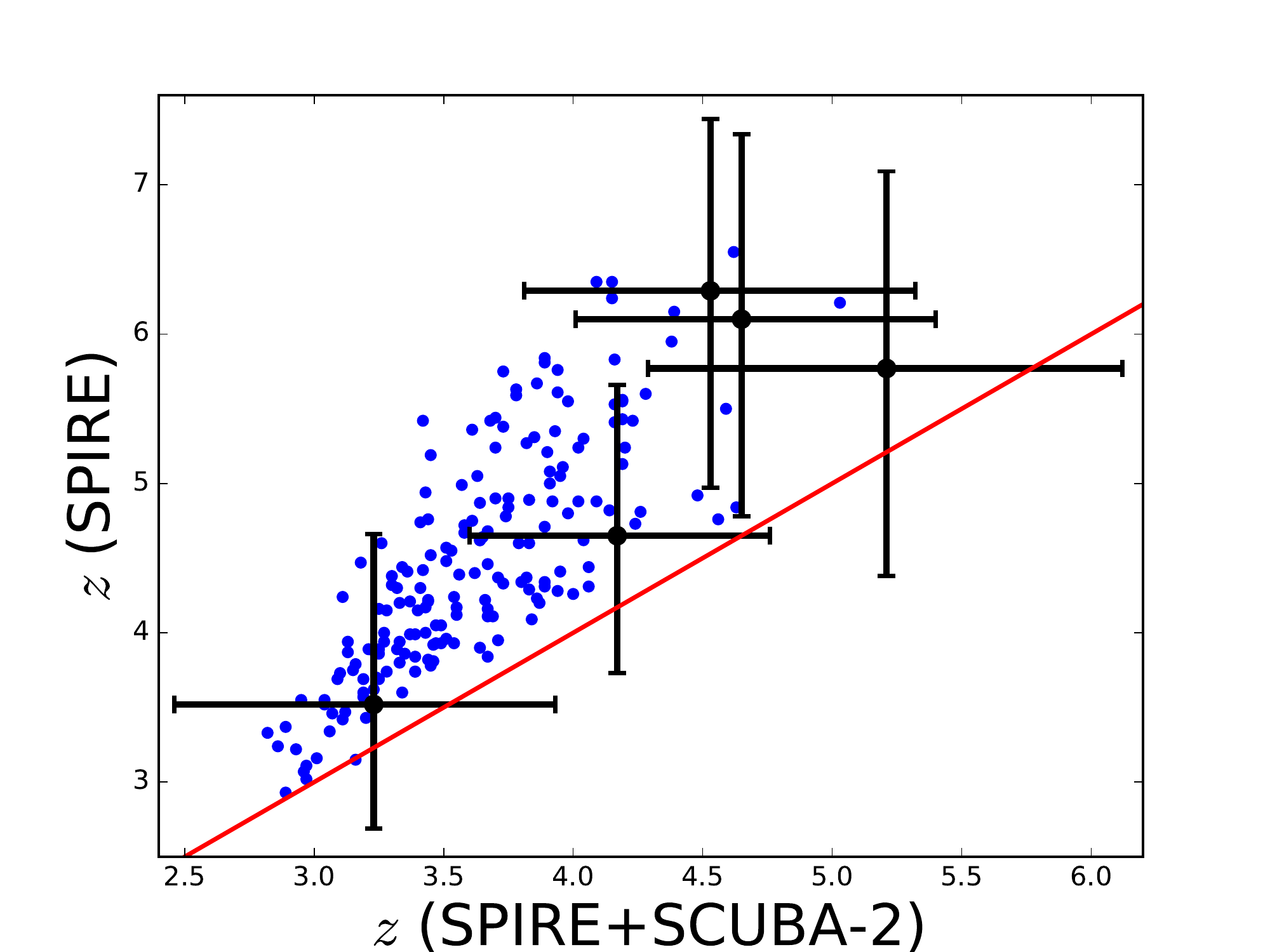}
  \caption{Redshift estimates from our SED fits using SPIRE photometry only vs. those where we include SCUBA-2 data. All points in blue, and in black a subset of representative error bars. The average uncertainty for the SPIRE-only dataset is larger, $\sigma_z/(1+z) = 0.21$, than the uncertainty with the additional SCUBA-2 data $\sigma_z/(1+z) = 0.15$. It is also clear that the SPIRE-only SED fits overestimate the redshift due to the lack of constraints on the peak of the FIR emission.} 
  \label{fig:comp_z}
\end{figure}

In Figure \ref{fig:comp_z} we show the improvement in photometric redshift by adding the longer wavelength SCUBA-2 data. The average uncertainty  (calculated from the variance of the estimated PDF from {\sc EAZY})  when we only use the SPIRE flux densities is larger, $\sigma_z/(1+z) = 0.21$, than the uncertainty with the additional SCUBA-2 data $\sigma_z/(1+z) = 0.15$.  This Figure also shows that we overestimate the photometric redshift when we only use the SPIRE data. 

\subsection{Physical parameters} \label{sec:phys}

Using \texttt{EAZY} we obtain the full PDF and the best fitting SED template for every galaxy. With this  template we compute the total infrared luminosity, $L_{\rm IR}$. The FIR luminosity is defined as the integral over the rest frame spectrum between 8$\,\mu$m and 1000$\,\mu$m, i.e. $L_{\rm IR}=\int_{8\mu{\rm m}}^{1000\mu{\rm m}} L_\nu\, d\nu$.  In practice we lack a good measurement of the flux in the rest frame mid-infrared (MIR) from 8 - 30$\,\mu$m. We therefore integrate between 30$\,\mu$m and 1000$\,\mu$m and use a correction factor for the potentially large amounts of missed flux in the MIR. We calculate the correction factor from the average fraction of the FIR luminosity contained in the MIR regime for 5 of our 6 templates. We exclude the HFLS3 template for this measurement due to a lack of constraints in the MIR. We obtain a correction factor of 1.17 and we multiply our measured integral by this factor to obtain the resulting $L_{\rm IR}$.  We also obtain an error on $L_{\rm IR}$ using both the errors on our flux density estimates and the scatter from our 1000 \texttt{EAZY} runs. 

The negative K-correction (for galaxies measured at longer wavelengths than the peak of their SED) counteracts (to some extent) the dimming with distance, and so these galaxies are relatively constant in brightness \citep[e.g.][]{2014PhR...541...45C}. Therefore our estimates of $L_{\rm IR}$ can be tightly constrained even with a large uncertainty in the redshift. 

Our $L_{\rm IR}$ can be translated into SFR estimates using \cite{1998ARA&A..36..189K} for a Salpeter IMF
\begin{equation} \label{eq:SED}
  \frac{\mbox{SFR}}{\mbox{M}_\odot \mbox{yr}^{-1}} = 1.96 \times 10^{-10} \frac{L_{\rm IR}}{\mbox{L}_\odot}  .
\end{equation}
Here the fraction of ultraviolet energy absorbed by dust has been assumed to be $\epsilon = 0.88$, for which we have no constraint. Our estimates for the SFR would be the same if we had used the \cite{1997MNRAS.289..490R} calibration factor with a $\epsilon = 2/3$. We assume no gravitational lensing (Section \ref{sec:blend}) and no contamination by AGN (Section \ref{sec:quas}) in our calculation of the SFR. The resulting SFRs should be multiplied by a factor 0.63 or 0.67 if assuming a   Chabrier or Krupa IMF \citep{2014ARA&A..52..415M}.

Our final catalogue is presented in Appendix \ref{ap:A}, where we list the positions, flux densities, redshifts and $L_{\rm IR}$ of all our galaxies observed with SCUBA-2.

\subsection{Testing the photometric redshifts}\label{sec:testz}

\cite{2016ApJ...832...78I} made a similar assumption with the selection of their templates, and tested their photometric redshift code against 25 red high-redshift DSFGs with spectroscopic redshift. Their photometric redshifts where found by finding the lowest $\chi^2$ value for their set of three templates. The main difference between our method is that \texttt{EAZY} not only fits the provided templates but also any linear combination of those templates. The results from \cite{2016ApJ...832...78I} show only a small offset in $(z_{\rm phot} - z_{\rm spec})/(1+z_{\rm spec})=-0.03$ with a scatter of 0.14.  

We compare our photometric redshift method ($z_{EAZY}$)  directly with \cite{2016ApJ...832...78I}, by running our code on their sample. We obtain a mean ($\mu$) offset in $(z_{\rm{EAZY}} - z_{\rm {Ivison}})/(1+z_{\rm{EAZY}})$ of 0.11 and a median ($\mu_{1/2}$) offset of 0.12. We note that this offset is smaller than the mean estimated error in our redshift ($\langle \sigma_z/(1+z_{\rm{EAZY}}) \rangle=0.15$).

The main difference between our method and that of \cite{2016ApJ...832...78I} is that they tested a set of 6 templates individually  with a sample of available spectroscopic redshifts, and discarded the ones with the poorest fit in $(z_{\rm{Ivison}} - z_{\rm {spec}})/(1+z_{\rm{spec}})$.  Two of the poorest fitting templates in their analysis were the Arp 220 and HFLS3, which are on the ``blue'' end of the range of FIR SEDs. If we discard our ``blue'' templates (M82, HFLS3 and Arp 220) we find that our photometric redshift estimates are very close to the \cite{2016ApJ...832...78I} estimates ($\mu$ = 0.024 and $\mu_{1/2}$ = 0.035).  However, we keep these ``blue'' templates in our analysis, to ensure conservative errors, noting that \texttt{EAZY} produces a full redshift PDF using all our templates (and all linear combinations of them) simultaneously. 

We can see how our results would change if we made a different choice of templates.  \cite{2016ApJ...822...80S} used a Monte Carlo method to sample a range of MBB from \cite{2012ApJ...756..101G} with dust temperature parameter sampled from a Gaussian with mean and standard deviation $39 \pm 10$ K. We use a similar full MCMC approach to fit using the FITIR module of the INTERROGATOR\footnote{\url{http://users.sussex.ac.uk/~sw376/Interrogator/}} code (Wilkins et al. in prep). With this method we can specify prior information about all free parameters. We consider both the MBB parameterisation of \citet{2012ApJ...756..101G} (which has two free parameters, the temperature $T$, and the emissivity $\beta$) and the parameterisation of \citet{2012MNRAS.425.3094C} (which has three free parameters: the temperature, emissivity $\beta$, and the slope of the near-IR power law $\alpha$). For the \citet{2012ApJ...756..101G} parameterisation we fix the emissivity $\beta=2.0$ and consider 3 different priors on the temperature $T$: fixed to $T=40$ K, a normal distribution centred at $T=39$ K with $\sigma=10$ K, and a uniform prior $T/\rm{K}=[20,40]$. For the \citet{2012MNRAS.425.3094C} we assume uniform prior on the temperature of  $T/\rm{K} =  [20, 60]$ and consider cases where both $\alpha$ and $\beta$ are fixed (to $2.0$ and $1.5$ respectively) and where they have a uniform prior: $\beta = [1,2]$ and $\alpha = [1.,2.5]$. 

The results are shown in Table~\ref{tab:specz} where we compare the output of each different template set to our chosen templates when applied to our sample. We compute a number of comparison statistics, the mean offset ($ \mu = \frac{z-z_{\rm this\ work}}{1+z_{\rm this\ work}})$, the rms scatter in $\mu$ ($\sigma$) and the $\chi^2$ in comparison with our three spectroscopic redshifts. For the normal distributed (T = 39 K) method we find a $\mu$ = $-$0.056 and a  $\chi^2$ = 1.35, for the uniform prior ($T/\rm{K}=[20,40]$) $\mu$ is $-$0.011 and the $\chi^2$ = 0.67 and for the single temperature model we find a $\chi^2$ = 54. From these results we can see that the Gaussian prior produces very similar results as our method, and that the flat 20-60 K prior models are consistent with the spectroscopic redshifts, but overestimate the size of the error bars ($\chi^2 \ll 1$). The single temperature model is insufficient in fitting photometric redshifts.    

\begin{table*}
  \begin{tabular}{lrrrrrr}
     Method & this work & Gaussian (39$\pm$10 K) &  Uniform (20-60 K) & Delta (40 K) & Casey (20-60 K)& Casey Wide (20-60 K)  \\ \hline
 $\langle z \rangle$ & 3.60 $\pm$ 0.43 & 3.34 $\pm$ 0.37 & 3.54 $\pm$ 0.40  & 3.24 $\pm$ 0.32 & 4.46 $\pm$ 0.54 & 4.79 $\pm$ 0.51    \\
  $\langle z_{\rm h}-z_{\rm l} \rangle$/2 & 0.67 & 1.04 & 1.16 & 0.32 & 2.03 & 1.80  \\ 
  $\mu$ & 0 &-0.056 & -0.011 & -0.078 & 0.187 & 0.260 \\
  $\sigma$ & 0 & 0.034 & 0.041 & 0.027 & 0.043 & 0.033 \\
  $\chi^2$ & 3.07 & 1.35 & 0.67 & 53.6 & 0.35 & 0.62
 \end{tabular}
  \caption{Comparison of templates for photometric redshift accuracy. Mean photometric redshift, 1$\sigma$ error, mean difference $(\mu)$ with the photometric redshift used is this work in $\frac{z-z_{\rm this\ work}}{1+z_{\rm this\ work}} $ and, the rms scatter ($\sigma$) in $\mu$ as function of different photometric redshift methods. The last row shows the sum of the $\chi^2$ in comparison with the three spectroscopic redshifts of our sample. The Gaussian (39$\pm$10 K) model produces comparable results compared to our method, but slightly overestimates the error bar size. The delta model is insufficient in fitting photometric redshifts, and the uniform models vastly overestimate the error bar size.}  \label{tab:specz}
\end{table*}

\begin{figure} 
  \centering  
  \includegraphics[width = 1.0\columnwidth]{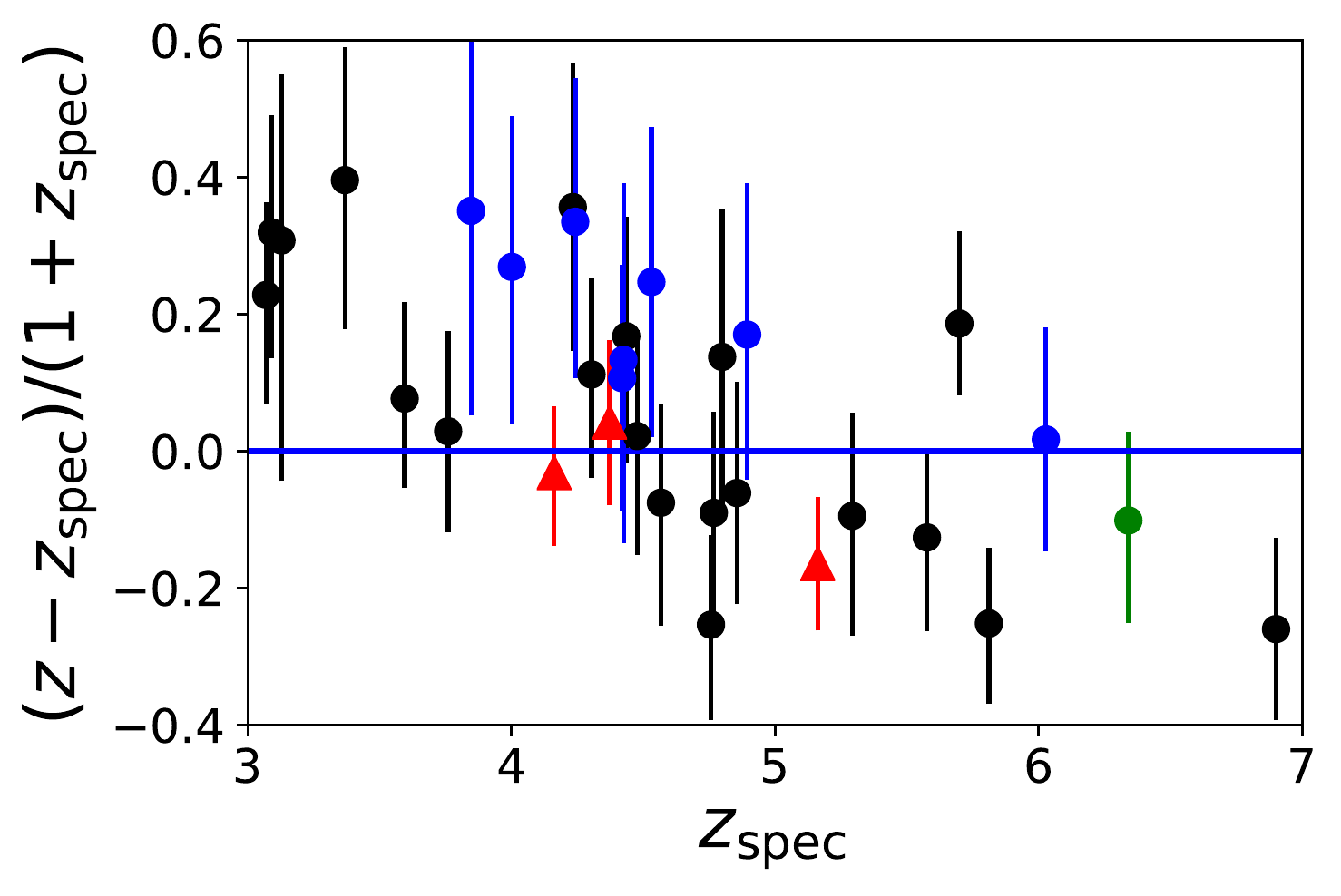} 
  \caption{Comparison with available $500 \mu$m riser spectroscopic redshifts at $z > 3$. In green we show HFLS3, in black the SPT sample and in blue the \textit{H}-atlas sample \citep{2013ApJ...767...88W,2016ApJ...822...80S,2017ApJ...842L..15S,2017MNRAS.472.2028F} and in red the spectroscopic redshifts for our sample. We obtain an offset  $\langle (z-z_{\rm spec})/(1+z_{\rm spec}) \rangle = 0.08 $ with a rms of 0.19 and a average $\chi^2$ per galaxy of 1.4. } 
  \label{fig:spec}
\end{figure}

The ultimate test is the comparison against spectroscopic redshifts.  We obtain a good total $\chi^2$ of 3.07 for our three spectroscopic redshifts. But due to the limited number of spectroscopic redshifts in our sample we also use the SPT detected DSFGs which fulfil our colour selection criteria \citep{2013ApJ...767...88W,2016ApJ...822...80S,2017ApJ...842L..15S}, HFLS3 and the \textit{H}-atlas 500 $\mu$m risers \cite{2017MNRAS.472.2028F}. The results are shown in Figure \ref{fig:spec}. We  estimate a bias of $\langle  (z-z_{\rm spec})/(1+z_{\rm spec})\rangle = 0.08 $ with a rms of 0.19 and a reduced $\chi^2$ of 1.4. The rms scatter in the bias (0.19), our average uncertainty per galaxy ($\sigma_z/(1+z) = 0.15$) and $\langle  |z-z_{\rm spec}|/(1+z_{\rm spec})\rangle = 0.17$ all have comparable values.

There is a visible trend in Figure \ref{fig:spec} that $(z-z_{\rm spec})/(1+z_{\rm spec})$ is decreasing with redshift, the reduced $\chi^2$ for linear decreasing model is 0.9 compared to 1.4 for the non-evolving model. This result indicates that we underestimate the redshift of high-redshift galaxies due to a rising dust temperature of our spectroscopic sample towards higher redshift \citep{2016ApJ...832...78I}. However, this same result could also arise from selection effects, where a warm HFLS3 type galaxy would not have made our selection criteria at $z < 4.6$ as it would not be a 500 $\mu$m riser (Figure \ref{fig:SFRz}). Another possible explanation for this trend is that higher redshift galaxies need to be brighter to fulfil our flux density selection criteria, and these brighter galaxies tend to be warmer \citep[e.g.][]{2013MNRAS.431.2317S,2017ApJ...843...71K}. 

Any fitting methods with a range of temperatures and no explicit prior on the temperatures is effectively assigning a uniform prior to the temperatures.  This is what our method does as do most photometric redshift fitting methods.  In the low signal-to-noise regime the prior has a stronger influence on the posterior and so there will be a trend to fit mid-range temperatures rather than high or low temperatures.  This naturally tempers the extremes of redshifts distributions based on the best redshift.  However, the redshift PDFs are a reasonable representation of the information available.

\section{Results} \label{sec:results}

\subsection{Statistical properties} \label{sec:sp}

\begin{figure}
  \centering  
  \includegraphics[trim = 0mm 0mm 0mm 0mm,width = 1.0\columnwidth]{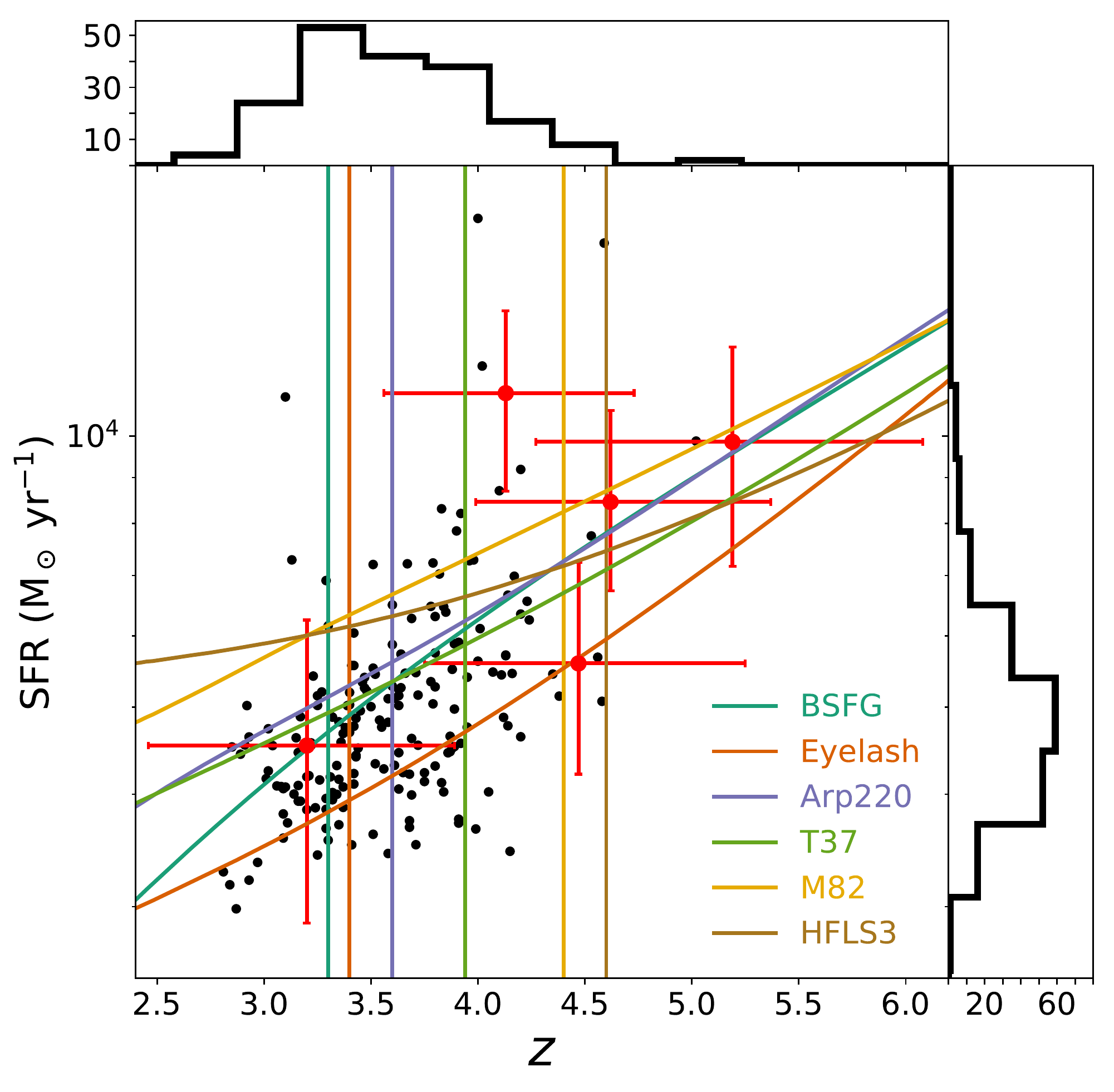}
  \caption{SFR vs. redshift for our 188 targets. Red represent a set of representative error bars. There are several objects which have a strong indication to lie at very high redshifts, but the bulk of our sample is expected to lie around z$\approx$3-4. The coloured lines represent the lower redshift limits for 500 $\mu$m riser galaxies and SFR tracks for our range of SED templates. } 
  \label{fig:SFRz}
\end{figure}

In Figure \ref{fig:SFRz} we show the SFR vs redshift distribution of our sources. Our sources have a median redshift of 3.6 $\pm 0.4$ and a median SFR  (uncorrected for flux boosting or the possible presence of gravitational lensing) of $5.2 \pm 1.9 \times 10^3$ M$_\odot$yr$^{-1}$.  All our galaxies could be classified as distant hyper-luminous infrared galaxies (HyLIRGS), i.e. with $L_{\rm IR}$ exceeding $10^{13}$ L$_\odot$ and a mean $L_{\rm IR}$ of 2.7 $\times 10^{13}$ L$_\odot$.

We find that 31.4 $\pm$ 4.7 per cent lie between redshifts of 4 and 6. This finding is consistent with \cite{2016ApJ...832...78I} who found 33 $\pm$ 6 per cent of their sample to lie within this redshift range. The inferred space density $(\rho_{obs})$ in this redshift range is $1.1 \times 10^{-8}$ Mpc$^{-3}$. Due to the predicted short lifetime for the starburst ($t_{\rm{burst}}$) phase we need to apply a duty-cycle correction to the observed space density to infer the actual underlying space density ($\rho$) for these type of galaxies
\begin{equation}
  \rho = t_{obs} / t_{\rm{burst}} \times \rho_{obs},
\end{equation} 
were $t_{obs}$ is the time between $ 4 <z< 6$. For $t_{\rm{burst}}$ we assume 100 Myr, which is in agreement with their expected gas depletion times \citep{2011MNRAS.412.1913I,2013MNRAS.429.3047B}. The final inferred space density estimate is thus $7 \times 10^{-8}$ Mpc$^{-3}$. The assumption of 100 Myr is the same as used by \cite{2016ApJ...832...78I} and while longer timescales (0.5-1.0 Gyr) have been postulated \citep[e.g.][]{2014ApJ...782...69L,2015ApJ...810...74A} these would result in an even lower space-density.   

The primary difference between the \cite{2016ApJ...832...78I} sample and our sample is that \cite{2016ApJ...832...78I} used a $S_{500} >30$ mJy selection where we use a $S_{500} >63$ mJy sample. Therefore our sample has a space density of about a factor of 10 lower than the \cite{2016ApJ...832...78I}  estimate of  $6 \times 10^{-7}$ Mpc$^{-3}$.

\begin{figure}
  \centering  
  \includegraphics[trim = 0mm 0mm 0mm 0mm,width = 1.0\columnwidth]{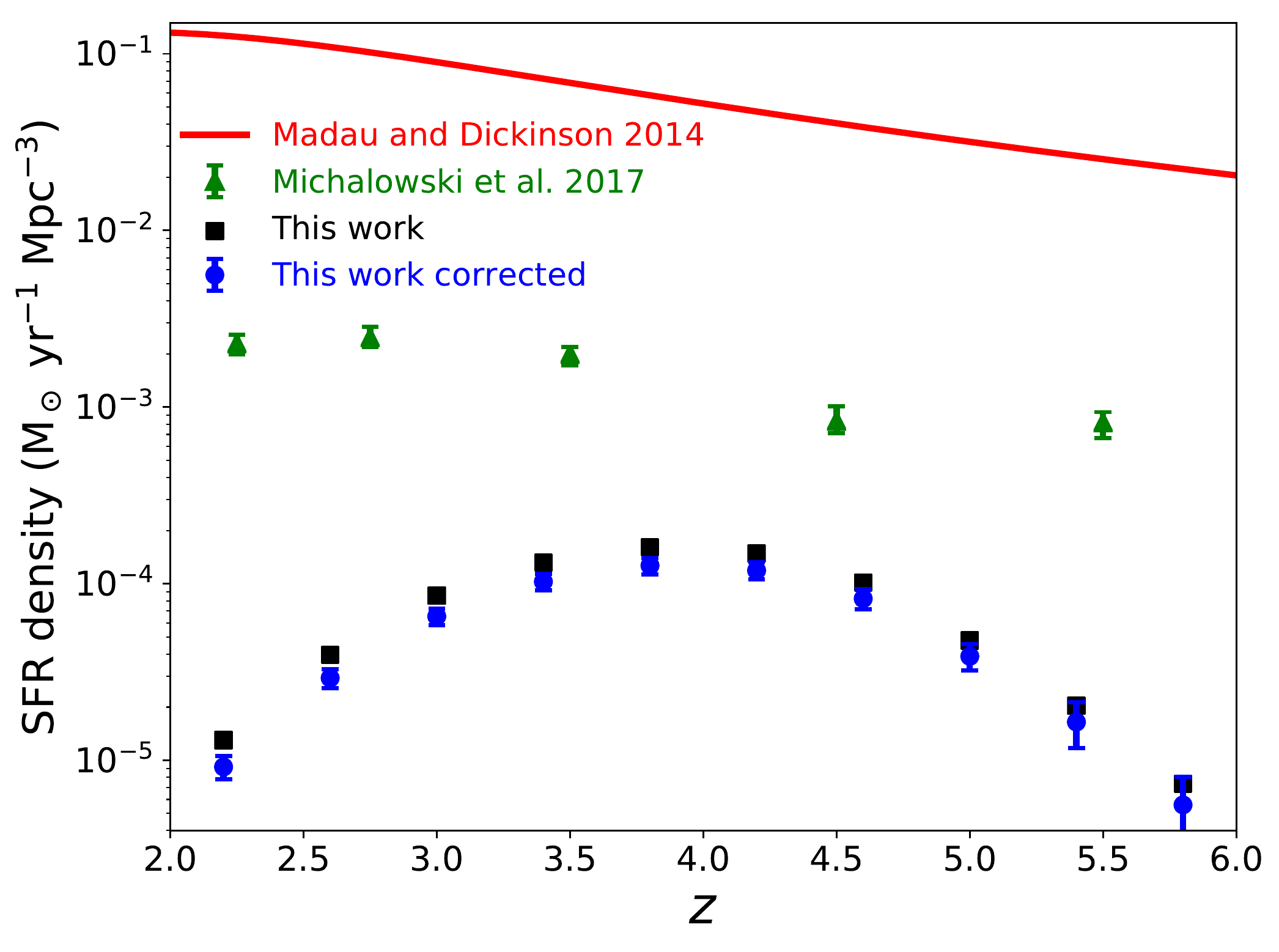}
  \caption{SFR density of sources with $S_{500}$ $>63$ mJy and $S_{500} > S_{350} > S_{250}$ in the HeLMS field in black squares, using the full redshift PDF. In blue is the corrected contribution of those sources, where contamination from AGN is removed and we corrected for flux boosting (see Section \ref{sec:quas} and \ref{sec:sp}). The red line is the  \citet{2014ARA&A..52..415M} SFRD estimates for all sources in the Universe. The green triangles are the \citet{2017MNRAS.469..492M} measurements of DSFGs with SFR $>$ 300 M$_\odot$ yr$^{-1}$ from two blank S2CLS fields.  The maximum contribution to the total SFRD is  0.3 per cent at \textit{z} $\simeq$ 4.2.} 
  \label{fig:SFRD}
\end{figure}

We use our sample to calculate the SFRD for bright 500 $\mu$m risers in the SPIRE bands as shown in Figure \ref{fig:SFRz}. The contribution to the overall SFRD is below 1 per cent at any redshift. For comparison we also show the SFRD results from the S2CLS $S_{850} \ge 4$ mJy selected sources, which is complete for galaxies with a SFR $>$ 300 M$_\odot$ yr$^{-1}$ \citep{2017MNRAS.469..492M}. The \cite{2017MNRAS.469..492M} result comes from 2 deg$^2$ blank fields, which observe the more common population of DSFGs and contribute more to the overall SFRD at any epoch.

\subsubsection{Luminosity function}

The SPIRE sources luminosity function and its evolution to z $\sim 4$ has been reported in \cite{2013MNRAS.432...23G}. We can use this luminosity function as a basis to predict the number of galaxies we expect in our sample. To get an accurate estimate for our incompleteness we need to know the relative distribution of different galaxy types at these high luminosities and redshifts. The intrinsic colours of different galaxy types can be used to determine whether or not they fulfil our selection criteria as a function of redshift.

Due to the lack of information on the distribution of galaxy types at high-redshift we have to extrapolate what we know about the distribution of SED shapes at lower redshift and luminosity to the redshifts and luminosities of our sample. We do this using the results from \cite{2013MNRAS.431.2317S}, who measured the correlation between average dust temperatures and infrared luminosities. They characterised the rising dust temperature with luminosity for a sample of $10^{11}<L_{\rm IR}/{\rm L_\odot}<10^{12.7}$ galaxies, and , to provide a simple phenomenological characterisation of this, we apply a linear fit in temperature vs. $\log{L_{\rm IR}} $ to predict the average temperature for $L_{\rm IR}/{\rm L_\odot} \ge 10^{12.5}$ galaxies. We also use the average value for the variance in the temperature for $L_{\rm IR}/{\rm L_\odot} > 10^{12}$ galaxies.

Using this temperature-luminosity-redshift distribution we draw 200 galaxies at every redshift between 1.5 and 8 ($\Delta z = 0.1$) and luminosities between $10^{12.5}<L_{\rm IR}<10^{15.0}$  ($\Delta \log{L_{\rm IR}}  = 0.1$) and then each galaxy is assigned a temperature drawn from a Gaussian with mean from the temperature-luminosity and a sigma of 6 K. This produces a mock catalogue of  325,000 galaxies, for which we have mock $T$, $z$ and $L_{\rm IR}$ values. We use the \cite{2012MNRAS.425.3094C} MBB to calculate the expected flux densities at SPIRE and SCUBA-2 wavelengths for each galaxy. The upper limit of $L_{\rm IR} = 10^{15.0}$ is used for practical reasons to simplify the drawing of a random luminosity.  It was not intended to indicate a realistic physical limit. However, the number density is dropping off very steeply at high luminosity so exactly where this cut is made makes little difference to the outcome.  

We add Gaussian noise with a mean of zero and a sigma of the mean instrumental error of our observations to simulate the variations caused by instrumental noise.  On top of the Gaussian noise we also draw a correlated confusion noise estimate for every source using the COSMOS map (see Section \ref{sec:noise}), and we add this correlated confusion to our mock observed flux density estimates. Our novel way of adding the correlated confusion noise is crucial as it partly conserves the colour of the source. The standard deviation of the confusion noise we added is 6.7, 7.1 and 6.8 mJy at 250, 350 and 500 $\mu$m, respectively and together with the instrumental noise of order 7 mJy this leads to 1$\sigma$ fluctuations of $\sim$10 mJy. It will therefore not be uncommon that sources of order 30 mJy at 500 $\mu$m will be boosted to the selection criteria of 63 mJy due to the noise and the steepness of luminosity function. 

We multiply the fraction of mock galaxies in every luminosity and redshift bin which fulfil our selection criteria by the expected space density for such galaxies \citep{2013MNRAS.432...23G} to obtain the number of galaxies we would expect in the HeLMS field. This results in a total sample of  $\sim260^{+180}_{-100}$ galaxies in our mock catalogue over an area of 274 square degrees. This is mildly larger than, but consistent with, the 200 galaxies we observed in the HeLMS field. The error bars are based on the large error on the normalisation of the luminosity function \citep{2013MNRAS.432...23G}. We do acknowledge that the consistency is partly due to the large error bars in this normalisation.

\begin{table}
  \begin{tabular}{l r} 
  model & number count \\ \hline
  Observed & 200 $\pm$ 14 \\
 \cite{beth} & 172 $\pm 18$ \\
  \cite{2013MNRAS.431.2317S} & 262$^{+184}_{-103}$\\  
  T $+$ 5 K & 54$^{+38}_{-21}$  \\ 
  T $+$ 4 K & 76$^{+54}_{-30}$ \\
  T $+$ 3 K & 85$^{+61}_{-34}$ \\
  T $+$ 2 K & 117$^{+83}_{-46}$  \\ 
  T $+$ 1 K & 170$^{+121}_{-57}$  \\
  T $-$ 1 K & 330$^{+234}_{-130}$  \\ 
  T $-$ 2 K & 373$^{+264}_{-147}$  \\
  T $-$ 3 K & 493$^{+349}_{-194}$\\
  T $-$ 4 K & 611$^{+433}_{-241}$\\
  T $-$ 5 K & 842$^{+597}_{-332}$
 \end{tabular}
  \caption{Red number counts from observations, from \citet{beth} and from our mock catalogue based on \citet{2013MNRAS.432...23G} and \citet{2013MNRAS.431.2317S}. We created additional mock catalogues with different average temperatures to show the dependency on temperature for the predicted number counts. Error bars on the mock catalogue come from the error in the normalisation of the \citet{2013MNRAS.432...23G} luminosity function, our observations error bars come from poison statistics. With the current large error bar sizes we can only exclude (difference $> 3\sigma$) the T+5 K model.}
  \label{tab:counts}
\end{table}



We make an additional 10 mock catalogues where we modify the mean temperature in the relations of \cite{2013MNRAS.431.2317S} to measure the effect of the average temperature of DSFGs on the observed number counts. In Table~\ref{tab:counts} we show the total number counts as function of  (mean) temperature. It is clear that the number of observed galaxies is a strong function of temperature and it is therefore important to get a better understanding of the distribution of galaxy types at high redshift to fully understand the number counts.   

\begin{figure*}
  \centering  
  \includegraphics[trim = 0mm 2.5mm 0mm 0mm,width = 2.\columnwidth]{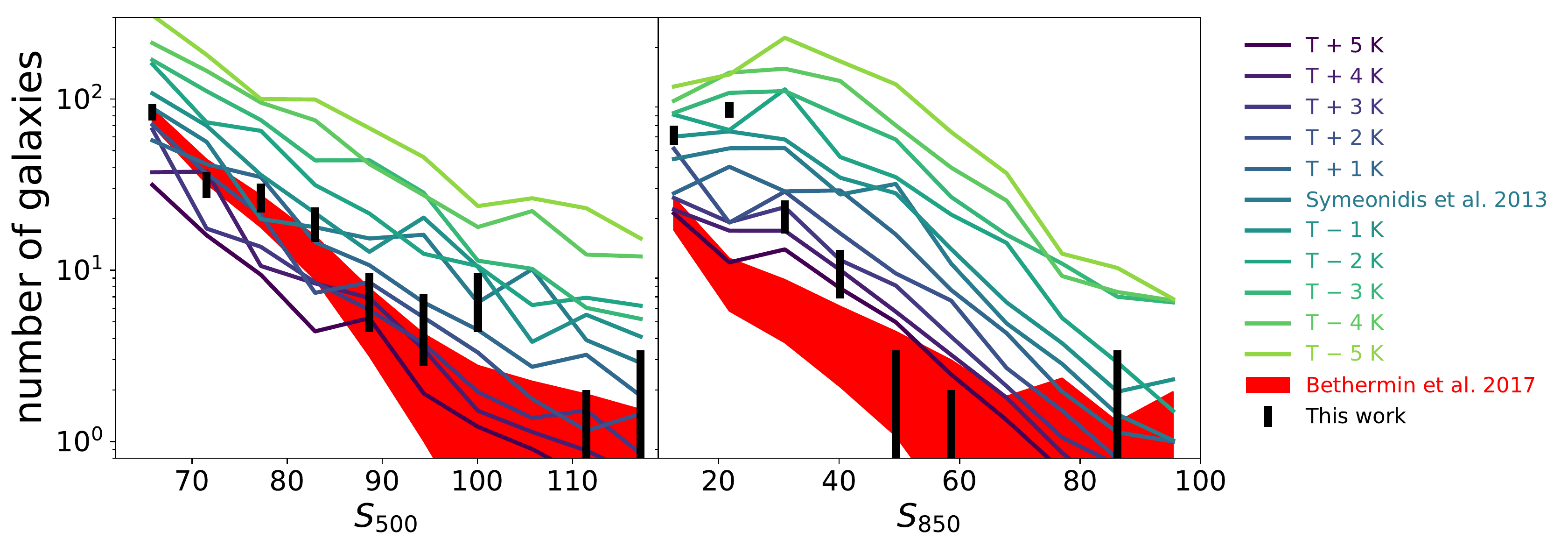}
  \caption{Number of galaxies which fulfil our selection criteria as function of 500 $\mu$m flux density on the left and as function of 850 $\mu$m on the right in black with Poisson error bars. In red the number of galaxies form the \texttt{SIDES} model \citep{beth} in combination with observational errors. The coloured lines represent the number of galaxies we expect from the \citet{2013MNRAS.432...23G} luminosity function in combination with the nominal mean temperature, and variations on that mean temperature from \citet{2013MNRAS.431.2317S}.} 
  \label{fig:mock_flux}
\end{figure*}

In Figure \ref{fig:mock_flux} we show the resulting $S_{500}$ and $S_{850}$ number counts for our mock catalogues show in Table~\ref{tab:counts}. Our mock catalogue is consistent at $S_{500}$ but over predicts the number of bright sources at $S_{850}$, even when we raise the temperature of our mock catalogues with 5~K we keep over predicting the number of sources at $S_{850}$ $>$ 50 mJy.

\begin{figure}
  \centering  
  \includegraphics[trim = 0mm 2.5mm 0mm 0mm,width = 1.\columnwidth]{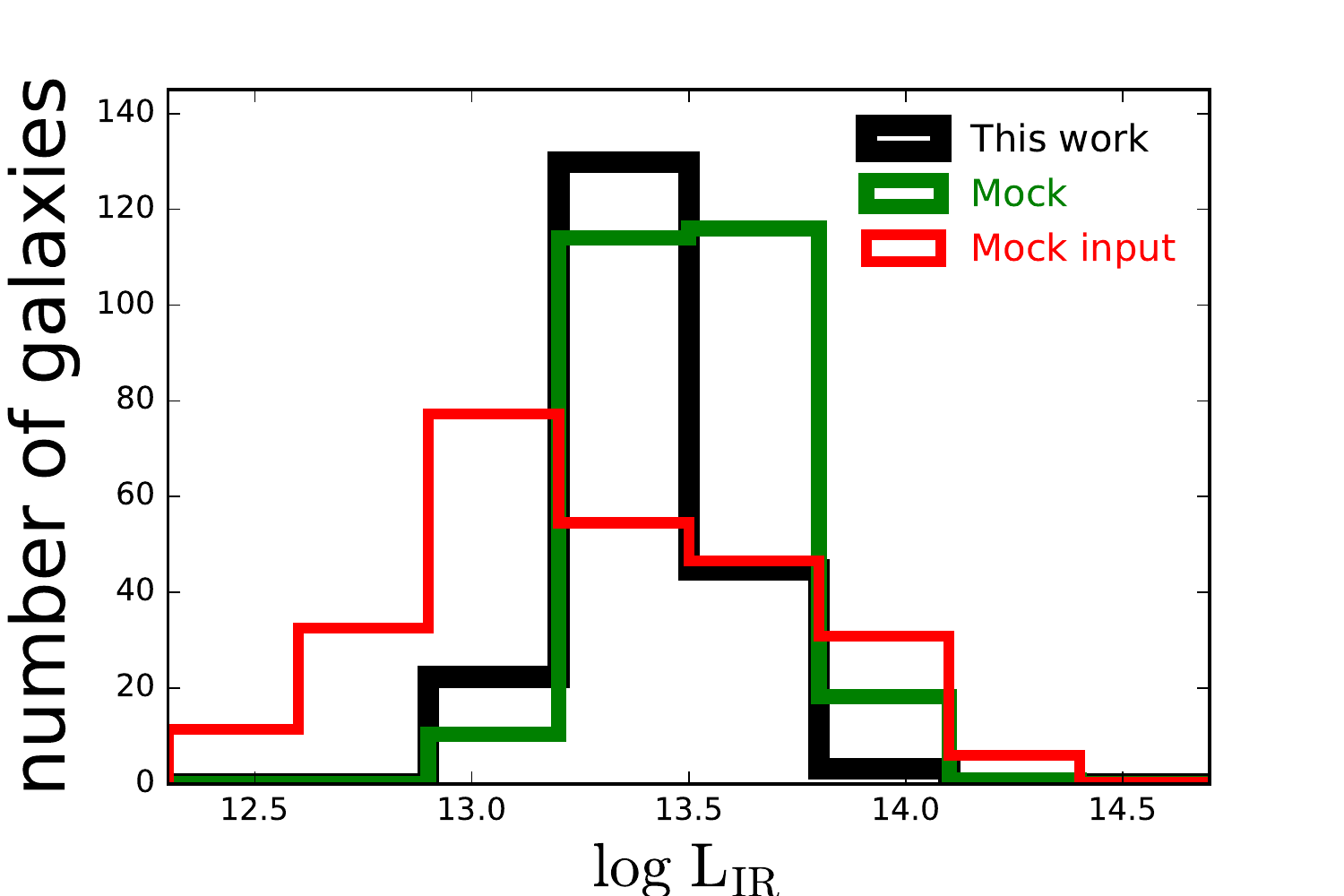}
  \caption{Luminosity histogram of 500 $\mu$m riser galaxies in the HeLMS field in black. In green we show the output from our pipeline for the mock catalogue obtained from sampling galaxies from the \citet{2013MNRAS.432...23G} luminosity function and adding observational uncertainties to them. In red, we show the input luminosities for the mock sample shown in green.} 
  \label{fig:L_func}
\end{figure}

We use our mock model as input for \texttt{EAZY} to predict the observed luminosity function using our method. On top of the 200 galaxies we have already drawn at every redshift and luminosity bin we draw an additional 100 galaxies for every very bright bin ($L_{\rm IR}/{\rm L_\odot} > 10^{13.5}$), an additional 300 galaxies for the $10^{13.1} < L_{\rm IR}/{\rm L_\odot} < 10^{13.5}$ bins and an additional 500 galaxies for the $L_{\rm IR}/{\rm L_\odot} < 10^{13.1}$ bins. these extra galaxies lead to a total mock size to test the luminosity function of 630,500 galaxies. These extra galaxies give us extra statistics on the lower end of the luminosity function, where galaxies are intrinsically not bright enough to be detected with our detection method but might be very occasionally scattered up by noise. In Figure \ref{fig:L_func} we compare the predicted luminosity with the calculated luminosities for our galaxies.

From Figure \ref{fig:L_func} we can see that the simulated galaxies are scattered up in luminosity due to confusion and instrumental noise. This is a flux boosting effect, well-known in sub-mm surveys \citep[e.g.][]{2005MNRAS.357.1022C,2006MNRAS.372.1621C}. From our mock catalogue we derive that 61 per cent of the mock galaxies which observational properties fulfil our selection criteria are intrinsically not bright enough and are scattered up due to confusion and instrumental noise. We use the average boosting factor (difference between input and output Luminosity of our Mock) to correct our SRFD in Figure \ref{fig:SFRD}.

\subsubsection{Comparison with simulations}

The Simulated Infrared Dusty Extragalactic Sky \citep[\texttt{SIDES}, ][]{beth} includes a 274 deg$^2$ simulation to match the size of the HeLMS field. The size of the model and its capability to simulate the observed FIR and submillimetre flux densities makes it ideal for comparison with our observations.   

The main \texttt{SIDES} model predicts the FIR and submillimetre emission in a 2 deg$^2$ light cone, which simulates clustering by using abundance matching to populate dark matter haloes with galaxies according to their star formation evolution model. This model is accurate in describing the number counts at 350 and 500 $\mu$m. This 2 deg$^2$ light cone is not a large enough volume to get accurate predictions for our rare sources. \cite{beth} tackled this problem by producing the 274 deg$^2$ simulation to predict number counts for much rarer (brighter) sources but this larger simulation does not contain any clustering estimates. 

The number of sources in the 274 deg$^2$ \texttt{SIDES} model which fulfil the \cite{2016MNRAS.462.1989A} criteria is 22, and all are strongly lensed. This number goes down to 11 in the case we use our $S_{500} > $ 63 mJy cut on top of the \cite{2016MNRAS.462.1989A} criteria. These numbers are an order of magnitude lower that the bright red sources found in the HeLMS field. 

Those results do not account for the effect of flux boosting by both instrumental and confusion noise. \cite{beth} calculated this effect of flux boosting by adding random (Gaussian) instrumental and confusion noise to the fluxes. This increased the number count to 114 sources which fulfil the \cite{2016MNRAS.462.1989A} criteria and 35 sources when we add $S_{500} > $ 63 mJy constraint. The 2 deg$^2$ \texttt{SIDES} model was used to calculate the effect of clustering on these number counts. They found that the confusion which arises from clustering increases the number of red sources by  a factor of 1.7$^{+1.9}_{-0.9}$. This leaves them with an estimate of 229$^{+258}_{-121}$ sources which is within $1\sigma$ of the 477 sources found in \cite{2016MNRAS.462.1989A}. This boosting factor of 1.7$^{+1.9}_{-0.9}$ is however not high enough to boost the 35 sources in the 274 deg$^2$ \texttt{SIDES} model to the 200 sources found in the HeLMS field. 

Our method of drawing correlated confusion noise estimates provides us with a different way of using the 274 deg$^2$ \texttt{SIDES} model to predict the number of sources in the HeLMS field. We do this by adding both random Gaussian instrumental noise, and our correlated confusion noise estimates to the \texttt{SIDES} 274 deg$^2$ catalogue. This noise increases the number of sources from 11 to 172$\pm 18$ (where the noise only accounts for different sets of random numbers and Poisson noise, and does not account for any other uncertainties in the \texttt{SIDES} model), which is very close to 200 sources which were detected with our selection criteria (see Table \ref{tab:counts} and Figure \ref{fig:mock_flux}). 

\begin{figure}
  \centering  
  \includegraphics[trim = 0mm 2.5mm 0mm 0mm,width = 1.0\columnwidth]{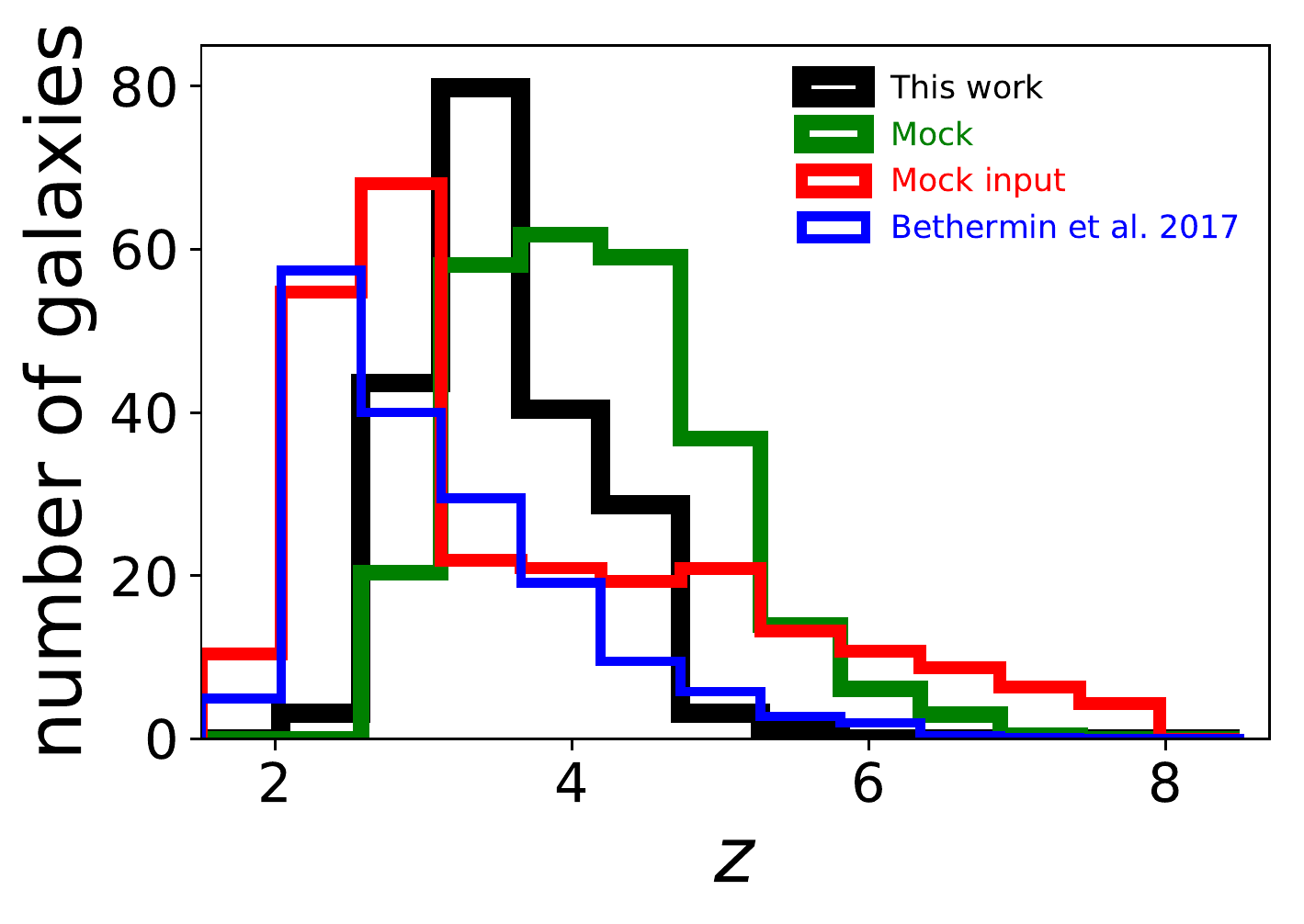}
  \caption{Redshift distributions of our observations (black), the mock (green) catalogue, the mock input (red) and the \citet{beth} model (blue).} 
  \label{fig:r_func}
\end{figure}

17 per cent of these 172 sources are strongly lensed and the mean redshift is $3.1 \pm 0.9$. Figure \ref{fig:r_func} shows the full redshift distribution of our data compared with the \texttt{SIDES} model and our mock catalogue.

From Figure \ref{fig:r_func} we can see that the redshift distribution of the mock has a larger tail to higher redshifts than our observations. We test if there is any significant net bias we calculate the mean of the observed mock and input mock redshifts, we calculate the error on this mean using jack-knife samples. We find  a different value for the mean redshift (4.17  $\pm$ 0.04 q.v. 3.69  $\pm$ 0.08), which is smaller than the RMS of the refshifts of 0.6, but nevertheless statistically significant. Flux boosting can happen at every wavelength band but because of our 500 $\mu$m riser selection we are biased towards selecting galaxies which are boosted  at 500 $\mu$m. These selected galaxies look therefore redder than they truly are, which results in a over estimate of the redshift. This argument mainly holds for galaxies which are intrinsically not red or bright enough to fulfil our selection criteria. For all galaxies we see the same trend as in Figure \ref{fig:spec}, where our redshifts are overestimated at high redshift and under estimated at low redshifts. As we stated in more detail in Section \ref{sec:testz} this trend is partly due to selection effects and due to the prior pushing us towards mean and not ``extreme"  redshift estimates.

The 274 deg$^2$ \texttt{SIDES} model model has a comparable high redshift tail, but this model peaks at lower redshift, causing the mean redshift to be lower (3.1 $\pm$ 0.07  q.v. 3.6 $\pm$ 0.04 from our observations).

\subsection{SDSS and WISE quasars} \label{sec:quas}

We cross-matched the 188 galaxies with the SDSS quasar catalogue \citep{2017A&A...597A..79P} and found two matches within 20 arcsec. We test the change on a random alignment with a a SDSS quasar by taking  50,000 random positions in the HeLMS field and see how many of these random positions match with a SDSS quasar within a 20 arcsec radius. The number of matches is 127, leading to a probability of 0.25 per cent that there is a random alignment within 20 arcsec. Using this statistics we would expect that there is a 38 per cent chance that at least one of our object is randomly aligned with a SDSS quasar and there is a probability of 8 per cent for at least two alignments. 

HELMS\_RED\_80 is located 3 arcsec from SDSS\_J005036.93+014449.1 which has a redshift of 3.4351$\pm 0.0003$. Our estimated photometric redshift is 3.65$^{+0.65}_{-0.7}$, which is within 1$\sigma$ agreement with the quasars spectroscopic redshift. The quasar is furthermore detected in WISE-1, WISE-2 and WISE-3. We use the intrinsic quasar SED derived in \cite{2016MNRAS.459..257S} in combination with the WISE magnitudes to calculate the AGN contribution to the FIR luminosity. This contribution is estimated at $\log (L_{FIR}) = 12.97^{+0.11}_{-0.12}$ and is a factor of $\sim$ 3 lower that our measured Luminosity. We thus conclude that it is likely that HELMS\_RED\_80 is associated with SDSS\_J005036.93+014449.1 and that the quasar contaminates our SFR estimate.

HELMS\_RED\_421 is located 12 arcsec away from SDSS\_J000127.11-010603.1 which has a redshift of 1.934$\pm 0.001$.  Our estimated photometric redshift is 2.95$^{+0.7}_{-0.8}$, which is in 1.3$\sigma$ tension with the quasars spectroscopic redshift. The separation of 12 arcsec is furthermore in 2$\sigma$ tension with our positions. The quasar is not detected in any WISE bands, but there is a nearby (z = 0.163) SDSS galaxy  9.0 arcsec away from our SPIRE detection which is detected in all 4 WISE bands $> 5 \sigma$ (WISE\_J000127.76-010607.5). Furthermore, WISE\_J000126.74-010612.2 is located 9.6 arcsec away and is detected in WISE-1 and WISE-2 and  WISE\_J000127.44-010626.6 is located 12.4 arcsec away and has besides a WISE-1 and WISE-2 detection a 3.3$\sigma$ detection in WISE-3. The location of our SPIRE source lies in the middles between those 4 WISE/SDSS sources, indicating that this source is likely contaminated by several of those galaxies. We tested the probabilistic de-blender XID+ \citep{2016MNRAS.tmp.1477H}using the default flat uniform flux prior ( as used for the HELP database) to disentangle the SPIRE flux densities over the four sources. XID+ with a uniform flux prior, assigns the flux evenly among them as they are all located at roughly the same distance from the centre of the SPIRE emission. We note XID+ can be run with more sophisticated priors, using both SED and redshift information, however this requires thorough analysis and so we leave the nature of this SPIRE detection for future work.

HELMS\_RED\_421 may be associated with  SDSS\_J000127.11-010603.1 but would be consistent with a spurious coincidence. The percentage of the FIR luminosity which is caused by the (potential) quasars is a function of the AGN luminosity \citep{2012A&A...545A..45R,2016MNRAS.459..257S,2017MNRAS.465.1401S}, which we do not know. We therefore exclude the source from our final corrected SFRD.   
 
\subsection{Sub-mm interferometry} \label{sec:summ}
  
\begin{figure*}
  \centering  
  \includegraphics[trim = 15mm 5mm 0mm 7mm,width = 2.\columnwidth]{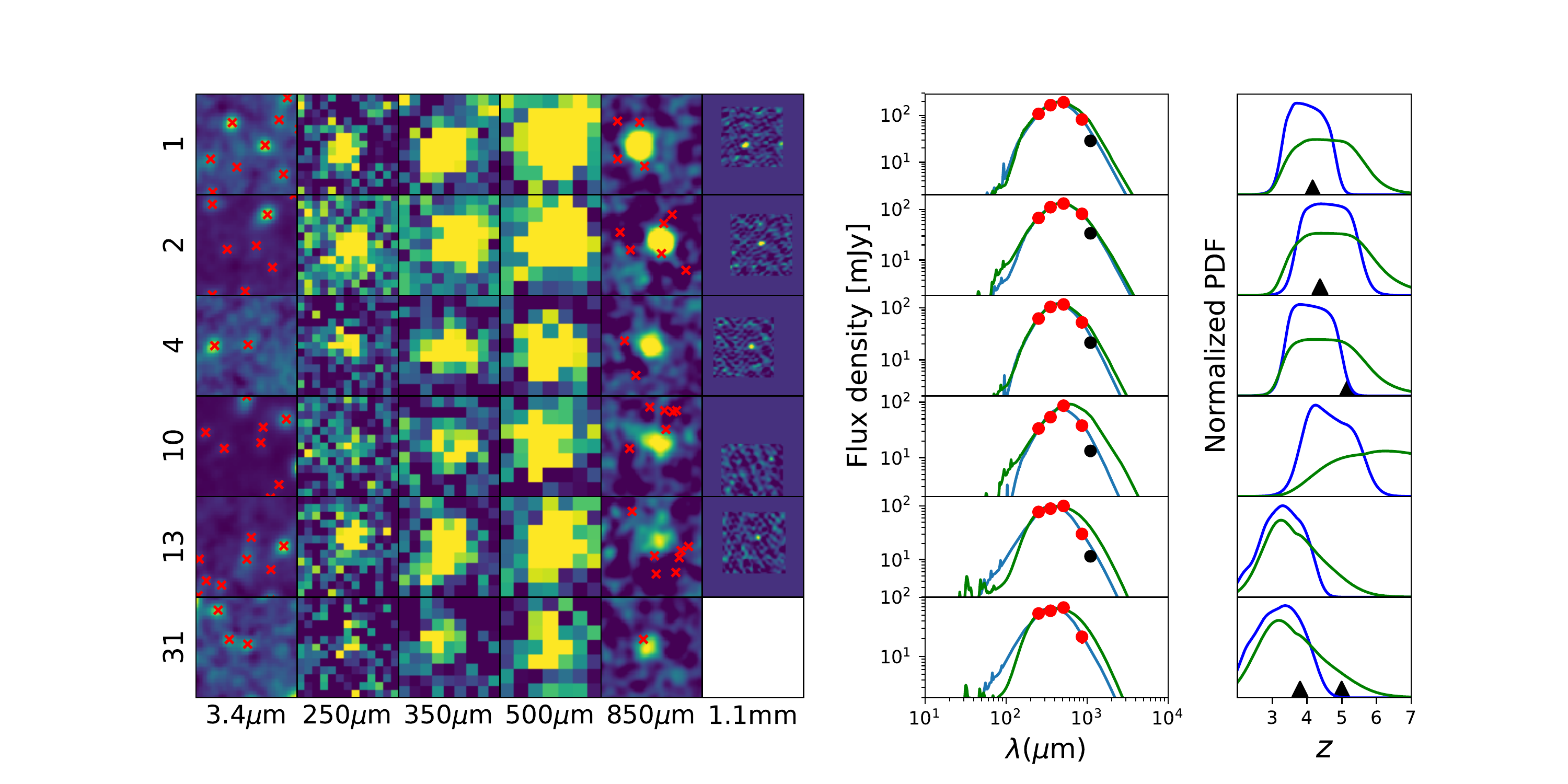}
  \caption{WISE-1 (3.4 $\mu$m), SPIRE (250, 350, 500 $\mu$m), SCUBA-2 (850 $\mu$m) and SMA (1.1~mm) 70 arcsec $\times$ 70 arcsec cut-outs of bright $S_{850}$ sources in the HeLMS field with ancillary sub-mm interferometry data. The wavelength of each image is noted on the bottom of the plot in $\mu$m and the source ID (see Appendix \ref{ap:A}) on the left. The second on the right shows the best-fit SED in blue, the best-fit SED using only SPIRE in green and the flux density from SPIRE and SCUBA-2 in red. The right panel shows the redshift PDF of our sample in blue, and the PDF if we exclude the SCUBA-2 data in green (showing the improvement in constraining the redshift by including longer wavelength data). The black triangles show the spectroscopic redshifts derived from ALMA and CARMA, where the two black triangles for HELMS\_RED\_31 show the redshift in the case the line detection is the CO(5-4) or the CO(4-3) line. The red crosses on top of the WISE bands show 5$\sigma$ source detections in WISE-1. On top of the SCUBA-2 image we overlay all SDSS-detected galaxies in red. } 
  \label{fig:SED_inf}
\end{figure*}

We use the high-resolution SMA data, the ALMA and the CARMA redshifts to more closely examine the properties of the subset of galaxies possessing this information. The images and SED fits of the 6 galaxies with  interferometry data are shown in Figure~\ref{fig:SED_inf}. We now discuss the sources individually below:
 \begin{itemize}
\item HELMS\_RED\_1: The photometric redshift of $4.0^{+0.55}_{-0.5}$ is consistent with the spectroscopic redshift of 4.163 which is obtained with CO(4-3) and CO(5-4) line detections (Riechers et al. in prep).  The 500 $\mu$m flux density of 192 mJy suggests that the object is lensed \citep[e.g.][]{2017MNRAS.465.3558N}. This source was also detected with ACT with flux densities of  $12.49\pm1.74$, $35.11\pm 2.62$ mJy and 72.32$\pm$6.26 mJy at 148 (2.0), 218 (1.4) and 278 (1.1) GHz (mm), respectively. Our best fit SED predicts flux densities of 7, 24 and 47 mJy at those frequencies, which are considerably lower. The SMA flux density at 1.1~mm is $28.6\pm2.3$ mJy, which is less than half that of the ACT value at 278 GHz which is observed at a similar wavelength but with a much larger beam. The predicted 1.1~mm flux density form our best fit SED is 46.6 mJy. The nearest WISE-1 or SDSS source near to the SMA position is 15.6 arcsec away. The SMA position is 3.9 arcsec away from the SCUBA-2 position.

\item HELMS\_RED\_2: The photometric redshift of $4.6^{+0.65}_{-0.65}$ is consistent with the spectroscopic redshift of 4.373, which is obtained with CO(4-3) and CO(5-4) line detections (Riechers et al. in prep). The 500 $\mu$m flux density of 132 mJy means the object is likely to be lensed.  The SMA flux density is $33.9\pm2.25$ and the predicted 1.1~mm flux density form our best fit SED is 53.9 mJy. The nearest WISE-1 or SDSS object near the SMA position is 2.0 arcsec away, the location of the source is J005258.53+061317.5 and has a WISE-1 AB magnitude of $17.5 \pm 0.2$. The SMA position is 2.0 arcsec away from the SCUBA-2 position.

\item HELMS\_RED\_4: The photometric redshift of $4.15^{+0.6}_{-0.6}$ is in 1.7$\sigma$ tension with the spectroscopic redshift of 5.162, which is obtained with CO(5-4) and CO(6-5) line detections \citep{2016MNRAS.462.1989A}. The 500 $\mu$m flux density of 116 mJy makes the object likely to be lensed. The SMA flux density is $21.3\pm1.9$ mJy and the predicted flux density at 1.1~mm from our best fit SED is 29.8 mJy. The nearest WISE-1 or SDSS object near the SMA position is 1.0 arcsec away, the location of the source is J002220.73-015520.2  and has a WISE-1 AB magnitude of $17.4 \pm 0.2$. The SMA position is 1.5 arcsec away from the SCUBA-2 position.

\item HELMS\_RED\_10: The photometric redshift is $4.6^{+0.75}_{-0.6}$. The SMA flux density of $13.3\pm2.8$ and the predicted 1.1~mm flux density form our best fit SED is 24.5 mJy. The nearest WISE-1 or SDSS object near the SMA position is 8.7 arcsec away.  The SMA observations are not centred on the SCUBA-2 position and the brightest peak is 4.7$\sigma$. The SMA position is 13.4 arcsec away from the SCUBA-2 position.  It is unclear if the SMA sources is the same source as our SPIRE/SCUBA-2 detection more detail of this sources will be provided in Greenslade et al., in prep. 

\item HELMS\_RED\_13: Our photometric redshift of $3.3^{+0.6}_{-0.65}$. The SMA flux density is $11.5\pm1.8$ mJy and the predicted flux density at 1.1~mm from our best fit SED is 19 mJy. The nearest WISE-1 or SDSS object near the SMA position is 3.6 arcsec away. The SMA position is 2.9 arcsec away from the SCUBA-2 position.

\item HELMS\_RED\_31:   This object has a single line detection which might be either the CO(5-4) or  the CO(4-3) transition \citep{2016MNRAS.462.1989A} suggesting a redshift of 3.798 or  4.997. The photometric redshift of $4.15^{+0.75}_{-0.75}$ is consistent with the lower  redshift from and in a small ($1.1\sigma$) tension with  $z=4.997$. The nearest WISE-1 or SDSS source is 4 arcsec away from the SCUBA-2 position.
\end{itemize}

\subsection{Extreme sources}

We isolate a subset of potentially high-redshift extremely bright galaxies. This subset consists of galaxies which have a clear detection with SCUBA-2 ($S_{850} \geq 5\sigma$) as well as a redshift PDF which has 50 per cent of its probability at $z > 4$. In total we find 21 galaxies fulfilling those conditions, which includes HELMS\_RED\_2 , 4, 10 and 31.  Figure \ref{fig:SED_all} shows the WISE-1, SPIRE and SCUBA-2 cut-outs of these sources, excluding the ones we already discussed in Section \ref{sec:summ}. 


These sources might contain some of the highest redshift DSFGs ever detected. Therefore this catalogue provides a high priority sample for spectroscopic follow-up with ALMA. High-resolution follow up observations are also required for accurately determining the blending fraction (see Section \ref{sec:blend}) for these types of sources. 

\begin{figure*}
  \centering  
  \includegraphics[trim = 15mm 20mm 0mm 7mm,width = 2.\columnwidth]{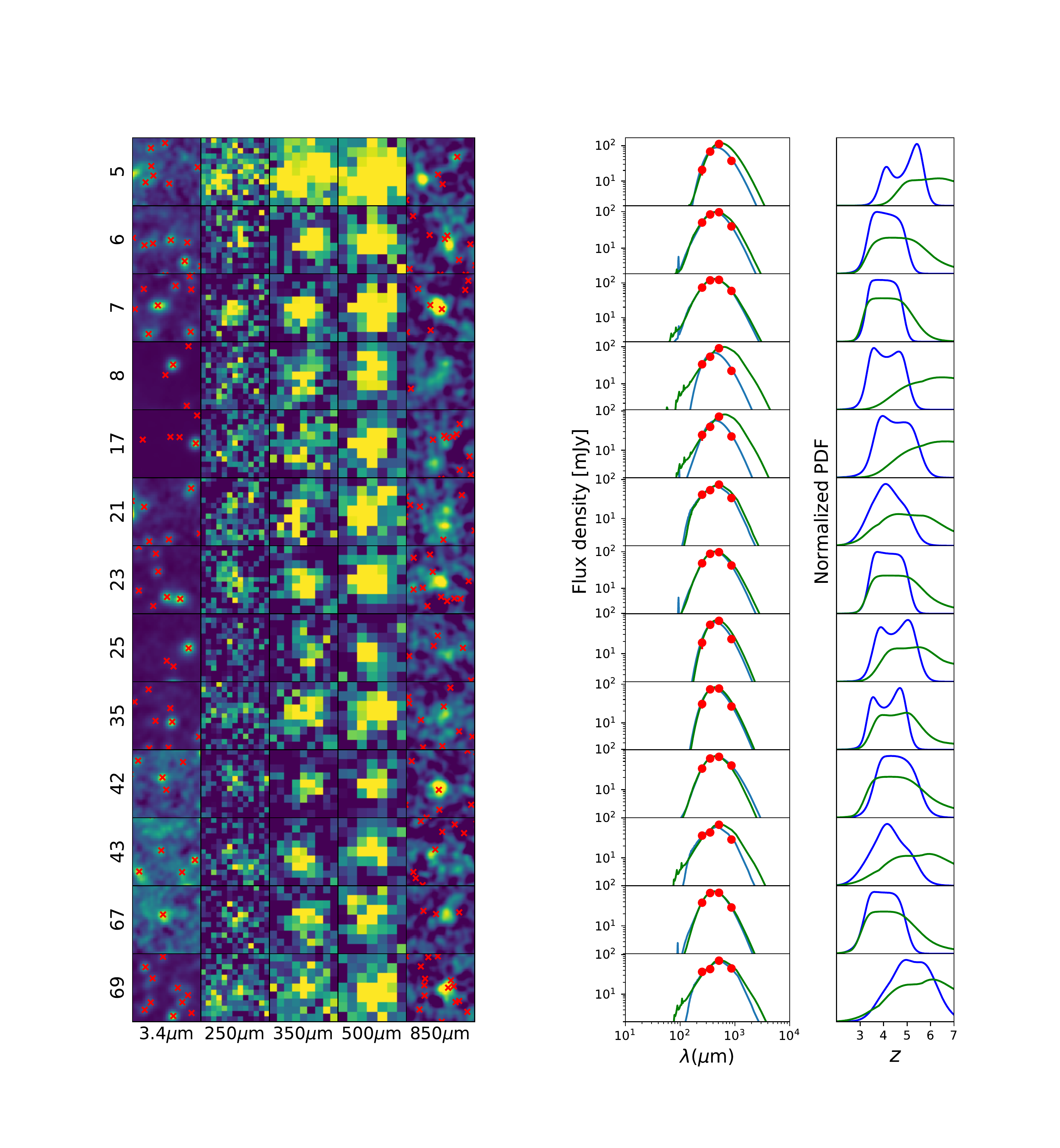}
  \caption{WISE-1 (3.4 $\mu$m), SPIRE (250, 350, 500 $\mu$m) and SCUBA-2 (850 $\mu$m) 70 arcsec $\times$ 70 arcsec cut-outs of bright $S_{850}$ sources in the HeLMS field. The wavelength of each image is noted on the bottom of the plot in $\mu$m and the source ID (see Appendix \ref{ap:A}) on the left. The second on the right shows the best-fit SED in blue, the best-fit SED using only SPIRE in green and the flux density from SPIRE and SCUBA-2 in red. The right panel shows the redshift PDF of our sample in blue, and the PDF if we exclude the SCUBA-2 data in green (showing the improvement in constraining the redshift by including longer wavelength data). The red crosses on top of the WISE bands show 5$\sigma$ source detections in WISE-1. We overlay all SDSS-detected galaxies in red over the SCUBA-2 images.} 
  \label{fig:SED_all}
\end{figure*}

\addtocounter{figure}{-1}
\begin{figure*} 
  \centering  
  \includegraphics[trim = 15mm 26mm 0mm 7mm,width = 2.\columnwidth]{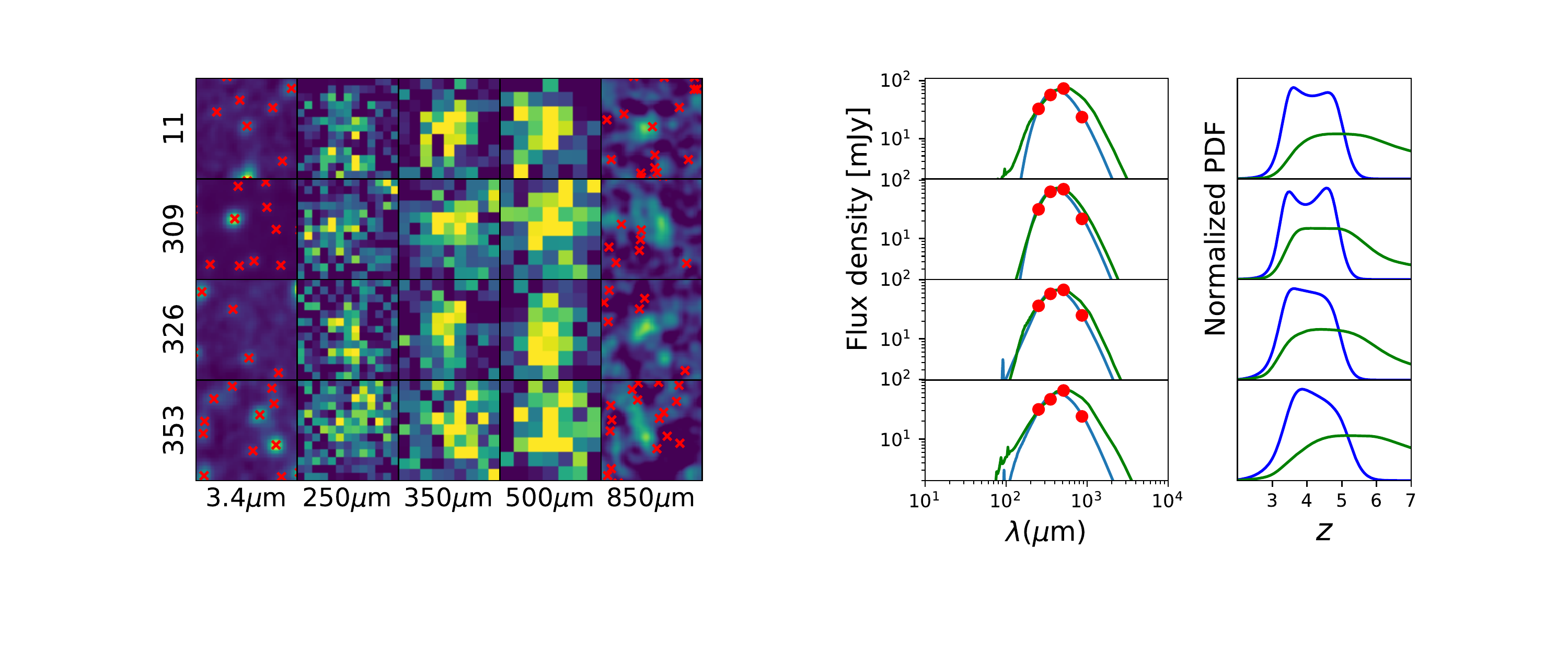}
  \caption{(Continued)} 
\end{figure*}

Our candidate with the highest chance of being a z $\ge$ 6 galaxy is HELMS\_RED\_69. Its redshift is estimated to be 5.19$^{+0.89}_{-0.92}$ and 19 per cent of its redshift PDF lies above a redshift of 6.  Another remarkable feature of HELMS\_RED\_69 is that its 500 $\mu$m flux density is 1.5 times higher than that of HFLS3. There is a possibility that this source has been lensed by a foreground galaxy as we find an SDSS counterpart at a distance of 3.0 arcsec.

\section{Discussion} \label{discussion}
\subsection{Blending and Lensing}  \label{sec:blend}

Due to the relatively large beam of the 500 $\mu$m data there is a high probability that in many cases some parts of the measured flux density comes from randomly aligned galaxies or companion galaxies of the main source \citep[confusion,][]{2010A&A...518L...5N}. \newline

ALMA observations \citep{2013MNRAS.432....2K,2013ApJ...768...91H} of bright LABOCA ($S_{870} >$ 12 mJy) sources in the 0.25 degree$^2$ LESS survey \citep{2009ApJ...707.1201W} showed that these sources contain emission from several fainter sources with an upper limit of 9 mJy per source, in later work this fraction of sources breaking up is found to be less significant \citep{2015ApJ...807..128S}. This indicates that there might be a maximum SFR for DSFGs of $10^3$ M$_\odot$ yr$^{-1}$ (Chabrier IMF). \cite{2015ApJ...812...43B} found that 20 out of 29 bright SPIRE sources ($S_{500}$ = 52-134 mJy) break down into multiple ALMA sources, and of the 9 isolated sources 5 have a magnification factor larger than 5. \cite{2015ApJ...807..128S} found that 61$^{+19}_{-15}$ per cent of their sample of bright galaxies (median $S_{850} \pm$ 0.4 mJy) consist of a blend of 2 or more sources in the ALMA maps. Their sample was selected to be representative of the bright end of the 1 degree$^2$ deep 850 $\mu$m S2CLS field. The brightest detection with ALMA had a flux density of 12.9 $\pm$ 0.6 mJy and is considerably brighter than the sources observed in \cite{2013MNRAS.432....2K}. \cite{2017MNRAS.469..492M} found that bright DSFGs found in SCUBA-2 blank fields (around 10 mJy) typically have a second component of about 1-2 mJy. Furthermore, they found that the bright end of the source counts is hardly affected by replacing from SCUBA-2  flux densities with those from ALMA. The survey was taken over an area of 2 deg$^2$. 

The bright end of the \cite{2013MNRAS.432....2K}, \cite{2015ApJ...807..128S} and \cite{2017MNRAS.469..492M} sources are fainter than 20 mJy, and are thus much fainter than our Group 1 galaxies. Hence it would be interesting to see if our brightest sources are also characterised by having a second component of about 1-2 mJy or a 61$^{+19}_{-15}$ per cent blending fraction.  

Prior-based source extraction \citep[XID+][]{2016MNRAS.tmp.1477H} to investigate  multiplicities of bright \textit{Herschel} sources at $250\ \mu$m in the COSMOS field show that the brightest component contributes roughly 40 per cent of the source flux density \citep{2016MNRAS.460.1119S}.

The multiplicity due to blending seen in these studies is a potential concern. Blending of objects at the same redshift will not seriously impact on the redshift determination, although we will determine the luminosity and star formation of the combined system, rather than a single object.  Blending of two (or more) objects at different redshifts will produce composite SEDs which are likely to elicit an intermediate redshift estimate. We derive from our mock observations that $\sim$ 60 per cent of our detected galaxies are likely to be scattered up to our selection criteria due to flux boosting partly caused by blending with foreground objects. Some of those boosting factors are as large as 0.5 dex, but can be explained by instrumental and confusion noise. An example of such a large effect might be HELMS\_RED\_421 where the SPIRE position is in the middle of three WISE sources and a SDSS detect quasar.

The advantage  with our sample is that we probed a much wider field, over 100 times wider than COSMOS and S2CLS and more than 1000 times bigger than the area targeted by the ALMA observations of the LESS field. Our sample is therefore expected to be comprised of a much rarer and more luminous  and less confused population of sources. However, a proper investigation of the blending of these objects is deferred until we are able to obtain high-resolution data of a significant sub-set.

HFLS3 has an observed flux density of 35.4$\pm 0.9$ mJy at 850 $\mu$m \citep{2014ApJ...793...11R}, which is comparable to our Group 1 galaxies. \cite{2013Natur.496..329R} found that HFLS3 is only marginally magnified by a factor of 1.2-1.5 by a foreground lens. This magnification factor was updated by \cite{2014ApJ...790...40C} to a factor of 2.2 $\pm$ 0.3, which yields a SFR of 1320 $M_\odot$ yr$^{-1}$.

Another explanation for the high SFRs in our sample is that the galaxies might be lensed. We do not expect to detect weak lensing from high-redshift lenses or even unambiguous confirmation of large magnifications from nearby lenses from our current data. We can however asses the likely incidence of lensing statistically by looking at the density of WISE-1 and SDSS sources near our SCUBA-2 detections. For this we use our Group 1 galaxies which have $>5\sigma$ SCUBA-2 detections. For this Group we can use the SCUBA-2 positions with a statistical positional accuracy of $\sigma_{pos} = 0.6 \times \frac{{\rm FWHM}}{S/N} $ \citep{2007MNRAS.380..199I} which is of the order of 2 arcsec, which is comparable with the JCMT pointing accuracy of 2-3 arcsec. We combine these 2 uncertainties ($\sigma_u^2 = \sigma_{pos}^2 + \sigma_{JCMT}^2$) and assume that either the optical/NIR counterpart or the lens should lie within a $\approx 2\sigma_u$ $\approx$ 7 arcsec annulus around our source position.

We use this 7 arcsec aperture to count the number of $5\sigma$ WISE-1 detected objects near our group 1 sources and find that 53 per cent have a nearby WISE galaxy. We calculate the significance of this number by using the same aperture at 64 (same number as Group 1 galaxies) random positions in the HeLMS field 1000 times. With these 1000 runs we can calculate both the expectation value and the 84.1, 97.8 and 99.9 percentiles.

\begin{figure*} 
  \centering  
  \includegraphics[width = 1.7 \columnwidth]{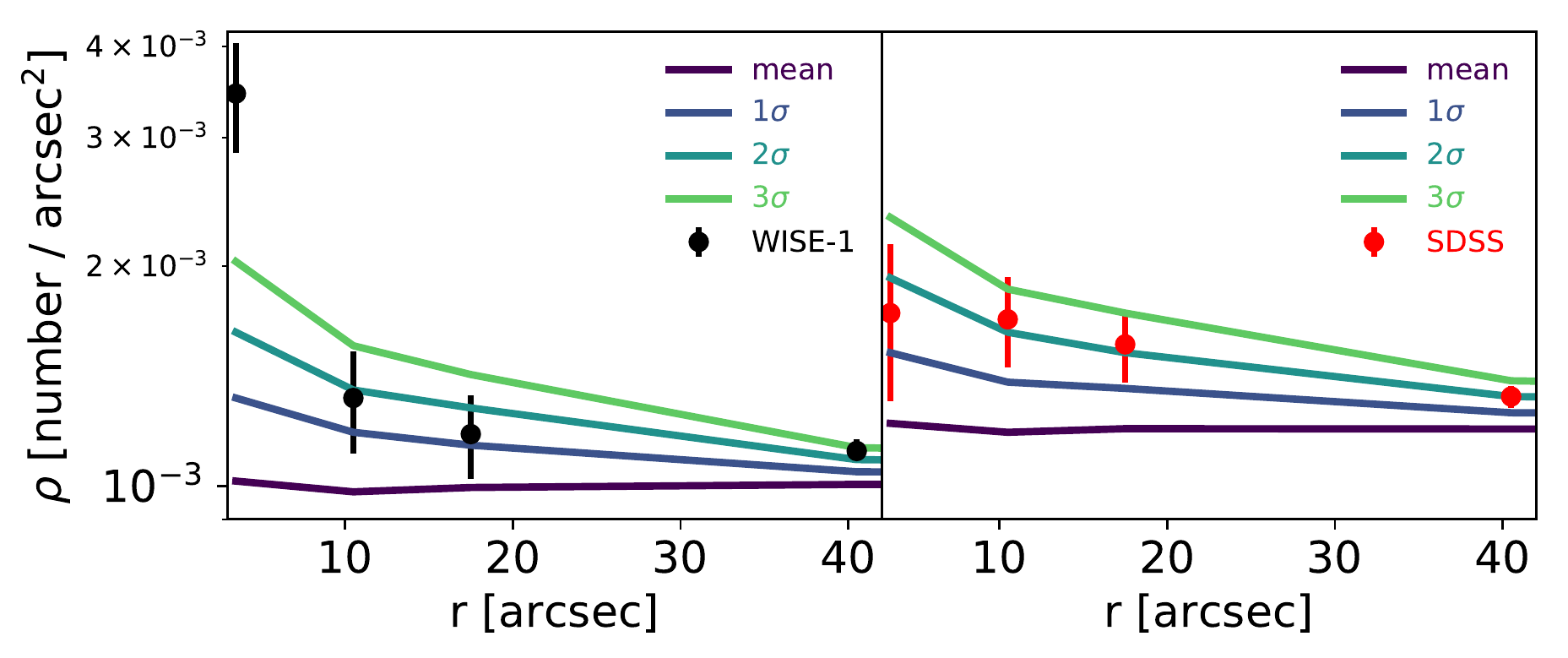} 
  \caption{Surface density of WISE-1 sources (left, black), and SDSS galaxies (right, red). There is a significant increase in low redshift sources (possible lenses) near our targets in comparison with average number counts (solid lines). The higher number density near our targets can be caused by an overlap of the redshift distribution of optical/NIR sources and sub-mm sources and/or a signature of foreground lensing.} 
  \label{fig:wise}
\end{figure*}

In Figure \ref{fig:wise} we show our results for our 7 arcsec aperture and several larger apertures. It is clear that there is a significant overdensity of WISE-1 sources near our Group 1 galaxies. The total space density of WISE-1 sources is a factor of 3 higher, which is a strong indication that part of this sample is lensed \citep{2011MNRAS.414..596W}. We perform the same measurement on a $S_{500} >$ 100 mJy subset and find that 7 of the 9 sources have a WISE-1 counterpart within the 7 arcsec annulus. This leads to an even higher WISE-1 space density compared to our Group 1 sample, but due to the low number of galaxies this is less significant than our Group 1 overdensity. \cite{2014MNRAS.442.2680G} found a similar result by cross-correlating SDSS and GAMA $0.2 \le z \le 0.6$ galaxies with $S_{350} \ge 30$ mJy \textit{H}-ATLAS sources. They found a $>10\sigma$ spatial correlation, which for non-overlapping redshift distributions can be explained by weak gravitation lensing, where the weak regime has a lensing factor smaller than 2.

We do, however, wish to stress that this is a statistical measurement, and we lack accurate enough positions and morphologies for our DSFGs to do a proper lensing analysis to yield a magnification estimate. Furthermore, we lack sufficiently deep optival data to find potential lenses at z $>$ 0.5. For our estimates of the $L_{\rm IR}$ and SFR we do not take this significant foreground source detection into account. 

An overlap between the redshift distribution of WISE-1 sources and 500 $\mu$m risers could provide another explanation for the excess WISE-1 sources seen near our high-redshift DSFGs. \cite{2013arXiv1303.4722M} used the cross-correlation between SDSS quasars and WISE sources to recover the redshift distribution of WISE sources. They found that there is a potential sub-sample of WISE sources with a redshifts $>$ 2, indicating that it is possible that our high-redshift sample could be detected in WISE.  However, their red WISE sources with z $>$ 2 are at least 1.2 magnitudes brighter in the WISE-2 band than in WISE-1 and several orders of magnitude brighter in the WISE-3 band.

We find that 34 of our 64 Group 1 galaxies have at least one WISE-1 source within the 7 arcsec aperture. Of these 33 closest WISE sources, only 2 have a WISE-2 magnitude 1.2 brighter than WISE-1. The remaining 31 are either undetected in WISE-2 (and we would have detected them if they were 1.2 magnitude brighter than in WISE-1) or they have a WISE-2 magnitude that is not bright enough to fall in the red sample. The non-red WISE sources have a mean redshift of 0.5 and are very unlikely to have a redshift above 1.5 \cite{2013arXiv1303.4722M}. 

High-resolution follow-up is required to properly assess the incidence of lensing and to resolve any blending issues and determine the merging, interacting or stable disk-like morphologies of the systems \citep[e.g.][Leung et al. in prep]{2017arXiv170505413D,2017ApJ...837..182O}.

\subsection{Space density}

\cite{2016ApJ...832...78I} compared their space density of DSFGs at $4<z< 6$ to the space density of \textit{UVJ} selected massive galaxies at 3.4 $< z <$ 4.2 \citep{2014ApJ...783L..14S} which are predicted to form their stellar mass around redshift 5. \cite{2014ApJ...783L..14S} found a space density for massive $M_{\rm stars} > 10^{11} M_\odot$ quiescent galaxies of $4 \times 10^{-6}$ Mpc$^{-3}$, which is an order of magnitude higher than \cite{2016ApJ...832...78I} and 2 orders of magnitude higher than our results of $7 \times 10^{-8}$ Mpc$^{-3}$. 

We can therefore make a similar conclusion to that of \cite{2016ApJ...832...78I} for \textit{H}-ATLAS ($S_{500}> 30$ mJy), i.e. that the HeLMS ($S_{500}> 63$ mJy) red sample cannot account for the massive quiescent galaxies found at $z \sim 3-4$.  This can be confirmed by our measured SFRs (5000 $M_\odot$yr$^{-1}$), which for a $t_{burst}$ of 100 Myr generate a higher stellar mass of $M_{\rm stars} \sim 5 \times 10^{11} M_\odot$yr$^{-1}$ than the \citep{2014ApJ...783L..14S} sample. This suggests that part of our sample might go through a short phase ($\ll$100 Myr) of extremely high star formation, or is lensed, or is more massive (and thus rarer) than the population probed by \cite{2014ApJ...783L..14S}.

\section{Conclusions} \label{sec:conclusion}
We have observed 188 high-redshift, dusty, star-forming galaxy candidates with the SCUBA-2 camera at the JCMT.  The sample had been selected to be very red and bright at \textit{Herschel} SPIRE wavelengths and was taken from the 274 deg$^2$ HerMES HeLMS field.  We achieve a 1$\sigma$ rms depth of $S_{850\mu{\rm m}}=$4.3 mJy and detected 87 per cent of our candidates with S/N $>$3.  

We developed a new method of incorporating correlated confusion noise into our SED fitting procedure. 

We applied \texttt{EAZY} with a range of galaxy templates to determine the full redshift PDF. The addition of the longer wavelength  850$\mu$m data  improves our photometric redshifts which are systematically lower than with SPIRE data alone and reduces the estimated uncertainties from $\sigma_z/(1+z)\approx0.25$ to 0.19. Our photometric redshifts are consistent with the four spectroscopic redshifts available in our sample.

With this final PDF we compute the redshift, FIR luminosity and SFRs of our sample. From these we computed the SFRD and showed that the population of 500 $\mu$m risers with $S_{500} >$ 63 mJy contribute less than 1 per cent of the total SFRD at any epoch.  The number density of 500 $\mu$m risers is consistent with a model extrapolated from the \cite{2013MNRAS.432...23G} FIR luminosity function and with the \cite{beth} empirical model, contradicting previous tensions with physically motivated models. consistency with the models arises from our novel way in adding both confusion and instrumental noise were $\sim$60 per cent of the galaxies are predicted to be scattered up to our selection criteria due to flux boosting.

The excess number of WISE-1 sources near our DSFGs gives a strong indication that some of our sample may be lensed or that there is a surprising overlap with the redshift distribution of WISE-1 sources.

High resolution SMA observations of 5 of our sources reveal that two of the sources have a WISE-1 source at the same position. In all cases the SMA flux density at 1.1~mm is lower than predicted from our best fit SED.

We identify a  subset of 21 excellent very high-redshift DSFGs candidates, of which two are already identified as z$>$4 DSFGs. This group is clearly detected by SCUBA-2 with a high probability that they lie above a redshift of 4. These 21 galaxies would be ideal targets for interferometric imaging and spectroscopy to get a better understanding of these high-redshift objects with extreme ($\gg 1000$ M$_\odot$yr$^{-1}$) star formation.    

Observing the high-redshift, dust-obscured Universe remains an important challenge for current day astronomy. Interferometry has the resolution and sensitivity to answer our questions about the SFR and nature of these obscured galaxies. With telescopes like ALMA it is still impractical to cover a large enough area of the sky to find a representative population of extreme star-forming sources. With our new catalogue we now possess an ideal target list for high-resolution and spectroscopic follow up.

\section{acknowledgements}

This project has received funding from the European Union’s Horizon 2020 research and innovation programme under grant agreement No. 607254; this publication reflects only the author’s view and the European Union is not responsible for any use that may be made of the information contained therein. S.D acknowledges support from the Science and Technology Facilities Council (grant number ST/M503836/1). S. O. and JMS acknowledges support from the Science and Technology Facilities Council (grant number ST/L000652/1). JMS acknowledges the support from the PATT travel grant. D.R. acknowledges support from the National Science Foundation under grant number AST-1614213 to Cornell University. J.L.W is supported by an STFC Ernest Rutherford fellowship, and acknowledges additional support from a European Union COFUND/Durham Junior Research Fellowship under EU grant agreement number 609412, and from STFC (ST/L00075X/1). H.D. acknowledges financial support from the Spanish Ministry of Economy and Competitiveness (MINECO) under the 2014 Ramón y Cajal program MINECO RYC-2014-15686. KC acknowledges support from the Science and Technology Facilities Council (grant number ST/M001008/1). MTS was supported by a Royal Society Leverhulme Trust Senior Research Fellowship (LT150041).
 
The James Clerk Maxwell Telescope is operated by the East Asian Observatory on behalf of The National Astronomical Observatory of Japan, Academia Sinica Institute of Astronomy and Astrophysics, the Korea Astronomy and Space Science Institute, the National Astronomical Observatories of China and the Chinese Academy of Sciences (Grant No. XDB09000000), with additional funding support from the Science and Technology Facilities Council of the United Kingdom and participating universities in the United Kingdom and Canada. 
 
The Herschel spacecraft was designed, built, tested, and launched under a contract to ESA managed by the Herschel/Planck Project team by an industrial consortium under the overall responsibility of the prime contractor Thales Alenia Space (Cannes), and including Astrium (Friedrichshafen) responsible for the payload module and for system testing at spacecraft level, Thales Alenia Space (Turin) responsible for the service module, and Astrium (Toulouse) responsible for the telescope, with in excess of a hundred subcontractors
 
SPIRE has been developed by a consortium of institutes led by Cardiff University (UK) and including University of Lethbridge (Canada); NAOC (China); CEA, LAM (France); IFSI, University of Padua (Italy); IAC (Spain); Stockholm Observatory (Sweden); Imperial College London, RAL, UCL-MSSL, UKATC, University of Sussex (UK); and Caltech, JPL, NHSC, University of Colorado (USA). This development has been supported by national funding agencies CSA (Canada); NAOC (China); CEA, CNES, CNRS (France); ASI (Italy); MCINN (Spain); SNSB (Sweden); STFC, UKSA (UK); and NASA (USA).

\appendix

\section[]{Table of source detection} \label{ap:A}

\begin{table*}
  \begin{tabular}{llrrrrcr}
   Source Name&name & $S_{250}$ & $S_{350}$ & $S_{500}$& $S_{850}$ & phot-$z$ &$\log (L_{FIR}/\rm{L}_\odot)$ \\ 
   &  &[mJy]&[mJy]&[mJy]&[mJy]\\ \hline
HerMES J004409.9+011823 & HELMS\_RED\_1 & 108.1 $\pm$ 6.9 & 166.5 $\pm$ 6.0 & 191.8 $\pm$ 8.2 & 82.0 $\pm$ 3.9 & 4.00$^{+0.55}_{-0.52}$ & 13.95$^{+0.09}_{-0.09}$ \\
HerMES J005258.9+061319 & HELMS\_RED\_2 & 68.2 $\pm$ 6.0 & 111.6 $\pm$ 5.9 & 131.7 $\pm$ 6.9 & 82.4 $\pm$ 4.9 & 4.59$^{+0.67}_{-0.65}$ & 13.92$^{+0.09}_{-0.10}$ \\
HerMES J003929.5+002424 & HELMS\_RED\_3 & 140.8 $\pm$ 6.5 & 152.6 $\pm$ 6.3 & 162.1 $\pm$ 7.3 & 52.3 $\pm$ 4.5 & 3.10$^{+0.58}_{-0.65}$ & 13.75$^{+0.12}_{-0.17}$ \\
HerMES J002220.8$-$015521 & HELMS\_RED\_4 & 62.2 $\pm$ 6.1 & 104.0 $\pm$ 5.8 & 116.3 $\pm$ 6.6 & 52.4 $\pm$ 4.4 & 4.13$^{+0.60}_{-0.57}$ & 13.76$^{+0.09}_{-0.11}$ \\
HerMES J005047.6+065720 & HELMS\_RED\_5 & 20.8 $\pm$ 6.0 & 68.2 $\pm$ 6.4 & 112.0 $\pm$ 6.8 & 37.2 $\pm$ 4.2 & 5.02$^{+0.54}_{-0.91}$ & 13.70$^{+0.07}_{-0.13}$ \\
HerMES J010053.9+030323 & HELMS\_RED\_6 & 50.1 $\pm$ 6.8 & 83.3 $\pm$ 6.1 & 96.1 $\pm$ 7.8 & 39.3 $\pm$ 5.4 & 4.10$^{+0.63}_{-0.61}$ & 13.65$^{+0.10}_{-0.11}$ \\
HerMES J003814.0+002250 & HELMS\_RED\_7 & 73.3 $\pm$ 5.5 & 119.0 $\pm$ 6.0 & 122.9 $\pm$ 6.7 & 58.8 $\pm$ 3.6 & 4.02$^{+0.57}_{-0.57}$ & 13.79$^{+0.09}_{-0.10}$ \\
HerMES J233802.0$-$011907 & HELMS\_RED\_8 & 33.6 $\pm$ 6.5 & 53.8 $\pm$ 6.1 & 90.9 $\pm$ 7.6 & 22.4 $\pm$ 3.4 & 4.13$^{+0.67}_{-0.65}$ & 13.46$^{+0.10}_{-0.12}$ \\
HerMES J002718.1+023946 & HELMS\_RED\_9 & 65.2 $\pm$ 5.9 & 76.4 $\pm$ 5.7 & 99.3 $\pm$ 6.9 & 18.2 $\pm$ 3.6 & 3.27$^{+0.59}_{-0.56}$ & 13.42$^{+0.12}_{-0.13}$ \\
HerMES J000304.4+024111 & HELMS\_RED\_10 & 33.6 $\pm$ 5.7 & 53.9 $\pm$ 6.5 & 86.5 $\pm$ 6.9 & 37.9 $\pm$ 4.4 & 4.62$^{+0.75}_{-0.63}$ & 13.63$^{+0.10}_{-0.09}$ \\
HerMES J004747.0+061444 & HELMS\_RED\_11 & 71.4 $\pm$ 5.9 & 112.0 $\pm$ 6.0 & 114.6 $\pm$ 7.8 & 20.4 $\pm$ 6.0 & 3.51$^{+0.54}_{-0.60}$ & 13.57$^{+0.10}_{-0.13}$ \\
HerMES J002115.6+013259 & HELMS\_RED\_12 & 58.5 $\pm$ 6.3 & 80.5 $\pm$ 6.7 & 81.5 $\pm$ 7.3 & 36.7 $\pm$ 4.8 & 3.67$^{+0.64}_{-0.64}$ & 13.57$^{+0.11}_{-0.13}$ \\
HerMES J002936.4+020706 & HELMS\_RED\_13 & 77.4 $\pm$ 6.4 & 89.3 $\pm$ 6.1 & 100.0 $\pm$ 6.7 & 30.1 $\pm$ 5.0 & 3.29$^{+0.62}_{-0.64}$ & 13.55$^{+0.12}_{-0.16}$ \\
HerMES J003847.0$-$021105 & HELMS\_RED\_14 & 61.0 $\pm$ 5.9 & 75.1 $\pm$ 5.6 & 100.5 $\pm$ 7.0 & 19.7 $\pm$ 3.9 & 3.42$^{+0.59}_{-0.57}$ & 13.45$^{+0.11}_{-0.13}$ \\
HerMES J011206.7+031417 & HELMS\_RED\_15 & 54.4 $\pm$ 5.8 & 78.7 $\pm$ 5.9 & 92.2 $\pm$ 7.4 & 17.8 $\pm$ 6.3 & 3.60$^{+0.61}_{-0.59}$ & 13.48$^{+0.11}_{-0.13}$ \\
HerMES J002959.4+032138 & HELMS\_RED\_16 & 47.5 $\pm$ 5.8 & 78.7 $\pm$ 6.3 & 100.0 $\pm$ 6.5 & 26.9 $\pm$ 5.7 & 3.98$^{+0.62}_{-0.61}$ & 13.57$^{+0.09}_{-0.12}$ \\
HerMES J005352.1+023916 & HELMS\_RED\_17 & 24.6 $\pm$ 6.6 & 39.7 $\pm$ 6.4 & 72.0 $\pm$ 6.7 & 22.5 $\pm$ 3.6 & 4.47$^{+0.78}_{-0.72}$ & 13.46$^{+0.11}_{-0.12}$ \\
HerMES J000727.1+015626 & HELMS\_RED\_19 & 53.9 $\pm$ 6.1 & 72.5 $\pm$ 6.4 & 81.6 $\pm$ 7.0 & 40.9 $\pm$ 3.4 & 3.90$^{+0.67}_{-0.65}$ & 13.60$^{+0.10}_{-0.13}$ \\
HerMES J003909.3+020247 & HELMS\_RED\_21 & 41.0 $\pm$ 5.9 & 53.5 $\pm$ 6.1 & 74.3 $\pm$ 7.5 & 33.6 $\pm$ 4.6 & 4.17$^{+0.74}_{-0.71}$ & 13.55$^{+0.11}_{-0.13}$ \\
HerMES J235411.8$-$082912 & HELMS\_RED\_22 & 40.9 $\pm$ 6.5 & 57.8 $\pm$ 6.2 & 71.3 $\pm$ 6.8 & 20.6 $\pm$ 5.4 & 3.80$^{+0.69}_{-0.66}$ & 13.43$^{+0.11}_{-0.13}$ \\
HerMES J004532.6+000121 & HELMS\_RED\_23 & 48.2 $\pm$ 6.7 & 87.6 $\pm$ 6.3 & 97.2 $\pm$ 7.4 & 42.1 $\pm$ 4.9 & 4.20$^{+0.63}_{-0.61}$ & 13.67$^{+0.10}_{-0.11}$ \\
HerMES J234805.1$-$052135 & HELMS\_RED\_25 & 18.7 $\pm$ 5.6 & 52.0 $\pm$ 6.3 & 65.2 $\pm$ 7.3 & 22.9 $\pm$ 3.5 & 4.56$^{+0.68}_{-0.78}$ & 13.46$^{+0.09}_{-0.12}$ \\
HerMES J003750.7+003323 & HELMS\_RED\_28 & 66.4 $\pm$ 5.6 & 85.3 $\pm$ 5.9 & 92.9 $\pm$ 6.6 & 16.2 $\pm$ 4.2 & 3.25$^{+0.55}_{-0.56}$ & 13.42$^{+0.11}_{-0.13}$ \\
HerMES J232404.6$-$055123 & HELMS\_RED\_30 & 51.0 $\pm$ 5.9 & 76.8 $\pm$ 6.5 & 79.1 $\pm$ 7.4 & 33.5 $\pm$ 4.3 & 3.82$^{+0.64}_{-0.61}$ & 13.55$^{+0.11}_{-0.12}$ \\
HerMES J002737.4$-$020801 & HELMS\_RED\_31 & 42.0 $\pm$ 6.9 & 49.4 $\pm$ 6.0 & 75.3 $\pm$ 6.9 & 31.9 $\pm$ 3.9 & 4.14$^{+0.76}_{-0.73}$ & 13.53$^{+0.11}_{-0.14}$ \\
HerMES J232133.3$-$040621 & HELMS\_RED\_32 & 34.3 $\pm$ 5.6 & 63.5 $\pm$ 6.3 & 77.7 $\pm$ 7.0 & 14.4 $\pm$ 4.8 & 3.89$^{+0.61}_{-0.69}$ & 13.40$^{+0.10}_{-0.14}$ \\
HerMES J004118.5+015537 & HELMS\_RED\_33 & 34.7 $\pm$ 6.0 & 39.9 $\pm$ 6.1 & 64.4 $\pm$ 7.0 & 17.6 $\pm$ 3.5 & 3.75$^{+0.76}_{-0.76}$ & 13.32$^{+0.12}_{-0.16}$ \\
HerMES J004302.6+011416 & HELMS\_RED\_35 & 30.7 $\pm$ 6.6 & 73.7 $\pm$ 6.0 & 77.4 $\pm$ 7.3 & 26.5 $\pm$ 3.5 & 4.23$^{+0.59}_{-0.71}$ & 13.52$^{+0.08}_{-0.13}$ \\
HerMES J001848.5$-$061051 & HELMS\_RED\_36 & 46.6 $\pm$ 5.6 & 61.0 $\pm$ 6.2 & 66.0 $\pm$ 6.9 & 17.4 $\pm$ 4.5 & 3.43$^{+0.66}_{-0.64}$ & 13.35$^{+0.12}_{-0.15}$ \\
HerMES J005254.9+032931 & HELMS\_RED\_37 & 55.4 $\pm$ 6.0 & 90.0 $\pm$ 6.0 & 91.3 $\pm$ 7.4 & 30.6 $\pm$ 4.5 & 3.79$^{+0.59}_{-0.58}$ & 13.57$^{+0.10}_{-0.12}$ \\
HerMES J000400.8$-$043103 & HELMS\_RED\_38 & 53.4 $\pm$ 5.3 & 59.5 $\pm$ 5.7 & 67.4 $\pm$ 6.9 & 21.6 $\pm$ 4.6 & 3.20$^{+0.69}_{-0.74}$ & 13.36$^{+0.14}_{-0.19}$ \\
HerMES J002822.0$-$021634 & HELMS\_RED\_39 & 39.8 $\pm$ 6.0 & 54.7 $\pm$ 6.1 & 64.9 $\pm$ 6.9 & 16.2 $\pm$ 4.6 & 3.61$^{+0.68}_{-0.65}$ & 13.34$^{+0.12}_{-0.14}$ \\
HerMES J234431.9$-$061852 & HELMS\_RED\_40 & 59.5 $\pm$ 5.6 & 89.5 $\pm$ 5.8 & 92.2 $\pm$ 7.3 & 39.8 $\pm$ 4.5 & 3.83$^{+0.60}_{-0.59}$ & 13.63$^{+0.10}_{-0.12}$ \\
HerMES J234647.8+000525 & HELMS\_RED\_41 & 74.6 $\pm$ 5.8 & 95.4 $\pm$ 5.8 & 100.5 $\pm$ 7.0 & 23.6 $\pm$ 3.4 & 3.30$^{+0.55}_{-0.54}$ & 13.50$^{+0.11}_{-0.13}$ \\
HerMES J002741.3$-$011650 & HELMS\_RED\_42 & 33.1 $\pm$ 5.4 & 59.0 $\pm$ 5.7 & 64.9 $\pm$ 6.5 & 39.5 $\pm$ 3.8 & 4.53$^{+0.75}_{-0.71}$ & 13.60$^{+0.10}_{-0.12}$ \\
HerMES J004656.1+013751 & HELMS\_RED\_43 & 36.1 $\pm$ 5.8 & 43.4 $\pm$ 5.9 & 67.6 $\pm$ 6.8 & 28.7 $\pm$ 5.3 & 4.24$^{+0.82}_{-0.76}$ & 13.50$^{+0.11}_{-0.13}$ \\
HerMES J004237.7+020457 & HELMS\_RED\_45 & 49.3 $\pm$ 6.2 & 66.1 $\pm$ 6.0 & 87.6 $\pm$ 7.3 & 17.3 $\pm$ 5.1 & 3.60$^{+0.63}_{-0.60}$ & 13.43$^{+0.11}_{-0.13}$ \\
HerMES J233247.6+003632 & HELMS\_RED\_46 & 46.3 $\pm$ 5.9 & 72.9 $\pm$ 6.0 & 75.8 $\pm$ 6.8 & 29.4 $\pm$ 3.4 & 3.84$^{+0.63}_{-0.60}$ & 13.52$^{+0.10}_{-0.12}$ \\
HerMES J003531.5+001536 & HELMS\_RED\_49 & 52.2 $\pm$ 6.3 & 75.1 $\pm$ 5.7 & 81.7 $\pm$ 6.7 & 26.9 $\pm$ 3.8 & 3.69$^{+0.62}_{-0.60}$ & 13.51$^{+0.10}_{-0.12}$ \\
HerMES J002937.6+002617 & HELMS\_RED\_50 & 50.3 $\pm$ 6.1 & 66.8 $\pm$ 6.2 & 67.5 $\pm$ 7.1 & 16.1 $\pm$ 3.6 & 3.31$^{+0.63}_{-0.61}$ & 13.33$^{+0.13}_{-0.14}$ \\
HerMES J232908.1$-$050653 & HELMS\_RED\_51 & 44.3 $\pm$ 6.4 & 69.1 $\pm$ 6.1 & 76.8 $\pm$ 7.0 & 23.2 $\pm$ 4.5 & 3.80$^{+0.63}_{-0.62}$ & 13.47$^{+0.11}_{-0.12}$ \\
HerMES J232342.0$-$035109 & HELMS\_RED\_53 & 34.3 $\pm$ 5.6 & 36.3 $\pm$ 6.2 & 65.9 $\pm$ 7.6 & 19.1 $\pm$ 5.0 & 3.87$^{+0.80}_{-0.83}$ & 13.36$^{+0.13}_{-0.16}$ \\
HerMES J234522.9+015601 & HELMS\_RED\_54 & 46.2 $\pm$ 6.4 & 75.1 $\pm$ 5.9 & 79.6 $\pm$ 6.9 & 19.4 $\pm$ 3.6 & 3.64$^{+0.59}_{-0.61}$ & 13.43$^{+0.10}_{-0.13}$ \\
HerMES J004600.3+065559 & HELMS\_RED\_56 & 70.6 $\pm$ 6.0 & 85.6 $\pm$ 6.9 & 98.1 $\pm$ 7.6 & 17.9 $\pm$ 4.5 & 3.23$^{+0.57}_{-0.56}$ & 13.44$^{+0.11}_{-0.15}$ \\
HerMES J001029.7$-$025524 & HELMS\_RED\_57 & 32.5 $\pm$ 5.6 & 56.3 $\pm$ 5.9 & 70.6 $\pm$ 7.2 & 15.1 $\pm$ 3.3 & 3.80$^{+0.62}_{-0.66}$ & 13.34$^{+0.10}_{-0.13}$ \\
HerMES J233943.0$-$013939 & HELMS\_RED\_58 & 51.0 $\pm$ 6.2 & 66.8 $\pm$ 6.2 & 81.2 $\pm$ 7.1 & 19.2 $\pm$ 3.6 & 3.50$^{+0.62}_{-0.59}$ & 13.41$^{+0.11}_{-0.13}$ \\
HerMES J232849.6+010843 & HELMS\_RED\_60 & 56.0 $\pm$ 5.7 & 95.3 $\pm$ 5.7 & 99.2 $\pm$ 6.9 & 35.3 $\pm$ 4.5 & 3.92$^{+0.58}_{-0.58}$ & 13.62$^{+0.10}_{-0.11}$ \\
HerMES J001432.9+014530 & HELMS\_RED\_61 & 60.7 $\pm$ 5.9 & 75.1 $\pm$ 5.8 & 81.5 $\pm$ 6.9 & 27.7 $\pm$ 3.4 & 3.42$^{+0.63}_{-0.63}$ & 13.49$^{+0.12}_{-0.14}$ \\
HerMES J002319.1+001557 & HELMS\_RED\_62 & 47.7 $\pm$ 5.9 & 60.9 $\pm$ 6.1 & 78.4 $\pm$ 7.3 & 20.9 $\pm$ 3.6 & 3.58$^{+0.66}_{-0.62}$ & 13.42$^{+0.11}_{-0.14}$ \\
HerMES J233755.3$-$053318 & HELMS\_RED\_64 & 48.4 $\pm$ 6.1 & 61.2 $\pm$ 6.5 & 66.1 $\pm$ 7.5 & 22.0 $\pm$ 6.0 & 3.45$^{+0.70}_{-0.72}$ & 13.40$^{+0.13}_{-0.16}$ \\
HerMES J000947.0+034432 & HELMS\_RED\_65 & 39.4 $\pm$ 5.6 & 57.1 $\pm$ 5.9 & 77.4 $\pm$ 6.8 & 22.9 $\pm$ 3.3 & 3.88$^{+0.66}_{-0.60}$ & 13.45$^{+0.10}_{-0.13}$ \\
HerMES J235922.9$-$043705 & HELMS\_RED\_67 & 37.8 $\pm$ 6.4 & 66.0 $\pm$ 6.0 & 67.0 $\pm$ 7.8 & 28.7 $\pm$ 4.1 & 4.01$^{+0.67}_{-0.65}$ & 13.49$^{+0.11}_{-0.12}$ \\
HerMES J235808.7+005553 & HELMS\_RED\_68 & 55.4 $\pm$ 5.6 & 73.9 $\pm$ 6.1 & 76.1 $\pm$ 6.5 & 32.7 $\pm$ 3.8 & 3.60$^{+0.63}_{-0.64}$ & 13.52$^{+0.11}_{-0.14}$ \\
HerMES J000900.6+050709 & HELMS\_RED\_69 & 36.6 $\pm$ 6.2 & 43.1 $\pm$ 6.0 & 70.2 $\pm$ 6.9 & 44.7 $\pm$ 4.8 & 5.19$^{+0.89}_{-0.92}$ & 13.70$^{+0.10}_{-0.13}$ \\
HerMES J004019.0+052714 & HELMS\_RED\_71 & 28.2 $\pm$ 5.6 & 49.8 $\pm$ 6.1 & 63.8 $\pm$ 7.1 & 10.0 $\pm$ 3.6 & 3.71$^{+0.65}_{-0.68}$ & 13.25$^{+0.11}_{-0.15}$ \\
HerMES J005227.0+020027 & HELMS\_RED\_72 & 66.9 $\pm$ 6.1 & 71.6 $\pm$ 6.2 & 84.1 $\pm$ 7.1 & 15.7 $\pm$ 4.6 & 3.04$^{+0.66}_{-0.67}$ & 13.36$^{+0.14}_{-0.19}$ \\
HerMES J001813.6+053159 & HELMS\_RED\_76 & 70.7 $\pm$ 6.0 & 75.6 $\pm$ 6.0 & 85.6 $\pm$ 6.2 & 17.5 $\pm$ 3.7 & 3.02$^{+0.63}_{-0.64}$ & 13.38$^{+0.14}_{-0.16}$ \\
HerMES J000056.0+010231 & HELMS\_RED\_77 & 70.4 $\pm$ 6.4 & 71.2 $\pm$ 6.1 & 84.4 $\pm$ 8.3 & 17.4 $\pm$ 4.9 & 2.89$^{+0.70}_{-0.80}$ & 13.35$^{+0.15}_{-0.24}$ \\
 \end{tabular}
  \caption{Flux densities with instrumental errors, redshifts and luminosities of our targets. }
  \label{tab:cm_v}
\end{table*}

\addtocounter{table}{-1}

\begin{table*}
  \begin{tabular}{llrrrrcr}
   Source Name&name & $S_{250}$ & $S_{350}$ & $S_{500}$& $S_{850}$ & phot-$z$ &$\log (L_{FIR}/\rm{L}_\odot)$ \\ 
   &  &[mJy]&[mJy]&[mJy]&[mJy]\\ \hline
HerMES J002552.3+031329 & HELMS\_RED\_79 & 46.6 $\pm$ 6.0 & 65.7 $\pm$ 6.1 & 73.6 $\pm$ 6.8 & 12.0 $\pm$ 3.5 & 3.32$^{+0.60}_{-0.60}$ & 13.30$^{+0.12}_{-0.14}$ \\
HerMES J005037.1+014449 & HELMS\_RED\_80 & 44.2 $\pm$ 5.7 & 59.8 $\pm$ 6.1 & 64.5 $\pm$ 7.8 & 25.1 $\pm$ 3.7 & 3.63$^{+0.68}_{-0.69}$ & 13.43$^{+0.11}_{-0.15}$ \\
HerMES J004724.4+010119 & HELMS\_RED\_82 & 47.4 $\pm$ 6.9 & 75.8 $\pm$ 6.0 & 76.2 $\pm$ 7.6 & 36.2 $\pm$ 5.0 & 3.96$^{+0.67}_{-0.65}$ & 13.57$^{+0.11}_{-0.13}$ \\
HerMES J235020.1$-$065224 & HELMS\_RED\_84 & 67.5 $\pm$ 6.1 & 82.4 $\pm$ 6.5 & 84.0 $\pm$ 7.3 & 10.3 $\pm$ 4.7 & 3.02$^{+0.57}_{-0.57}$ & 13.34$^{+0.13}_{-0.15}$ \\
HerMES J233823.1$-$042924 & HELMS\_RED\_86 & 48.9 $\pm$ 6.5 & 58.1 $\pm$ 6.1 & 63.2 $\pm$ 7.5 & 22.5 $\pm$ 5.8 & 3.37$^{+0.73}_{-0.78}$ & 13.38$^{+0.14}_{-0.18}$ \\
HerMES J002058.4+002114 & HELMS\_RED\_88 & 36.3 $\pm$ 5.9 & 54.2 $\pm$ 6.0 & 63.8 $\pm$ 6.8 & 24.8 $\pm$ 3.5 & 3.95$^{+0.70}_{-0.66}$ & 13.44$^{+0.11}_{-0.13}$ \\
HerMES J234940.0$-$025551 & HELMS\_RED\_89 & 46.0 $\pm$ 6.3 & 58.5 $\pm$ 6.5 & 85.2 $\pm$ 6.5 & 7.4 $\pm$ 5.0 & 3.37$^{+0.65}_{-0.63}$ & 13.32$^{+0.12}_{-0.14}$ \\
HerMES J010040.6+051550 & HELMS\_RED\_95 & 52.0 $\pm$ 6.4 & 62.6 $\pm$ 6.1 & 78.9 $\pm$ 7.6 & 18.3 $\pm$ 3.8 & 3.40$^{+0.65}_{-0.64}$ & 13.38$^{+0.12}_{-0.15}$ \\
HerMES J001533.3$-$054652 & HELMS\_RED\_96 & 37.8 $\pm$ 6.3 & 51.6 $\pm$ 6.0 & 63.9 $\pm$ 7.1 & 19.1 $\pm$ 3.8 & 3.72$^{+0.71}_{-0.68}$ & 13.36$^{+0.13}_{-0.14}$ \\
HerMES J232014.8$-$045552 & HELMS\_RED\_98 & 20.4 $\pm$ 6.5 & 34.4 $\pm$ 6.0 & 74.2 $\pm$ 7.4 & 17.8 $\pm$ 4.9 & 4.58$^{+0.82}_{-0.79}$ & 13.41$^{+0.11}_{-0.13}$ \\
HerMES J235221.4$-$043114 & HELMS\_RED\_101 & 33.7 $\pm$ 5.7 & 58.3 $\pm$ 5.4 & 63.7 $\pm$ 7.3 & 17.8 $\pm$ 4.4 & 3.87$^{+0.65}_{-0.66}$ & 13.37$^{+0.11}_{-0.13}$ \\
HerMES J004526.1+031638 & HELMS\_RED\_104 & 48.6 $\pm$ 5.8 & 69.4 $\pm$ 5.6 & 76.6 $\pm$ 6.9 & 29.9 $\pm$ 4.3 & 3.78$^{+0.64}_{-0.63}$ & 13.52$^{+0.10}_{-0.13}$ \\
HerMES J005134.1+053502 & HELMS\_RED\_105 & 51.9 $\pm$ 5.9 & 61.2 $\pm$ 6.3 & 73.4 $\pm$ 7.6 & 24.8 $\pm$ 5.4 & 3.46$^{+0.68}_{-0.73}$ & 13.43$^{+0.13}_{-0.16}$ \\
HerMES J232033.8$-$020958 & HELMS\_RED\_106 & 56.9 $\pm$ 6.3 & 72.9 $\pm$ 6.5 & 74.8 $\pm$ 7.3 & 23.7 $\pm$ 6.4 & 3.41$^{+0.66}_{-0.67}$ & 13.45$^{+0.12}_{-0.15}$ \\
HerMES J233052.3$-$060958 & HELMS\_RED\_107 & 52.0 $\pm$ 6.3 & 59.6 $\pm$ 6.2 & 80.1 $\pm$ 7.3 & 21.9 $\pm$ 4.7 & 3.48$^{+0.68}_{-0.71}$ & 13.42$^{+0.12}_{-0.17}$ \\
HerMES J233554.3$-$054408 & HELMS\_RED\_108 & 46.6 $\pm$ 6.0 & 72.3 $\pm$ 6.6 & 86.1 $\pm$ 7.5 & 16.6 $\pm$ 4.2 & 3.61$^{+0.59}_{-0.62}$ & 13.42$^{+0.11}_{-0.13}$ \\
HerMES J000208.8$-$015521 & HELMS\_RED\_110 & 42.9 $\pm$ 5.7 & 57.3 $\pm$ 6.3 & 71.8 $\pm$ 6.8 & 20.9 $\pm$ 4.7 & 3.72$^{+0.67}_{-0.64}$ & 13.42$^{+0.12}_{-0.12}$ \\
HerMES J235003.0$-$015825 & HELMS\_RED\_114 & 41.9 $\pm$ 6.2 & 65.6 $\pm$ 6.0 & 77.8 $\pm$ 6.7 & 23.3 $\pm$ 4.9 & 3.89$^{+0.65}_{-0.62}$ & 13.48$^{+0.10}_{-0.13}$ \\
HerMES J010631.8+015002 & HELMS\_RED\_117 & 42.9 $\pm$ 6.4 & 46.8 $\pm$ 5.7 & 63.4 $\pm$ 7.4 & 19.0 $\pm$ 4.0 & 3.42$^{+0.77}_{-0.83}$ & 13.32$^{+0.15}_{-0.19}$ \\
HerMES J003943.5+003955 & HELMS\_RED\_118 & 32.7 $\pm$ 6.1 & 57.0 $\pm$ 6.1 & 73.7 $\pm$ 7.0 & 23.5 $\pm$ 3.9 & 4.13$^{+0.67}_{-0.65}$ & 13.46$^{+0.10}_{-0.12}$ \\
HerMES J233208.3$-$022211 & HELMS\_RED\_119 & 34.7 $\pm$ 6.0 & 57.5 $\pm$ 6.0 & 75.4 $\pm$ 7.0 & 25.7 $\pm$ 6.4 & 4.20$^{+0.71}_{-0.66}$ & 13.51$^{+0.10}_{-0.12}$ \\
HerMES J005708.2+023637 & HELMS\_RED\_123 & 35.1 $\pm$ 5.9 & 55.3 $\pm$ 5.8 & 79.7 $\pm$ 6.8 & 9.7 $\pm$ 4.7 & 3.68$^{+0.66}_{-0.67}$ & 13.33$^{+0.11}_{-0.15}$ \\
HerMES J000000.7$-$054310 & HELMS\_RED\_124 & 51.8 $\pm$ 6.6 & 62.2 $\pm$ 6.2 & 66.9 $\pm$ 6.4 & 21.8 $\pm$ 4.3 & 3.35$^{+0.69}_{-0.73}$ & 13.39$^{+0.13}_{-0.18}$ \\
HerMES J233521.4$-$040227 & HELMS\_RED\_126 & 39.6 $\pm$ 6.2 & 44.6 $\pm$ 6.9 & 63.9 $\pm$ 7.7 & 21.7 $\pm$ 4.6 & 3.69$^{+0.76}_{-0.81}$ & 13.37$^{+0.13}_{-0.17}$ \\
HerMES J010433.0+044510 & HELMS\_RED\_127 & 51.7 $\pm$ 6.1 & 62.1 $\pm$ 7.0 & 75.0 $\pm$ 8.7 & 19.7 $\pm$ 4.5 & 3.38$^{+0.68}_{-0.69}$ & 13.38$^{+0.12}_{-0.17}$ \\
HerMES J235712.0$-$041341 & HELMS\_RED\_134 & 53.6 $\pm$ 5.8 & 60.8 $\pm$ 5.5 & 70.4 $\pm$ 8.0 & 20.6 $\pm$ 4.0 & 3.21$^{+0.70}_{-0.76}$ & 13.36$^{+0.14}_{-0.19}$ \\
HerMES J235833.6$-$042150 & HELMS\_RED\_135 & 62.4 $\pm$ 6.2 & 71.8 $\pm$ 6.0 & 81.8 $\pm$ 7.3 & 12.3 $\pm$ 4.1 & 3.06$^{+0.61}_{-0.59}$ & 13.32$^{+0.13}_{-0.16}$ \\
HerMES J004700.2+004214 & HELMS\_RED\_136 & 46.6 $\pm$ 5.5 & 63.3 $\pm$ 6.0 & 63.8 $\pm$ 7.5 & 18.5 $\pm$ 3.7 & 3.43$^{+0.64}_{-0.64}$ & 13.35$^{+0.11}_{-0.15}$ \\
HerMES J004434.7+070159 & HELMS\_RED\_137 & 35.2 $\pm$ 6.2 & 42.3 $\pm$ 6.2 & 66.0 $\pm$ 7.5 & 17.3 $\pm$ 4.1 & 3.75$^{+0.76}_{-0.76}$ & 13.33$^{+0.12}_{-0.16}$ \\
HerMES J011130.9+041443 & HELMS\_RED\_139 & 44.2 $\pm$ 6.8 & 63.4 $\pm$ 6.2 & 64.2 $\pm$ 7.9 & 23.6 $\pm$ 4.6 & 3.63$^{+0.70}_{-0.69}$ & 13.42$^{+0.12}_{-0.15}$ \\
HerMES J003651.3$-$015617 & HELMS\_RED\_140 & 29.8 $\pm$ 6.3 & 51.8 $\pm$ 5.7 & 65.4 $\pm$ 7.6 & 13.5 $\pm$ 4.0 & 3.84$^{+0.66}_{-0.70}$ & 13.31$^{+0.11}_{-0.14}$ \\
HerMES J233832.1$-$040953 & HELMS\_RED\_142 & 27.3 $\pm$ 6.2 & 42.4 $\pm$ 6.1 & 66.7 $\pm$ 7.7 & 22.9 $\pm$ 4.7 & 4.35$^{+0.79}_{-0.72}$ & 13.44$^{+0.11}_{-0.12}$ \\
HerMES J000407.6$-$050014 & HELMS\_RED\_143 & 33.5 $\pm$ 6.6 & 49.4 $\pm$ 6.3 & 73.8 $\pm$ 7.4 & 15.1 $\pm$ 4.4 & 3.86$^{+0.70}_{-0.67}$ & 13.36$^{+0.11}_{-0.13}$ \\
HerMES J005213.2+000447 & HELMS\_RED\_146 & 54.3 $\pm$ 5.6 & 80.6 $\pm$ 6.2 & 81.3 $\pm$ 7.8 & 19.2 $\pm$ 5.4 & 3.52$^{+0.59}_{-0.59}$ & 13.44$^{+0.11}_{-0.12}$ \\
HerMES J003512.0+010758 & HELMS\_RED\_153 & 59.0 $\pm$ 6.0 & 72.6 $\pm$ 6.2 & 91.4 $\pm$ 7.1 & 12.5 $\pm$ 5.4 & 3.32$^{+0.60}_{-0.59}$ & 13.40$^{+0.12}_{-0.13}$ \\
HerMES J235157.2$-$044058 & HELMS\_RED\_154 & 29.4 $\pm$ 5.7 & 40.2 $\pm$ 6.7 & 63.4 $\pm$ 6.8 & 20.2 $\pm$ 4.4 & 4.12$^{+0.78}_{-0.70}$ & 13.40$^{+0.12}_{-0.13}$ \\
HerMES J235752.2$-$040711 & HELMS\_RED\_155 & 25.4 $\pm$ 5.6 & 40.3 $\pm$ 5.9 & 68.3 $\pm$ 6.7 & 20.1 $\pm$ 4.5 & 4.38$^{+0.77}_{-0.70}$ & 13.42$^{+0.10}_{-0.12}$ \\
HerMES J233623.2+000108 & HELMS\_RED\_160 & 54.7 $\pm$ 6.2 & 58.9 $\pm$ 6.4 & 69.2 $\pm$ 7.5 & 10.8 $\pm$ 3.5 & 2.93$^{+0.67}_{-0.73}$ & 13.21$^{+0.14}_{-0.22}$ \\
HerMES J232847.2$-$053724 & HELMS\_RED\_161 & 46.6 $\pm$ 6.5 & 65.7 $\pm$ 5.6 & 67.4 $\pm$ 8.0 & 33.1 $\pm$ 4.5 & 3.80$^{+0.71}_{-0.69}$ & 13.51$^{+0.11}_{-0.14}$ \\
HerMES J010906.7+052709 & HELMS\_RED\_163 & 39.2 $\pm$ 6.4 & 51.9 $\pm$ 6.0 & 63.5 $\pm$ 8.3 & 18.6 $\pm$ 5.0 & 3.63$^{+0.74}_{-0.73}$ & 13.36$^{+0.12}_{-0.16}$ \\
HerMES J004909.5+005712 & HELMS\_RED\_165 & 25.6 $\pm$ 5.9 & 43.9 $\pm$ 5.5 & 63.1 $\pm$ 6.8 & 11.7 $\pm$ 3.9 & 3.91$^{+0.71}_{-0.71}$ & 13.28$^{+0.11}_{-0.14}$ \\
HerMES J235924.0$-$075406 & HELMS\_RED\_169 & 30.5 $\pm$ 6.8 & 54.6 $\pm$ 6.3 & 65.0 $\pm$ 7.3 & 17.2 $\pm$ 4.0 & 3.92$^{+0.68}_{-0.69}$ & 13.37$^{+0.11}_{-0.13}$ \\
HerMES J004623.3+000425 & HELMS\_RED\_173 & 27.4 $\pm$ 5.8 & 31.1 $\pm$ 5.9 & 64.7 $\pm$ 7.1 & 16.2 $\pm$ 3.6 & 4.05$^{+0.84}_{-0.78}$ & 13.31$^{+0.13}_{-0.15}$ \\
HerMES J000326.9$-$041214 & HELMS\_RED\_174 & 19.4 $\pm$ 5.9 & 56.1 $\pm$ 6.3 & 68.4 $\pm$ 7.5 & 12.8 $\pm$ 4.5 & 4.20$^{+0.63}_{-0.81}$ & 13.37$^{+0.10}_{-0.15}$ \\
HerMES J233254.6+001616 & HELMS\_RED\_179 & 62.5 $\pm$ 6.3 & 66.3 $\pm$ 6.3 & 75.4 $\pm$ 7.8 & 10.8 $\pm$ 3.2 & 2.81$^{+0.63}_{-0.68}$ & 13.22$^{+0.15}_{-0.20}$ \\
HerMES J233927.1$-$052258 & HELMS\_RED\_180 & 51.4 $\pm$ 6.1 & 59.4 $\pm$ 5.9 & 66.9 $\pm$ 7.9 & 18.1 $\pm$ 4.3 & 3.21$^{+0.71}_{-0.76}$ & 13.33$^{+0.14}_{-0.19}$ \\
HerMES J004120.1+015220 & HELMS\_RED\_183 & 44.4 $\pm$ 6.0 & 65.3 $\pm$ 6.4 & 67.5 $\pm$ 7.2 & 21.5 $\pm$ 3.6 & 3.63$^{+0.64}_{-0.63}$ & 13.41$^{+0.10}_{-0.14}$ \\
HerMES J235818.3$-$081029 & HELMS\_RED\_188 & 45.8 $\pm$ 5.5 & 76.4 $\pm$ 5.6 & 77.0 $\pm$ 6.5 & 26.9 $\pm$ 5.6 & 3.85$^{+0.62}_{-0.60}$ & 13.51$^{+0.10}_{-0.12}$ \\
HerMES J010733.0+042228 & HELMS\_RED\_191 & 46.1 $\pm$ 6.1 & 69.3 $\pm$ 5.7 & 80.7 $\pm$ 8.1 & 15.7 $\pm$ 4.3 & 3.55$^{+0.61}_{-0.61}$ & 13.38$^{+0.11}_{-0.13}$ \\
HerMES J003846.3$-$033526 & HELMS\_RED\_196 & 54.7 $\pm$ 5.9 & 63.4 $\pm$ 5.7 & 68.5 $\pm$ 6.6 & 16.3 $\pm$ 3.4 & 3.16$^{+0.66}_{-0.67}$ & 13.32$^{+0.13}_{-0.17}$ \\
HerMES J235320.4$-$054743 & HELMS\_RED\_202 & 35.3 $\pm$ 5.7 & 49.6 $\pm$ 6.3 & 66.3 $\pm$ 7.6 & 13.7 $\pm$ 4.5 & 3.69$^{+0.69}_{-0.66}$ & 13.31$^{+0.12}_{-0.15}$ \\
HerMES J232711.4$-$051505 & HELMS\_RED\_206 & 57.2 $\pm$ 6.0 & 64.8 $\pm$ 6.2 & 75.4 $\pm$ 6.8 & 11.7 $\pm$ 4.2 & 3.09$^{+0.63}_{-0.62}$ & 13.29$^{+0.12}_{-0.17}$ \\
HerMES J010510.0+044223 & HELMS\_RED\_212 & 41.3 $\pm$ 6.5 & 54.5 $\pm$ 5.8 & 72.0 $\pm$ 7.5 & 27.5 $\pm$ 4.3 & 3.91$^{+0.72}_{-0.70}$ & 13.48$^{+0.12}_{-0.14}$ \\
HerMES J001134.9+002738 & HELMS\_RED\_219 & 46.9 $\pm$ 5.6 & 69.4 $\pm$ 6.2 & 69.7 $\pm$ 6.4 & 23.4 $\pm$ 4.4 & 3.66$^{+0.63}_{-0.61}$ & 13.44$^{+0.11}_{-0.13}$ \\
HerMES J234106.3$-$061457 & HELMS\_RED\_223 & 56.5 $\pm$ 6.7 & 76.7 $\pm$ 6.4 & 83.3 $\pm$ 7.3 & 17.8 $\pm$ 4.2 & 3.40$^{+0.59}_{-0.59}$ & 13.41$^{+0.11}_{-0.14}$ \\
HerMES J235900.9$-$062939 & HELMS\_RED\_224 & 48.2 $\pm$ 5.7 & 54.9 $\pm$ 6.2 & 67.0 $\pm$ 7.1 & 14.9 $\pm$ 4.0 & 3.24$^{+0.69}_{-0.72}$ & 13.29$^{+0.14}_{-0.19}$ \\
HerMES J235647.0$-$023312 & HELMS\_RED\_226 & 29.6 $\pm$ 5.7 & 58.3 $\pm$ 5.5 & 69.2 $\pm$ 7.6 & 20.9 $\pm$ 5.1 & 4.16$^{+0.67}_{-0.69}$ & 13.44$^{+0.10}_{-0.13}$ \\
HerMES J233838.8+000032 & HELMS\_RED\_228 & 58.4 $\pm$ 5.9 & 61.9 $\pm$ 5.8 & 67.7 $\pm$ 6.9 & 10.6 $\pm$ 3.4 & 2.84$^{+0.65}_{-0.71}$ & 13.21$^{+0.15}_{-0.21}$ \\
  \end{tabular}
  \caption{}
  \label{tab:cm_v2}
\end{table*}

\addtocounter{table}{-1}

\begin{table*}
  \begin{tabular}{llrrrrcr}
   Source Name&name & $S_{250}$ & $S_{350}$ & $S_{500}$& $S_{850}$ & phot-$z$ &$\log (L_{FIR}/\rm{L}_\odot)$ \\ 
   &  &[mJy]&[mJy]&[mJy]&[mJy]\\ \hline
HerMES J002012.1$-$044523 & HELMS\_RED\_232 & 44.0 $\pm$ 6.2 & 57.9 $\pm$ 6.2 & 67.4 $\pm$ 7.1 & 20.4 $\pm$ 4.1 & 3.58$^{+0.68}_{-0.67}$ & 13.39$^{+0.11}_{-0.15}$ \\
HerMES J234707.6+021633 & HELMS\_RED\_235 & 55.7 $\pm$ 6.4 & 59.4 $\pm$ 6.2 & 67.5 $\pm$ 6.6 & 23.4 $\pm$ 3.5 & 3.15$^{+0.72}_{-0.74}$ & 13.37$^{+0.14}_{-0.20}$ \\
HerMES J233123.5+000631 & HELMS\_RED\_241 & 60.3 $\pm$ 6.4 & 71.2 $\pm$ 6.8 & 71.5 $\pm$ 7.1 & 14.3 $\pm$ 4.5 & 3.08$^{+0.63}_{-0.64}$ & 13.32$^{+0.12}_{-0.17}$ \\
HerMES J234247.3$-$024555 & HELMS\_RED\_242 & 60.0 $\pm$ 6.6 & 80.7 $\pm$ 5.9 & 81.0 $\pm$ 7.6 & 13.3 $\pm$ 4.6 & 3.22$^{+0.58}_{-0.59}$ & 13.37$^{+0.12}_{-0.14}$ \\
HerMES J235512.7$-$045840 & HELMS\_RED\_249 & 39.1 $\pm$ 5.9 & 69.2 $\pm$ 6.5 & 73.3 $\pm$ 7.7 & 17.8 $\pm$ 4.8 & 3.79$^{+0.61}_{-0.65}$ & 13.41$^{+0.10}_{-0.13}$ \\
HerMES J002057.1+051242 & HELMS\_RED\_251 & 42.5 $\pm$ 6.1 & 46.0 $\pm$ 6.0 & 65.4 $\pm$ 7.0 & 16.4 $\pm$ 3.7 & 3.37$^{+0.75}_{-0.80}$ & 13.30$^{+0.14}_{-0.20}$ \\
HerMES J003743.6$-$011423 & HELMS\_RED\_255 & 48.0 $\pm$ 6.2 & 53.1 $\pm$ 5.9 & 64.4 $\pm$ 7.4 & 16.8 $\pm$ 3.4 & 3.20$^{+0.73}_{-0.80}$ & 13.29$^{+0.15}_{-0.21}$ \\
HerMES J001618.9$-$040118 & HELMS\_RED\_258 & 48.7 $\pm$ 6.3 & 68.0 $\pm$ 6.1 & 90.6 $\pm$ 7.1 & 11.4 $\pm$ 4.4 & 3.44$^{+0.59}_{-0.61}$ & 13.36$^{+0.11}_{-0.14}$ \\
HerMES J001936.8+025855 & HELMS\_RED\_262 & 51.6 $\pm$ 5.9 & 63.7 $\pm$ 6.3 & 67.3 $\pm$ 7.3 & 16.6 $\pm$ 3.6 & 3.26$^{+0.64}_{-0.66}$ & 13.33$^{+0.13}_{-0.16}$ \\
HerMES J005557.4+063518 & HELMS\_RED\_264 & 54.6 $\pm$ 6.5 & 61.7 $\pm$ 6.3 & 73.5 $\pm$ 7.3 & 6.8 $\pm$ 6.1 & 3.11$^{+0.67}_{-0.67}$ & 13.28$^{+0.14}_{-0.17}$ \\
HerMES J000831.4+035303 & HELMS\_RED\_266 & 44.6 $\pm$ 6.3 & 65.7 $\pm$ 5.5 & 69.9 $\pm$ 7.1 & 11.0 $\pm$ 3.2 & 3.29$^{+0.58}_{-0.61}$ & 13.27$^{+0.11}_{-0.14}$ \\
HerMES J001732.5+031559 & HELMS\_RED\_267 & 46.5 $\pm$ 5.9 & 59.2 $\pm$ 5.8 & 64.6 $\pm$ 8.2 & 17.5 $\pm$ 3.7 & 3.35$^{+0.68}_{-0.69}$ & 13.33$^{+0.13}_{-0.17}$ \\
HerMES J004919.4+012439 & HELMS\_RED\_268 & 53.9 $\pm$ 6.2 & 54.4 $\pm$ 6.1 & 69.7 $\pm$ 6.7 & 21.3 $\pm$ 5.7 & 3.16$^{+0.75}_{-0.79}$ & 13.36$^{+0.15}_{-0.21}$ \\
HerMES J234220.9$-$045604 & HELMS\_RED\_269 & 50.9 $\pm$ 6.4 & 55.3 $\pm$ 5.9 & 67.1 $\pm$ 6.8 & 26.8 $\pm$ 4.5 & 3.39$^{+0.74}_{-0.75}$ & 13.41$^{+0.14}_{-0.18}$ \\
HerMES J235830.9+005631 & HELMS\_RED\_270 & 47.8 $\pm$ 5.7 & 62.3 $\pm$ 5.8 & 63.8 $\pm$ 7.0 & 31.5 $\pm$ 4.8 & 3.64$^{+0.72}_{-0.69}$ & 13.47$^{+0.12}_{-0.15}$ \\
HerMES J003819.5+064505 & HELMS\_RED\_272 & 47.7 $\pm$ 5.8 & 61.5 $\pm$ 6.0 & 64.4 $\pm$ 6.5 & 13.7 $\pm$ 3.9 & 3.29$^{+0.63}_{-0.62}$ & 13.29$^{+0.12}_{-0.16}$ \\
HerMES J002943.2+010330 & HELMS\_RED\_277 & 53.0 $\pm$ 5.8 & 63.7 $\pm$ 6.3 & 67.3 $\pm$ 8.0 & 17.6 $\pm$ 3.6 & 3.20$^{+0.67}_{-0.70}$ & 13.33$^{+0.13}_{-0.18}$ \\
HerMES J010231.1+005416 & HELMS\_RED\_279 & 55.1 $\pm$ 6.3 & 61.6 $\pm$ 6.2 & 83.2 $\pm$ 7.5 & 10.5 $\pm$ 4.6 & 3.16$^{+0.65}_{-0.64}$ & 13.30$^{+0.13}_{-0.17}$ \\
HerMES J232606.3$-$023610 & HELMS\_RED\_283 & 34.6 $\pm$ 6.8 & 61.5 $\pm$ 5.8 & 65.6 $\pm$ 7.6 & 11.4 $\pm$ 4.6 & 3.63$^{+0.63}_{-0.68}$ & 13.32$^{+0.10}_{-0.15}$ \\
HerMES J004811.1+000810 & HELMS\_RED\_287 & 47.9 $\pm$ 6.1 & 48.6 $\pm$ 6.4 & 63.8 $\pm$ 6.9 & 18.4 $\pm$ 4.1 & 3.17$^{+0.76}_{-0.82}$ & 13.30$^{+0.16}_{-0.21}$ \\
HerMES J234046.8$-$051205 & HELMS\_RED\_288 & 38.5 $\pm$ 6.7 & 44.9 $\pm$ 6.3 & 67.9 $\pm$ 7.2 & 16.8 $\pm$ 4.3 & 3.65$^{+0.76}_{-0.76}$ & 13.33$^{+0.13}_{-0.16}$ \\
HerMES J002625.4+024405 & HELMS\_RED\_290 & 31.3 $\pm$ 6.3 & 56.7 $\pm$ 6.1 & 65.8 $\pm$ 7.2 & 16.6 $\pm$ 3.7 & 3.88$^{+0.65}_{-0.68}$ & 13.36$^{+0.10}_{-0.14}$ \\
HerMES J002148.7+013522 & HELMS\_RED\_293 & 30.8 $\pm$ 6.1 & 58.1 $\pm$ 5.9 & 68.5 $\pm$ 7.1 & 15.4 $\pm$ 3.9 & 3.89$^{+0.63}_{-0.69}$ & 13.36$^{+0.10}_{-0.14}$ \\
HerMES J233159.8$-$025408 & HELMS\_RED\_301 & 26.6 $\pm$ 7.2 & 48.4 $\pm$ 6.2 & 64.5 $\pm$ 7.2 & 18.2 $\pm$ 4.5 & 4.14$^{+0.74}_{-0.72}$ & 13.39$^{+0.11}_{-0.14}$ \\
HerMES J003706.2+011634 & HELMS\_RED\_309 & 31.8 $\pm$ 5.9 & 63.5 $\pm$ 6.3 & 70.3 $\pm$ 7.0 & 21.8 $\pm$ 3.6 & 4.07$^{+0.63}_{-0.67}$ & 13.45$^{+0.10}_{-0.12}$ \\
HerMES J010151.9+000822 & HELMS\_RED\_314 & 42.1 $\pm$ 6.4 & 68.4 $\pm$ 6.5 & 69.5 $\pm$ 6.7 & 21.2 $\pm$ 5.3 & 3.78$^{+0.64}_{-0.63}$ & 13.43$^{+0.10}_{-0.13}$ \\
HerMES J004808.8+040359 & HELMS\_RED\_315 & 47.4 $\pm$ 5.8 & 65.3 $\pm$ 5.8 & 66.4 $\pm$ 7.1 & 13.5 $\pm$ 4.3 & 3.34$^{+0.62}_{-0.61}$ & 13.31$^{+0.11}_{-0.14}$ \\
HerMES J000154.4$-$031845 & HELMS\_RED\_318 & 33.0 $\pm$ 5.4 & 47.5 $\pm$ 5.6 & 64.4 $\pm$ 7.0 & 8.7 $\pm$ 4.5 & 3.58$^{+0.68}_{-0.67}$ & 13.24$^{+0.11}_{-0.15}$ \\
HerMES J232656.9$-$043112 & HELMS\_RED\_319 & 33.6 $\pm$ 6.6 & 42.5 $\pm$ 6.1 & 63.6 $\pm$ 7.7 & 20.7 $\pm$ 4.6 & 3.95$^{+0.80}_{-0.79}$ & 13.38$^{+0.12}_{-0.16}$ \\
HerMES J232658.4$-$021900 & HELMS\_RED\_320 & 60.5 $\pm$ 6.2 & 60.9 $\pm$ 6.0 & 77.1 $\pm$ 7.1 & 16.3 $\pm$ 4.8 & 3.01$^{+0.71}_{-0.82}$ & 13.33$^{+0.15}_{-0.23}$ \\
HerMES J003527.5+002227 & HELMS\_RED\_323 & 51.2 $\pm$ 5.8 & 61.8 $\pm$ 6.2 & 70.3 $\pm$ 7.1 & 27.3 $\pm$ 5.8 & 3.51$^{+0.68}_{-0.73}$ & 13.45$^{+0.13}_{-0.16}$ \\
HerMES J232856.6$-$041652 & HELMS\_RED\_324 & 75.5 $\pm$ 6.3 & 80.0 $\pm$ 5.8 & 80.6 $\pm$ 6.8 & 22.4 $\pm$ 4.5 & 2.92$^{+0.67}_{-0.76}$ & 13.41$^{+0.15}_{-0.22}$ \\
HerMES J234656.1+002246 & HELMS\_RED\_326 & 36.3 $\pm$ 5.9 & 58.1 $\pm$ 5.9 & 67.8 $\pm$ 7.4 & 25.0 $\pm$ 3.7 & 4.00$^{+0.68}_{-0.64}$ & 13.46$^{+0.11}_{-0.12}$ \\
HerMES J233254.7$-$060301 & HELMS\_RED\_331 & 42.1 $\pm$ 7.4 & 60.0 $\pm$ 5.6 & 64.0 $\pm$ 8.5 & 10.7 $\pm$ 4.7 & 3.35$^{+0.66}_{-0.67}$ & 13.28$^{+0.12}_{-0.16}$ \\
HerMES J232414.9$-$025250 & HELMS\_RED\_333 & 67.4 $\pm$ 5.5 & 70.8 $\pm$ 6.0 & 74.3 $\pm$ 7.2 & 20.4 $\pm$ 4.8 & 2.91$^{+0.68}_{-0.77}$ & 13.36$^{+0.15}_{-0.22}$ \\
HerMES J232057.2$-$044412 & HELMS\_RED\_335 & 50.9 $\pm$ 6.4 & 63.8 $\pm$ 6.2 & 70.0 $\pm$ 7.1 & 23.7 $\pm$ 5.1 & 3.47$^{+0.68}_{-0.70}$ & 13.43$^{+0.13}_{-0.16}$ \\
HerMES J001242.5$-$042634 & HELMS\_RED\_336 & 53.8 $\pm$ 5.9 & 57.4 $\pm$ 5.8 & 64.4 $\pm$ 6.6 & 9.1 $\pm$ 3.6 & 2.87$^{+0.67}_{-0.72}$ & 13.18$^{+0.16}_{-0.21}$ \\
HerMES J005008.5+024618 & HELMS\_RED\_339 & 52.1 $\pm$ 6.3 & 55.6 $\pm$ 5.8 & 64.5 $\pm$ 8.1 & 19.7 $\pm$ 4.9 & 3.10$^{+0.75}_{-0.80}$ & 13.32$^{+0.15}_{-0.22}$ \\
HerMES J003306.4+030116 & HELMS\_RED\_342 & 39.5 $\pm$ 6.0 & 53.4 $\pm$ 5.9 & 68.2 $\pm$ 7.0 & 8.7 $\pm$ 4.5 & 3.41$^{+0.66}_{-0.64}$ & 13.25$^{+0.12}_{-0.15}$ \\
HerMES J235955.2$-$032724 & HELMS\_RED\_348 & 44.5 $\pm$ 5.6 & 50.4 $\pm$ 5.9 & 68.4 $\pm$ 6.4 & 10.8 $\pm$ 3.8 & 3.25$^{+0.67}_{-0.66}$ & 13.24$^{+0.13}_{-0.16}$ \\
HerMES J000742.7+051438 & HELMS\_RED\_350 & 49.7 $\pm$ 5.7 & 59.4 $\pm$ 5.8 & 65.1 $\pm$ 6.8 & 12.4 $\pm$ 3.3 & 3.13$^{+0.64}_{-0.63}$ & 13.57$^{+0.12}_{-0.17}$ \\
HerMES J002223.9+025047 & HELMS\_RED\_353 & 31.6 $\pm$ 6.0 & 46.7 $\pm$ 6.5 & 65.8 $\pm$ 7.2 & 24.1 $\pm$ 4.1 & 4.15$^{+0.75}_{-0.69}$ & 13.25$^{+0.11}_{-0.13}$ \\
HerMES J003446.0+045549 & HELMS\_RED\_368 & 44.6 $\pm$ 6.4 & 55.4 $\pm$ 6.2 & 74.0 $\pm$ 7.5 & 24.4 $\pm$ 3.7 & 3.71$^{+0.69}_{-0.70}$ & 13.44$^{+0.12}_{-0.15}$ \\
HerMES J234723.5$-$015213 & HELMS\_RED\_369 & 41.7 $\pm$ 5.5 & 49.1 $\pm$ 5.8 & 63.4 $\pm$ 7.1 & 18.0 $\pm$ 4.7 & 3.47$^{+0.72}_{-0.75}$ & 13.44$^{+0.14}_{-0.18}$ \\
HerMES J003931.4+014822 & HELMS\_RED\_373 & 53.6 $\pm$ 5.9 & 70.6 $\pm$ 6.2 & 79.7 $\pm$ 7.5 & 18.5 $\pm$ 3.7 & 3.42$^{+0.61}_{-0.58}$ & 13.33$^{+0.12}_{-0.13}$ \\
HerMES J005016.4+055923 & HELMS\_RED\_377 & 51.5 $\pm$ 5.6 & 61.8 $\pm$ 6.5 & 71.0 $\pm$ 7.2 & 24.1 $\pm$ 4.7 & 3.43$^{+0.67}_{-0.71}$ & 13.39$^{+0.12}_{-0.18}$ \\
HerMES J233755.0$-$051000 & HELMS\_RED\_379 & 60.3 $\pm$ 6.5 & 82.0 $\pm$ 6.3 & 85.3 $\pm$ 7.8 & 19.6 $\pm$ 4.9 & 3.40$^{+0.59}_{-0.58}$ & 13.42$^{+0.11}_{-0.14}$ \\
HerMES J232933.2+003149 & HELMS\_RED\_385 & 26.7 $\pm$ 5.7 & 34.8 $\pm$ 6.1 & 68.7 $\pm$ 7.5 & 15.6 $\pm$ 4.5 & 4.11$^{+0.81}_{-0.73}$ & 13.44$^{+0.12}_{-0.13}$ \\
HerMES J232101.9$-$033260 & HELMS\_RED\_387 & 37.5 $\pm$ 5.9 & 49.5 $\pm$ 5.9 & 64.6 $\pm$ 6.9 & 11.5 $\pm$ 4.8 & 3.56$^{+0.69}_{-0.67}$ & 13.34$^{+0.12}_{-0.14}$ \\
HerMES J001016.5$-$032131 & HELMS\_RED\_389 & 40.8 $\pm$ 6.1 & 61.3 $\pm$ 6.0 & 65.0 $\pm$ 7.2 & 20.0 $\pm$ 3.7 & 3.68$^{+0.65}_{-0.64}$ & 13.28$^{+0.11}_{-0.14}$ \\
HerMES J235712.3+022917 & HELMS\_RED\_390 & 47.0 $\pm$ 6.3 & 55.0 $\pm$ 6.3 & 68.0 $\pm$ 7.9 & 19.5 $\pm$ 5.1 & 3.42$^{+0.72}_{-0.77}$ & 13.39$^{+0.14}_{-0.18}$ \\
HerMES J001251.7+061210 & HELMS\_RED\_399 & 78.2 $\pm$ 5.8 & 79.6 $\pm$ 6.4 & 81.8 $\pm$ 6.6 & 23.3 $\pm$ 4.4 & 2.85$^{+0.67}_{-0.75}$ & 13.36$^{+0.15}_{-0.23}$ \\
HerMES J234602.1+001736 & HELMS\_RED\_402 & 46.9 $\pm$ 6.6 & 48.1 $\pm$ 6.2 & 64.6 $\pm$ 7.2 & 18.9 $\pm$ 3.4 & 3.25$^{+0.76}_{-0.83}$ & 13.41$^{+0.14}_{-0.22}$ \\
HerMES J010735.0+032259 & HELMS\_RED\_403 & 49.1 $\pm$ 6.7 & 52.8 $\pm$ 6.1 & 64.1 $\pm$ 7.7 & 15.0 $\pm$ 5.4 & 3.14$^{+0.77}_{-0.86}$ & 13.31$^{+0.15}_{-0.23}$ \\
HerMES J234612.3$-$054812 & HELMS\_RED\_405 & 23.0 $\pm$ 5.5 & 47.6 $\pm$ 6.0 & 64.2 $\pm$ 8.1 & 8.7 $\pm$ 4.6 & 3.91$^{+0.68}_{-0.75}$ & 13.28$^{+0.11}_{-0.15}$ \\
HerMES J004414.7+002550 & HELMS\_RED\_420 & 27.4 $\pm$ 6.0 & 55.6 $\pm$ 5.9 & 68.5 $\pm$ 7.6 & 16.2 $\pm$ 3.6 & 3.99$^{+0.64}_{-0.70}$ & 13.27$^{+0.10}_{-0.13}$ \\
HerMES J000127.4$-$010614 & HELMS\_RED\_421 & 65.4 $\pm$ 6.3 & 69.1 $\pm$ 5.9 & 72.5 $\pm$ 7.3 & 18.5 $\pm$ 4.8 & 2.93$^{+0.70}_{-0.79}$ & 13.37$^{+0.15}_{-0.23}$ \\
HerMES J004055.2+021131 & HELMS\_RED\_423 & 45.7 $\pm$ 6.1 & 57.5 $\pm$ 5.9 & 66.0 $\pm$ 7.1 & 22.1 $\pm$ 3.5 & 3.52$^{+0.69}_{-0.71}$ & 13.34$^{+0.13}_{-0.16}$ \\
  \end{tabular}
  \caption{}
  \label{tab:cm_v3}
\end{table*}

\addtocounter{table}{-1}

\begin{table*}
  \begin{tabular}{llrrrrcr}
   Source Name&name & $S_{250}$ & $S_{350}$ & $S_{500}$& $S_{850}$ & phot-$z$ &$\log (L_{FIR}/\rm{L}_\odot)$ \\ 
   &  &[mJy]&[mJy]&[mJy]&[mJy]\\ \hline
HerMES J003136.0$-$011856 & HELMS\_RED\_428 & 39.9 $\pm$ 6.0 & 52.6 $\pm$ 6.0 & 64.5 $\pm$ 6.4 & 14.9 $\pm$ 4.1 & 3.54$^{+0.68}_{-0.65}$ & 13.39$^{+0.12}_{-0.14}$ \\
HerMES J002414.5+035239 & HELMS\_RED\_430 & 47.8 $\pm$ 6.1 & 52.1 $\pm$ 6.3 & 64.9 $\pm$ 7.2 & 12.1 $\pm$ 3.8 & 3.09$^{+0.72}_{-0.77}$ & 13.32$^{+0.14}_{-0.22}$ \\
HerMES J233857.1$-$034441 & HELMS\_RED\_434 & 59.2 $\pm$ 5.6 & 65.1 $\pm$ 5.8 & 65.7 $\pm$ 7.0 & 16.4 $\pm$ 4.4 & 2.97$^{+0.69}_{-0.76}$ & 13.23$^{+0.15}_{-0.22}$ \\
HerMES J002000.9$-$060219 & HELMS\_RED\_440 & 49.4 $\pm$ 6.2 & 71.7 $\pm$ 5.8 & 76.9 $\pm$ 7.2 & 10.0 $\pm$ 4.3 & 3.29$^{+0.59}_{-0.60}$ & 13.30$^{+0.12}_{-0.14}$ \\
HerMES J232909.1+003450 & HELMS\_RED\_441 & 38.0 $\pm$ 6.4 & 50.6 $\pm$ 5.9 & 76.0 $\pm$ 7.9 & 20.2 $\pm$ 3.3 & 3.83$^{+0.71}_{-0.66}$ & 13.32$^{+0.11}_{-0.14}$ \\
HerMES J232249.3$-$024437 & HELMS\_RED\_443 & 52.1 $\pm$ 5.9 & 54.8 $\pm$ 6.3 & 66.9 $\pm$ 7.4 & 21.0 $\pm$ 4.7 & 3.17$^{+0.73}_{-0.78}$ & 13.40$^{+0.15}_{-0.20}$ \\
HerMES J002824.0$-$013329 & HELMS\_RED\_447 & 47.2 $\pm$ 5.5 & 60.1 $\pm$ 5.8 & 69.5 $\pm$ 6.6 & 14.0 $\pm$ 3.9 & 3.34$^{+0.63}_{-0.61}$ & 13.34$^{+0.11}_{-0.14}$ \\
HerMES J000823.4+012423 & HELMS\_RED\_448 & 43.7 $\pm$ 6.2 & 58.1 $\pm$ 6.7 & 63.9 $\pm$ 6.6 & 12.5 $\pm$ 3.3 & 3.32$^{+0.63}_{-0.62}$ & 13.31$^{+0.13}_{-0.14}$ \\
HerMES J234547.9$-$054412 & HELMS\_RED\_449 & 52.0 $\pm$ 6.7 & 59.4 $\pm$ 6.2 & 70.7 $\pm$ 7.7 & 19.4 $\pm$ 5.2 & 3.30$^{+0.73}_{-0.78}$ & 13.26$^{+0.14}_{-0.20}$ \\
HerMES J232619.7$-$050855 & HELMS\_RED\_451 & 49.2 $\pm$ 6.5 & 61.3 $\pm$ 6.4 & 65.3 $\pm$ 8.1 & 20.0 $\pm$ 4.6 & 3.36$^{+0.70}_{-0.73}$ & 13.37$^{+0.13}_{-0.17}$ \\
HerMES J003753.1+050029 & HELMS\_RED\_452 & 47.5 $\pm$ 7.1 & 61.5 $\pm$ 6.0 & 68.2 $\pm$ 7.2 & 12.1 $\pm$ 3.4 & 3.22$^{+0.63}_{-0.63}$ & 13.36$^{+0.13}_{-0.15}$ \\
HerMES J233351.1$-$035745 & HELMS\_RED\_458 & 54.9 $\pm$ 6.2 & 61.3 $\pm$ 6.3 & 79.6 $\pm$ 7.1 & 8.0 $\pm$ 4.7 & 3.09$^{+0.64}_{-0.63}$ & 13.26$^{+0.13}_{-0.17}$ \\
HerMES J001638.5+042328 & HELMS\_RED\_463 & 31.8 $\pm$ 6.2 & 46.5 $\pm$ 6.3 & 63.5 $\pm$ 6.8 & 12.0 $\pm$ 3.7 & 3.68$^{+0.69}_{-0.68}$ & 13.27$^{+0.11}_{-0.15}$ \\
HerMES J010438.2+002613 & HELMS\_RED\_472 & 42.4 $\pm$ 6.0 & 55.8 $\pm$ 6.3 & 64.7 $\pm$ 7.1 & 16.9 $\pm$ 3.8 & 3.51$^{+0.68}_{-0.66}$ & 13.27$^{+0.13}_{-0.15}$ \\
  \end{tabular}
  \caption{}
  \label{tab:cm_v4}
\end{table*}

\newpage
\textit{\\
{\small
$^{1}$Astronomy Centre, Department of Physcis and Astronomy, University of Sussex, Brighton BN1 9QH\\
$^{2}$Astrophysics Group, Imperial College, Blackett Laboratory, Prince Consort Road, London SW7 2AZ, UK\\
$^{3}$Department of Astronomy, Cornell University, Space Sciences Building,Ithaca, NY 14853, USA\\
$^{4}$Aix-Marseille Universite, CNRS, LAM (Laboratoire d’ Astrophysique de Marseille) UMR 7326, F-13388 Marseille, France\\
$^{5}$Department of Physics and Atmospheric Science, Dalhousie University,Halifax NS B3H 4R2, Canada\\
$^{6}$Dept. of Physics \& Astronomy, University of California, Irvine, CA 92697, USA\\
$^{7}$Centre for Astrophysics Research, University of Hertfordshire, Hatfield, AL10 9AB, UK\\
$^{8}$Instituto de Astrofísica de Canarias (IAC), E-38205 La Laguna, Tenerife, Spain\\
$^{9}$Universidad de La Laguna, Dpto. Astrofísica, E-38206 La Laguna, Tenerife, Spain\\
$^{10}$INAF-Osservatorio Astronomico di Padova, Vicolo dell Osservatorio 5, I-35122 Padova, Italy \\
$^{11}$Institute for Astronomy, University of Edinburgh, Royal Observatory, Edinburgh EH9 3HJ, UK\\
$^{12}$School of Physics \& Astronomy, Cardiff University, Queen's Buildings, The Parade, Cardiff CF24 3AA, UK\\
$^{13}$School of Sciences, European University Cyprus, Diogenes Street, Engomi, 1516 Nicosia, Cyprus\\
$^{14}$Department of Physics, Virginia Tech, Blacksburg, VA 24061, USA\\
$^{15}$UK Astronomy Technology Centre, Royal Observatory,Blackford Hill, Edinburgh, EH9 3HJ, UK\\
$^{16}$European Southern Observatory, Karl-Schwarzschild-Stra{\ss}e 2, D-85748 Garching, Germany\\
$^{17}$Department of Astronomy, University of Cape Town, Private Bag X3, Rondebosch, 7701, South Africa\\
$^{18}$Department of Physics and Astronomy, University of the Western Cape, Robert Sobukwe Road, 7535 Bellville, Cape Town, South Africa \\
$^{19}$INAF - Istituto di Radioastronomia, via Gobetti 101, 40129 Bologna, Italy\\
$^{20}$Harvard-Smithsonian Center for Astrophysics, Cambridge, MA 02138, UK\\
$^{21}$Department of Physics and Astronomy, University of British Columbia, 6224 Agricultural Road, Vancouver, BC V6T-1Z1, Canada\\
$^{22}$Mullard Space Science Laboratory, University College London, Holmbury St. Mary, Dorking, Surrey RH5 6NT, UK\\
$^{23}$Department of Astronomy and Department of Physics, University of Illinois, 1002 West Green St., Urbana, IL 61801\\
$^{24}$SRON Netherlands Institute for Space Research, Landleven 12, 9747 AD, Groningen, The Netherlands\\
$^{25}$Kapteyn Astronomical Institute, University of Groningen, Postbus 800, 9700 AV, Groningen, The Netherlands\\
$^{26}$Centre for Extragalactic Astronomy, Department of Physics, Durham University,South Road, Durham DH1 3LE, UK\\
$^{27}$Center for Detectors, School of Physics and Astronomy, Rochester Institute of Technology, Rochester, NY 14623, USA
}}
\label{lastpage}

\end{document}